\documentclass[oneside,reqno,12pt,notitlepage]{amsproc}

\usepackage{color}
\newtheorem{thm}{\bf Theorem}[section]
\newtheorem{prop}[thm]{\bf Proposition}
\newtheorem{corol}[thm]{\bf Corollary}

\newtheorem{lemma}[thm]{\bf Lemma}

\numberwithin{equation}{section}

\def\data{\number\day/\ifcase\month\or gennaio \or febbraio \or marzo \or
aprile \or maggio \or giugno \or luglio \or agosto \or
settembre \or ottobre \or novembre \or dicembre
\fi/\number\year}

\newcommand{\notes}[1]{}

\newcount\pgn
\def\foglio{\veroparagrafo:\number\pgn
\global\advance\pgn by 1}

\usepackage{latexsym}

\usepackage{amsfonts,amssymb}

\usepackage{enumerate}

\usepackage{a4wide}

\newcommand{\red}{\textcolor{red}}

    \definecolor{turck}{rgb}{0,0.5,0.5}

\newcommand{\blue}[1]{\textcolor{blue}{#1}}

    \definecolor{greena}{rgb}{0.2,0.8,0}
\definecolor{greenb}{rgb}{0,0.5,0.2}
\definecolor{greenc}{rgb}{0.1,0.7,0.3}
\definecolor{greend}{cmyk}{0.7,0,0.3,0}
\definecolor{greene}{cmyk}{0.7,0,0.3,0.1}

               \newcommand{\tobias}[2]{{#1}}
	       \newcommand{\tobi}[1]{#1}
	       \newcommand{\tobii}[2]{#2}
	       \newcommand{\tobbi}[2]{#1}

\newcommand{\sumtwo}[2]{\sum_{\substack{#1 \\ #2}}} 
\newcommand{\inttwo}[2]{\int_{\substack{#1 \\ #2}}}     

\newcommand{\dis}{\displaystyle}

       \newcommand{\q}{\varpi}
       \newcommand{\Q}{\Upsilon}
\newcommand{\eps}{\epsilon}

\newcommand{\La}{\Lambda}
\newcommand{\la}{\lambda}
\newcommand{\ga}{\gamma}

\newcommand{\Ga}{\Gamma}

\newcommand{\om}{\omega}
\newcommand{\si}{\sigma}
\newcommand{\Si}{\Sigma}

\newcommand{\und}{\underline}

\newcommand{\Rr}{\mathbb{R}}

\newcommand{\ssp}{\mathrm{sp}} 

\newcommand{\ba}{\begin{array}}
\newcommand{\ea}{\end{array}}
\newcommand{\bq}{\begin{eqnarray}}
\newcommand{\eq}{\end{eqnarray}}
\newcommand{\bqw}{\begin{eqnarray*}}
\newcommand{\eqw}{\end{eqnarray*}}
\title[Quantum phase transitions]{A liquid vapor  phase transition in quantum
statistical mechanics}

\author{F. Baffioni}
\address{Francesco Baffioni,
    Liceo Cannizzaro, Viale Civiltà del Lavoro 2/d, 
00144 Roma, Italy}
\email{francesco.baffioni@lcannizzaro.edu.it} 
\thanks{}

\author{T. Kuna}
\address{Tobias Kuna,
DISIM, Universit\`a dell'Aquila \newline \parbox{0cm}{}  \hspace{3cm} and Fondazione Gran Sasso Tech, 67100 L'Aquila, Italy}
\email{tobias.kuna@univaq.it}
 \thanks{}

\author{I. Merola}
\address{Immacolata Merola,
DISIM, Universit\`a dell'Aquila
, 67100 L'Aquila, Italy}
\email{merola@univaq.it}
\thanks{}

\author{E. Presutti$^\dagger$ \footnote{$\dagger$(posthumous, deceased Dec. 2024)}}
\address{Errico Presutti,
GSSI, 67100 L'Aquila, Italy}
\thanks{Research partially supported by
MURST, NATO Grant PST.CLG.976552, NSF Grant DMR 01-279-26, DAAD PostDoc and
A.v. Humboldt Feodor-Lynen fellowship.}

\begin{document}

\begin{abstract}  We prove a liquid vapor phase transition for a quantum system of particles
 in the continuum.  The system is the quantum version with Boltzmann statistics
  of the point particles
 model introduced in \cite{LMP} by Lebowitz, Mazel and Presutti.

 \end{abstract}

\maketitle

 \vskip .5cm \noindent


\pagenumbering{arabic}    

\setcounter{equation}{0}

\section{{\bf Introduction}}
       \label{sec:a1}

A mathematically rigorous proof of the existence of phase transitions
for particle systems in $\mathbb R^d$ is a  fundamental and essential problem
in equilibrium statistical mechanics, but despite its importance very
few progress has  been  made  so
far. The purpose of this paper is to
prove that the quantum version of the so-called  ``LMP
model'', described in more details below, has a phase
transition as well.  We start the introduction with a short
survey on phase transitions in the continuum, first in the classical and then in the quantum frameworks.

The first  result  was obtained by D.~Ruelle, \cite{ruelleWR},
who proved that the two component Widom-Rowlinson model
has a phase transition. As the density increases the system undergoes
a transition  from a disordered  state (where the two components are equally mixed)
to ordered states (where a component has  density higher than the
other). This was done adapting Peierls' argument developed in the analysis of
the Ising spin system.

The first proof of phase transitions for identical particles
is  by  K.~Johansson, \cite{johansson}, who  has  studied a system of
hard rods in one dimension with long range attractive pair potential,
i.e. with  infinite second moment. In this case, the order parameter is the density,  
and at the transition there is coexistence between a low and a
high density state (the vapor and the liquid phase). For the lattice much more is known, see e.g. \cite{FV}.
To extend the techniques used in the lattice case, like spin flip
symmetry, stochastic inequalities the FK representation, and the
connection with percolation problems, has in general had only limited success. 

The existence of a vapor-solid phase transition for Lennard-Jones like
potentials in $d=2$ dimensions was proved by F.~Theil, \cite{Theil},
at zero temperature. In \cite{DC26} a first-order phase transition in an extremal regime 
was proven for a class of potentials excluding usual pair potentials but including the area model among others.

The theory of the liquid-vapor transition goes back to van der Waals;
a derivation in the context of statistical mechanics  was first
addressed by M.~Kac, G.~Uhlenbeck and P. C.~Hemmer, see
\cite{KUH1}, \cite{KUH2}, \cite{KUH3}, and then
J.~Lebowitz and O.~Penrose, \cite{LP}, who reframed the
analysis 
by introducing  coarse graining techniques and  by  relating the
analysis to variational problems for non local free energy
functionals.

The original van der Waals phase transition was rigorously
derived in this way, but the result involved a scaling limit where
the ``Kac scaling parameter'' $\ga$  tends to zero.
More precisely, the attractive part of the
interaction is given by a  Kac interaction i.e. a
potential that has range proportional to $\ga^{-1}$ and intensity
proportional to $\ga^d$ ($d$ the space dimensions),
so that the interaction energy of each
particle with the others remains bounded.
In the aforementioned papers
of M.~Kac et al. and \cite{LP}
the model differs from the LMP as the particles
obey a hard ball  repulsion. As  Kac interactions  have no limit as
$\ga\to 0$  the above results, in fact, do not prove that there is a well
defined interacting particle system which has a phase
transition. They only show that the phase diagram of a system with $\ga$ small is
close to one with a phase transition. It may very well be, however, that
for each $\ga>0$ the phase diagram does not show any phase
transition. This observation is not just of technical nature, because
for $d=1$ it is known by general results that the system for finite
$\ga$ has no phase transition.

The problem of realizing the Kac program without taking the limit
$\ga\to 0$ was first solved by J.L.~Lebowitz, A.~Mazel and
E.~Presutti, \cite{LMP}. In the following we will refer to
  the model considered in this paper as ``the LMP
model'' which is a system
of point particles in $\mathbb R^d$, $d\ge 2$, which interact via two
and four body  Kac potentials. The two body interaction is attractive
(as in  aforementioned papers by M.~Kac and al.) while the  four body
interaction is repulsive thus ensuring stability, which in the
original Kac model was provided by the hard core condition.

The proof is based on a variation of the Pirogov-Sinai
scheme, \cite{PS1}, \cite{PS2}, \cite{zahr84}, perturbing mean field
instead of  ground states and with
$\ga$ as small parameter rather than small temperature. A
significant  complication in comparison to the Pirogov-Sinai theory for
lattice systems comes from the complexity of the so called
``restricted ensemble'' which on the lattice  is just  a gas of
polymers with small activity while in the continuum  it is a gas of
polymers coupled to a gas of particles with constraints on their
empirical density  which in turn depend on the polymers. J.~Bricmont, K.~Kuroda and J.L.~Lebowitz,
\cite{BKL1} and \cite{BKL2}, had previously shown
that the Pirogov-Sinai scheme can be used if conditions  are fulfilled
sufficient  for cluster expansion  to hold  in the restricted
ensemble.  In the LMP model such conditions  have not been proved, but
instead it is possible to show   that a sort of Dobrushin uniqueness
condition holds, \cite{LMP}, which  allows to carry through the
Pirogov-Sinai scheme.
 
Further properties of the phase transition have been studied. An
analysis of the curve  of coexistence of phases  in the LMP model has
been obtained in  \cite{BMPZ} while in Chapters 10--12 of \cite{errico} a
more detailed analysis of the structure of the DLR measures in
  the case of coexistence can be found.

The same approach has been used by   A.~De Masi, I.~Merola,
E.~Presutti and Y.~Vignaud, \cite{DMPV08}, \cite{DMPV09}, to study
the Potts model in $\mathbb R^d$, $d\ge 2$ with Kac
potentials. In more details,  the
model consists of point particles where each particle
posses a spin with $k\ge 2$ values,
particles with different spins repel each other via a Kac
potential. The main result of that paper is that
if $k\ge 3$ at the critical point there
are at least $k+1$  co-existing extremal DLR states,
$k$ of them are ``ordered'' and one disordered (ordered meaning
in this context that there is one spin with a higher
  density than the others).
Besides its intrinsic interest the result is of technical relevance,
because the proof shows that the Dobrushin uniqueness in the
restricted ensemble, used in the proof, may be relaxed by
weaker conditions in the spirit of the  Dobrushin-Shlosman approach to uniqueness.
For other works on LMP like models see \cite{errico}. For particle systems in the continuum, let us mention that in \cite{ElenaThesis} and  \cite{ElenaDimitris16} on top of the LMP interaction a usual repulsive pair potential is added acting at a length scale of order one, i.e. $\gamma^0$, but the strength and the radius of the hard core are sufficiently small. This requires a coarse graining in the canonical ensemble, see  \cite{ElenaDimitris12}.   In \cite{Jauslin} a variant of the LMP has been considered, where the repulsive part of the interaction is replaced by a usual pair potential but the model is boxed, that is, on a partition of the scale of the weak long-range interaction, i.e. $\gamma^{-1}$ all cross box interaction of the short range repulsive potential are neglected. 

The situation gets richer in the quantum case
where besides the Boltzmann statistics (the  direct  analogue of the
classical  case) one has as well the Fermi-Dirac  as
the Bose-Einstein statistics which give rise to completely new effects.

Fermions have been much studied in a
field theory  context.
The Bose-Einstein condensation is well understood for free particles
and in mean field, see e.g. \cite{BCMP}. The interacting case at zero
temperature in the low density regime has been much studied in the recent years,
see for instance \cite{LSY}.\notes{Check that in the lecture notes of Cetraro}
However, outside of the superfluid
regime, bosons should undergo also ``normal'' phase transitions like
a vapor-liquid transition as in the classical case. For such type of
phase transitions  the Bose statistics  should not have great
relevance and  therefore  from now on we shall restrict to the
Boltzmann statistics.

The quantum extension of the Ruelle's proof of phase transitions for
the Widom-Rowlinson model has been carried out by
M.~Cassandro and P.~Picco, \cite{casspicco}, and using the FK
representation by, J.~Chayes, L.~Chayes and  R.~Koteck\'y, \cite{CCK},
and by D.~Ioffe, \cite{ioffe}.

In this paper we shall consider the quantum version of the LMP
model with Boltzmann statistics.
We shall use the Ginibre representation, \cite{ginibre},
(as in the above papers on the quantum Widom-Rowlinson model) which
allows for an essentially classical
description of the model. In this representation, point
particles are replaced by Brownian loops. For a system with in the Bose statistics
several particles may join their loops leading to Brownian bridge with larger time parameters which will give higher probability to longer
  loops; leading even to loops with infinite time in the regime of Bose
  condensation. To avoid all that we restrict ourself to the Boltzmann statistics. 
  As the parameter
regime we consider is far away from Bose condensation the result
should be virtually the same. However, it may or may not be technically
non-trivial to extend the results to Bose-statistics.
Also in the Maxwell-Boltzmann statistics loops may be arbitrarily long 
(with exponentially small probability, though) as in the Bose statistic.
Hence in both cases the quantum version of LMP has to deal with interactions which have 
effectively infinite range due to the
extended loops which may bring distant particles within interaction range.

The Pirogov-Sinai scheme (which is the central technique in the analysis
of the classical LMP model) consists mainly out of two steps.
Before that, the art is to introduce a suitable notion of
contours. This is where quantum effects become most relevant and 
difficult to handle. Our contours in
fact must take into account both local deviations of the density from the
liquid and vapor densities (as in LMP), but also the
occurrence of exceptionally extended loops. Contours should not interact which each other
in our approach. 

The combination of the two mechanisms could in principle
produce dangerous domino effects. Extended loops may connect
separated regions of local deviations which then for technical
reasons have to be considered as one contour. The heart of our proof
is to exclude that this effect leads to a break-down of the
  technique. Extended  loops, in the
end, will give the largest contribution to the weight of
the contours.

With the notion of contours established, the
first stage of the P-S scheme is the reduction to
restricted ensembles. It is at this point that  the correct
value of the chemical potential for phase coexistence is
found by equalizing the pressures in the plus and minus
restricted ensembles. Beside this 
we also prove some free energy estimates for all densities, which constitute
the main term in the Peierls' bounds leading to the
  estimate of the probability of a contour. They are derived by
solving variational problems for the non local, free energy
functional of the Lebowitz-Penrose limit.

The second stage (of the P-S scheme) involves the swap of a bubble
from one restricted ensemble to the other. To estimate the effect of
this swap one has to analyse
finite volume surface corrections to the pressure. The bulk
  terms coincide due to the choice of the chemical potential.
Typically in the Pirogov-Sinai techniques this is proven by
cluster expansion, but, as in the classical LMP model, we
do not know whether a
cluster expansion is valid in the range of parameters we
are using. In LMP, cluster expansion has been replaced by
Dobrushin uniqueness technique and here we follow the same approach,
but with a quite different proof. The reduction to
restricted ensembles effectively places the problem in a 
phase-transition free setting, such that one can establish that
the (high temperature) Dobrushin uniqueness condition is satisfied,
but not for all boundary conditions.  The ones for which the
  condition is violated,
however, have small probability, hence the necessity to develop a
relativized Dobrushin uniqueness theorem. In a companion
paper by the same authors, \cite{BKMP2}, the issue is
discussed in a more general setting and a uniqueness
criterion is first established and then its usefulness is demonstrated
by checking the criterion in a
variety of models including the present one. With the help
of this result, the surface corrections to the pressure are
estimated and the Peierls' bounds proved.  Phase transition
is then established  in the usual way. These bounds are also
strong enough to establish the quantum states in the thermodynamic limit.
In itself a non-trivial problem.

The paper is organized as follows. In Section~\ref{sec:a2},
we define the model and state the main theorems, Theorem~\ref{thm2.1}
and Theorem~\ref{thm2.1q}.
In Section~\ref{sec:a3}, we introduce the
fundamental notion of contours and  in Section \ref{sec:a4}
we rewrite the $\pm$, diluted (in the sense of Pirogov-Sinai)
partition functions in terms of contours, introducing then
the $\pm$ restricted ensembles for cutoff contour weights.
In Section \ref{sec:b4} we choose the value of the chemical
potential, by equating the pressures of the plus and of the
minus ensembles. In Section \ref{sec:a6}, we divide the
proof of the Peierls bounds in a sequence of intermediate
estimates which are then proved in the remaining sections
and in the appendices.

In Section~\ref{sec:11}, we collect a few consequences
of the Peierls' estimates remaining
steps to conclude the proof of the main theorems and in particular construct two quantum states.

\setcounter{equation}{0}

\section{{\bf Model and main results}}
       \label{sec:a2}

\subsection{Classical model}

The phase space of a system of identical point particles in classical statistical
mechanics, denoted here by $Q$, is the set of all particle configurations. A particle
configuration is either a sequence or a vector $q=(q_i)_{i \in I}$ with $q_i\in
\mathbb{R}^d$ (for $I$ a finite or a countably infinite index set), such that in any
bounded subset of $\mathbb{R}^d$ one finds only
finitely many of the $q_i$.  The labeling of the elements in
the sequence is of no physical significance and hence physical observables are
represented by symmetric functions.
Accordingly, a configuration with $n$ elements has a statistical
weight $1/n!$, see \eqref{2.a1} below.

Let $\La$ be a measurable subset of $\mathbb{R}^d$. We will denote by
$\tobi{q_\Lambda:=} q\cap \La$ the configuration of $q$
obtained by discarding  the elements of $q$ which are not
in $\La$; we denote by $|q|$ the cardinality of $q$, therefore
%
%
  $|q\cap \La|$ is the number
  of particles in the configuration $q$ which are in $\La$.
We denote by 
 $Q_{\text{\rm
fin}}:=\{ q\in Q: |q|<\infty\}$, by $Q^\La:= \{ q  \in Q : q =
q\cap \La \}$, and by $Q^\La_n:=\{q
\in Q^\La : |q|=n\}$, for $n$ a non negative integer.
Obviously, one has that $Q^\La= \bigcup_{n\geq 0} Q^\La_n$, whenever $\La$ is bounded.

In the model considered here, the LMP model, the energy of a configuration $q\in
Q_{\text{\rm fin}}$ is given by
      \begin{equation}
          \label{2.a2}
h_{\ga,\la}(q) := \int_{\mathbb{R^d}}  e_\la\left(j_\ga *
q(r)\right)dr,
    \end{equation}
where $\ga>0$ is the so-called Kac scaling parameter and
     \begin{equation}
      \label{2.a3}
j_\ga * q(r) := \sum_{q_i\in q}j_\ga (r,q_i),\qquad j_\ga
(r,r'):= \ga^d j(\ga r, \ga r')
     \end{equation}
with $j$ a bounded, symmetric, translation-invariant
probability kernel on $\mathbb{R}^d$ such that $j(0,\cdot)$ is
supported within the unit ball, and $j$ fulfils some further  minimal regularity
properties  described in Subsection \ref{subsec:appd0.1} of
Appendix \ref{app:d0}. The chemical
potential is denoted by $\la \in \mathbb{R}$ and the function $e_\la: [0, \infty) \rightarrow \mathbb{R}$ is given
by
     \begin{equation}
     \label{2.a4}
      e_\la(x):= -\la x - \frac{x^2}{2!} + \frac{x^4}{4!}.
     \end{equation}
This particular form of the energy give rise to a convenient
expression for the mean-field functional, cf. \cite{errico}. The energy $h_{\ga,\la}(q)$ of a configuration
 defined in \eqref{2.a2} can be rewritten as a sum of contributions by one up to four body
interactions. Indeed, defining
\begin{eqnarray*}
\label{la}
\la(q_i)&:=&\la q_i+\int dr \left( - \tobi{ \frac{1}{2!} } j^2_\ga(r,q_i)+   \tobi{ \frac{1}{4!} } j^4_\ga(r,q_i) \right)
\\
      \label{V2}
V_2(q_i,q_j)&:=& \int dr \left( -   \tobi{ \frac{1}{2!} } j_\ga(r,q_i) j_\ga(r,q_j)+
\tobii{6}{\frac{1}{4}} j^2_\ga(r,q_i) j^2_\ga(r,q_j) + \tobii{4}{\frac{1}{3!}} j^3_\ga(r,q_i) j_\ga(r,q_j) \right)
\\
 \label{V3}
V_3(q_i,q_j,q_k)&:=&
\tobii{6}{ \frac{1}{4} }  \int dr j^2_\ga(r,q_i) j_\ga(r,q_k)j_\ga(r,q_j)
     \\
 \label{V4}
V_4(q_i,q_j,q_k,q_\ell)&:=& \tobi{ \frac{1}{4!} }  \int dr j_\ga(r,q_i) j_\ga(r,q_j)j_\ga(r,q_k)j_\ga(r,q_\ell),
     \end{eqnarray*}
we can write  $h_{\ga,\la}$ for $q=(q_i)_{i\in I}$ as: 
\begin{eqnarray}
\label{V}
h_{\ga,\la}(q):=\sum_{i\in I}\la(q_i)+\sum_{i,j \in I, i \neq
  j}V_2(q_i,q_j)+
\sum_{\stackrel{i,j,k\in I,}{\scriptstyle  i \neq j \neq k}}V_3(q_i,q_j,q_k)+\sum_{\stackrel{i,j,k,\ell\in
  I,}{\scriptstyle i\neq j \neq k \neq \ell} }V_4(q_i,q_j,q_k,q_\ell) .
\end{eqnarray}
The conditional energy of the configuration $q$ given
$\bar{q} $ for $q,\bar q\in Q_{\text{\rm fin}}$  is defined as 
   \begin{equation}
        \label{2.a5}
h_{\ga,\la}(q|\bar{q}):=
h_{\ga,\la}(q,\bar{q})-h_{\ga,\la}(\bar{q}),
    \end{equation}
where by $(q,\bar{q})$ we denote the configuration which collects all the
particles of $q$ and $ \bar{q}$.

The free measure  $\nu_\La$ of the model for  $\La$ a bounded measurable
set, is the Lebesgue-Poisson measure on $Q^\La$, defined for bounded measurable
functions $G:Q^\La \rightarrow {\mathbb R}$ by
     \begin{equation}
     \label{2.a1}
\int_{Q^\La} G(q) \nu_\La(dq) := \sum_{n=0}^\infty
\frac{1}{n!} \int_{\La^n} G(q_1,\ldots,q_n)dq_1\ldots
dq_n.
      \end{equation}

\vskip 1cm

\subsection{Euclidean model\label{susecEuc}}

The model introduced in this subsection, the so-called ``Euclidean model'', is the main focus of the paper. It is analogous to the classical one considered in the previous section, but based on
configurations of loops rather than configurations of
particles.
In Subsection~\ref{susecAlg}, we will demonstrate that this Euclidean model is nothing else than the
Feynman-Kac representation of the quantum mechanical version of the
LMP model. This type of representation was introduced by J.~Ginibre, see
e.g. \cite{ginibre2,ginibre}, the concept of Euclidean Gibbs measures was
introduce in \cite{KLRRS}.

A configuration of loops is a sequence $\q= \tobii{\{\q_i\}}{( \q_i)_{i \in I}}$,
of loops $\q_i$, that is a
continuous function $\q_i:[0,\beta]\rightarrow \mathbb{R}^d$
with $\q_i(0)=\q_i(\beta)$, and the configuration of the starting points fulfils  $\tobii{\{\q_i\}(0)}{( \q_i(0))_{i \in I}} \in Q$. For short, we will call them also loop configurations. We will denote by $\Q$ the
space of loop configurations, by $\Q_{\text{\rm fin}}$ we denote the
subset of $\Q$ consisting of all  loop configurations which consists of
finite many loops, by
$\Q_n$ we denote the subset configurations of $n$ loops. By $\Q^{\La}$ for $\La\subset
\mathbb{R}^d$ measurable, we denote the subset of all loop configurations whose
loops have starting
point in $\La$, by $\Q^\La_n$ we denote the subset of $\Q^\La$ consisting of all
configurations which contain exactly
$n$ loops. By $\Q(q_1, \ldots ,q_m ; q'_1, \ldots q'_m)$, for
$q_1, \ldots ,q_m , q'_1, \ldots q'_m \in \mathbb{R}^d$, we denote the
space of all configurations consisting of $m$ continuous paths $\q=\tobii{\{\q_i\}}{(\q_i)_{i\in \{ 1, \ldots, m\} }}$ (not necessarily loops)
  such that for each $i$ the path
$\q_i$ has the property that $\q_i(0)=q_i$ and $\q_i(\beta)=q'_i$ for $i=1,
\ldots m$. 
For two configurations $\q$ and $\q'$, we denote by $(\q,\q')$ the
sequence which collects the loops of the two configurations $\q$  and
$\q'$. For measurable subsets $\Lambda' \subset \mathbb{R}^d$ with $\Lambda
\subset \Lambda'$, we denote by $\Q_{\Lambda'}(q_1, \ldots ,q_m ; q'_1, \ldots
q'_m)$, or $\Q^\Lambda_{\Lambda'}$ respectively, the subset of $\Q(q_1, \ldots ,q_m ; q'_1, \ldots q'_m)$ consisting of configurations whose paths stay inside of $\Lambda'$ for all $0\leq t \leq \beta$, the subset of $\Q^\Lambda$ consisting of configuration whose loops  stay inside of $\Lambda'$ for all $0\leq t \leq \beta$ respectively.

In the Euclidean model, the energy $H_{\beta,\ga,\la}$ is defined for configurations
$\q\in \Q_{\text{\rm fin}}$ and $\q' \in \Q^\La(q_1, \ldots ,q_m ;
q'_1, \ldots q'_m)$, by
\begin{equation}
     \label{2.a7}
H_{\beta,\ga,\la}(\q,\q'):=\int_0^\beta h_{\ga,\la}(\q(t),\q'(t))\, dt
\end{equation}
and the corresponding conditional energy for $\q\in \Q_{\text{\rm
    fin}}$, $\q' \in \Q^\La(q_1, \ldots ,q_m ; q'_1, \ldots q'_m)$ and
{$\bar{\q} \in \Q_{\text{\rm
    fin}}$} is
\begin{equation}
     \label{2.a8}
H_{\beta,\ga,\la}(\q,\q'|\bar\q):=H_{\beta,\ga,\la}(\q,\q',
\bar\q)-H_{\beta,\ga,\la}( \bar\q).
      \end{equation}

The free measure for this model is based on the conditional Wiener measure (Brownian bridge)
$W_{x|y}^\beta $ for
$x,y \in \mathbb{R}^d$, which is the probability measure on $\Q_1(x;y)$ 
, the space of continuous paths $[0, \beta]\rightarrow \mathbb{R}^d$ from $x$ to $y$, which is defined via the values it takes on cylinder sets. For the cylinder set
    \[
A:=\left\{ \q_1 \in \Q_1(x;y)
\left|\, \q_1(t_1)\in B_1,\ldots ,\q_1(t_n)\in B_n\right.
\right\}
    \]
(with $0<t_1<\ldots <t_n<\beta $ and $B_1,\ldots ,B_n$
measurable subsets of $\mathbb{R}^d)$ the conditional Wiener measure (Brownian bridge) takes by definition the values
     \begin{eqnarray}
{W_{x|y}^\beta (A)} &:=& (2\pi\beta)^{d/2}
\int_{B_1}\ldots
\int_{B_n} p_{t_1}(r_1-x)\cdot p_{t_2-t_1}(r_2-r_1) \nonumber\\
&&\cdot \ldots \cdot p_{t_n-t_{n-1}}(r_n-r_{n-1})\cdot
p_{\beta -t_n}(y-r_n) dr_1 \ldots dr_n,
     \label{2.a9}
     \end{eqnarray}
where
   \[
p_t(r):=\left( \frac 1{2\pi t}\right) ^{d/2}\exp \left(
-\frac{r^2}{2t} \right).
     \]
The associated Lebesgue-Poisson measure $\nu_{\beta,\La}(d\q)$ on
$\Q^\La$ for $\La$ a bounded measurable set, is defined for
any bounded measurable function $G:\Q^\La \rightarrow {\mathbb R}$ by
     \begin{eqnarray}
&&\int_{\Q^\La} G(\q) \nu_{\beta,\La}(d \q)  :=
\sum_{n=0}^\infty \frac{1}{n!} \int_{\La^n}\int_{(\Q_1)^n}
G(\q_1,\ldots ,\q_n)\nonumber \\&& \hskip2cm \times
W_{x_1|x_1}(d\q_1) \ldots W_{x_n|x_n}(d\q_n)dx_1\ldots
dx_n.
    \label{2.a10}
      \end{eqnarray}

\vskip1cm

\subsection{{\bf Gibbs and DLR measures}}
 \label{subsec:2.3}

In this subsection, we recall the definition of classical Gibbs measures and introduce measures on loop configurations in
direct analogy to classical Gibbs measures. Their relation with the usual
quantum states considered in quantum statistical mechanics are explained in Subsection~\ref{susecAlg}.

In  Appendix \ref{app:d0}, Subsection \ref{subsec:d0.2}, it
is proved that for any $q\in Q_{\text{\rm fin}}$ and $\bar
q \in Q$, one has
     \begin{equation}
         \label{2.a11}
h_{\ga,\la}(q|\bar{q}) \geq b \ |q|, \qquad b:= \inf_{x\geq
0} e'_\la(x).
       \end{equation}
       {The lower bound for $h_{\ga,\la}(q|\bar{q})$ in }
       \eqref{2.a11} is a stability property of the Hamiltonian, which is
much stronger than usual stability because of its
uniformity in the conditioning; this is a  very peculiar
feature of the LMP model, in many respects unessential in
\cite{LMP} and the present paper, but technically quite \tobii{important}{handy} in the proofs.

Let us first introduce the definition of Gibbs measures for a classical particle system. Recall that for $q \in Q$, we will denote by $q_\Delta$ the collection of particles of
$q$ which are in $\Delta$, where $\Delta$ is a bounded measurable subset
of $\mathbb{R}^d$. The Gibbs measure in the
bounded, measurable region $\La$ with boundary conditions
$\bar q \in \tobii{Q_{\La^c}}{Q}$ is the probability on $Q$ defined by
     \begin{equation}
     \label{2.a6}
 \mu_{\beta,\ga,\la,\La}(dq|\bar q) := \frac{e^{-\beta h_{\ga,\la}(q_\La|\bar q)}}
 {Z_{\beta,\ga,\la}(\La|\bar q)} \;\nu_\La(dq_\La) \delta_{\bar
 q_{\La^c}}(dq_{\La^c}),
      \end{equation}
where $\delta_q$ denotes the Delta measure at $q \in Q$.  The expression on
the right hand side of \eqref{2.a6} is well defined for any $\bar q
\in Q$ by \eqref{2.a11}.

We will now generalize the previous construction to configurations of
loops. In analogy, we denote by $\q_{\Delta}$ the configuration of
loops in the configuration $\q$ which have the starting point in
$\Delta$. Note that  $\q_{\Delta}$ is only a restriction of the starting point, the loop can extend throughout of all $\mathbb{R}^d$.
In the Euclidean model,   the Gibbs measure
in $\La$ with boundary conditions $\bar \q \in \Q$ is defined as
     \begin{equation}
     \label{2.a12}
 \mu_{\beta,\ga,\la,\tobias{\La}{}}(d\q|\bar \q) := \frac{e^{- H_{\beta,\ga,\la}(\q_\La|\bar \q)}}
 {Z_{\beta,\ga,\la,\La}(\La|\bar \q)} \;\nu_{\beta,\La}(d\q_\La)
 \ \delta_{\bar \q_{\La^c}}\!(d\q_{\La^c})
      \end{equation}
whenever $\bar \q \tobii{ \in \Q_{\La^c}}{}$ satisfies the property that in each ball
of $\mathbb R^d$ the number of particles in the configuration $\bar\q(t)$ is
bounded uniformly in $t$. By  \eqref{2.a11}, then the measure
  $\mu_{\beta,\ga,\la,\La}(d\q|\bar \q)$ in \eqref{2.a12}
is well defined.

A DLR measure $\mu$ for the classical, the Euclidean model respectively,  for given values of $\beta,\ga,\la$ is
a probability on $Q$ for the classical model, on $\Q$ for the Euclidean model respectively, such that, for any
bounded, measurable region $\La$, the conditional
probability of $\mu$ with respect to the $\si$-algebra
$\mathcal F_{\La^c}$, is $\mu$-almost surely equal to \eqref{2.a6},
 \eqref{2.a12} respectively.  In the classical model the
$\si$-algebra $\mathcal F_{\La^c}$ is defined as the $\si$-algebra
generated by all functions which depend on
$q_{\La^c}$, and in
the Euclidean model generated by all functions which depend on $\q_{\La^c}$ , i.e.\ on configuration consisting of loops which have starting points 
in $\La^c$. In the Euclidean case, one has to require additionally for consistency of the above
definition, that for any $n$ holds
     \begin{equation}
     \label{2.a13}
\mu\Big(\q: \sup_{0\le t\le \beta} |\q(t)\cap B_n|
<\infty\Big)=1,\qquad  B_n=\{r\in \mathbb R^d: |r|\leq n\}.
      \end{equation}
In the context of the DLR measures, one says that a system has a phase
transition at $\beta,\ga,\la$ if there exists at least two distinct DLR measures for these parameter values $\beta,\ga,\la$.

\subsection{Algebraic description and Quantum states\label{susecAlg}}

As already mentioned in Subsection~\ref{susecEuc} we will use in this paper an
functional integration (Euclidean) representation for quantum
statistical systems, which we call the Euclidean model.
The aim of this subsection is to describe how the DLR-measures in the
Euclidean model, cf. Subsection~\ref{subsec:2.3}, can be used to represent
quantum states in the standard
algebraic framework of quantum statistical mechanics and how phase transitions can be described. In contrast to Quantum Field Theory, the relation between the Euclidean
representation and the algebraic description is direct without any reference to an analytical
continuation. 
It is essential in order to use the Pirogov-Sinai-scheme in this paper that we use the Euclidean model and its DLR-structure prescribing conditional probabilities.

Let $\Lambda \subset \mathbb{R}^d$ be a bounded open subset. As
Hilbert space $\mathcal{F}(L^{2}(\Lambda))$ of the system in the finite volume
$\Lambda$ we consider the Fock space of $L^{2}(\Lambda)$ with the
scalar product $\langle \cdot , \cdot \rangle := \sum_{n=0}^\infty \frac{1}{n!}
\langle \cdot , \cdot \rangle_{L^2(\Lambda)^{\otimes n}}$. Note that we neither require the functions to be symmetric or anti-symmetric.
In the $n$-particle sector
of the Fock space, we consider  Hamiltonians $\mathcal{H}_{\gamma,\lambda'}$ for the
chemical potential $\lambda'$ of the form
\[
\mathcal{H}_{\gamma,\lambda', \Lambda} := \sum_{i=1}^n -\frac{\Delta_i}{2} + \int_{\mathbb{R}^d} e_{\lambda'}(\sum_{i=1 }^n j_\gamma(r,q_i))dr,
\]
where $\Delta_i$ denotes the Laplacian in the variable $q_i$ with
Dirichlet boundary condition at the boundary of $\Lambda$.
Notice that, as for the classical LMP model (see \eqref{V}), this expression can be written as a sum of one-body up to four-body potential contributions.

The aim of this paper is to study phase transitions with the density as order
parameter. We imagine our quantum system as being in contact with an
other system at the boundary, which we will refer to as the outside system.
\tobi{We will identify parameters $\beta,\ga,\la$ at which the system becomes sensitive to the boundary influence and will give rise to two extremal states corresponding to different densities. For any other nearby value of $\la$ one expect a unique state and continuous dependence on the boundary condition. Our considerations are quite independent of the precise structure of
the boundary influence. }
Technically, the two most straight forward way to model such an
influence in a physically reasonable way are the following: first,
one adds an additional potential term which acts only near the
boundary. This term is chosen in such a way to incite a particular
density near the boundary of the system. Second, one could imagine an outside
system of particle configuration $\bar{q} \in Q_{\Lambda^c}$ chosen with a
probability distribution which give rise to a particular density.

Hence the Hamiltonian
$\mathcal{H}_{\gamma, \lambda', \Lambda}$ is defined for $q = (q_1, \ldots , q_n) \in \tobi{Q_\La}$ via
\begin{equation} \label{260611_1}
\mathcal{H}_{\gamma, \lambda', \Lambda} := \sum_{i=1}^n -\frac{\Delta_i}{2} + h_{\gamma,
  \lambda'}(q|\bar{q}),
\end{equation}
or as
\begin{equation} \label{260611_1a}
\mathcal{H}_{\gamma, \lambda', \Lambda} := \sum_{i=1}^n
-\frac{\Delta_i}{2} + h_{\ga,\la'}\big(q|\rho_\beta^{\pm}\text{\bf 1}_{\La^c}\big),
\end{equation}
where we define
 \begin{eqnarray}
h_{\ga,\la}\big(q|\rho_\beta^{\pm}\text{\bf
1}_{\La^c}\big):= \int e_\la\big(j_\ga
*[q +\rho_\beta^{\pm}\text{\bf 1}_{\La^c}]\big)-
e_\la\big(\rho_\beta^{\pm}j_\ga*\text{\bf 1}_{\La^c}\big)\;
dr . \label{230512_1}
 \end{eqnarray}
For a particle configuration $q$ and for a measurable (non negative)
function $\rho(r)$, $r\in \mathbb R^d$, we denote by
 \begin{equation}
      \label{a5.2}
j_\ga*(q+\rho)(r) = j_\ga* q (r) + j_\ga* \rho(r) =
\sum_{q_i\in q} j_\ga(r,q_i) + \int j_\ga(r,r')\rho(r')dr' .
   \end{equation}

For concreteness, we concentrate \tobi{in this paper}\notes{True that we concentrate on \eqref{260611_1} in this paper??} on the case \eqref{260611_1}.
\tobi{In order to choose the boundary conditions $\bar{q}$ and $\rho_\beta^{\pm}$ suitably to incite particular densities as the boundary, we would have to anticipate the introduction of several further structures, which
are not relevant for the discussion in this subsection. The precise
structure of the actually outside system considered is discussed in
Section~\ref{sec:11}.}

The corresponding equilibrium state in the grand canonical ensemble  for Maxwell-Boltzmann statistics in the volume $\Lambda \subset \mathbb{R}^d$ for the inverse temperature
$\beta$ (the chemical potential $\lambda'$ is already
incorporated in the definition of $h_{\gamma, \lambda'}$)  is defined
on the $C^*$-algebra of all bounded linear operators on
$\mathcal{F}(L^{2}(\Lambda))$, namely $B(\mathcal{F}(L^{2}(\Lambda)))$ via
\[
\omega_{\Lambda,\bar{q}} (A) := \frac{\mathrm{Tr}(A e^{-\beta \mathcal{H}_{\gamma,\lambda', \Lambda}})}{\mathrm{Tr}(e^{-\beta \mathcal{H}_{\gamma,\lambda', \Lambda}})},
\]
where $A \in B(\mathcal{F}(L^{2}(\Lambda)))$.

\tobi{As discussed the aim of this paper is to prove the non-trivial dependence on the
boundary influences of these
systems even for large volumes. Hence we want to perform a thermodynamic limit.} Denote by $\Delta \subset \mathbb{R}^d$
a bounded open subset, which describes the area in which we observe
our system. Then we consider $B(\mathcal{F}(L^{2}(\Delta)))$ as a
sub-algebra of $B(\mathcal{F}(L^{2}(\Lambda)))$ via the imbedding $A
\mapsto A \otimes I$, where $I$ denotes the identity operator on
$B(\mathcal{F}(L^{2}(\Lambda \setminus \Delta)))$
using that $\mathcal{F}(L^{2}(\Lambda)) = \mathcal{F}(L^{2}(\Delta)) \otimes
\mathcal{F}(L^{2}(\Lambda \setminus \Delta))$ (for this identity the choice of the
scalar product in $\mathcal{F}(L^{2}(\Lambda))$ was essential). So we can define the projection of
$\omega_{\Lambda,\bar{q}}$ to the volume $\Delta$ as the following state on $B(\mathcal{F}(L^{2}(\Delta)))$
\[
\omega_{\Delta,\Lambda,\bar{q}} (A) := \frac{\mathrm{Tr}((A \otimes I) e^{-\beta
    \mathcal{H}_{\gamma,\lambda', \Lambda}})}{\mathrm{Tr}(e^{-\beta
    \mathcal{H}_{\gamma,\lambda', \Lambda}})}.
\]
The state of the system in the thermodynamics limit observed in  $\Delta$
is the limit of $\omega_{\Delta,\Lambda,\bar{q}}$ for $\Lambda
\uparrow \mathbb{R}^d$, if the limit exists.

Following J.~Ginibre, see e.g. \cite{ginibre2,ginibre} we will derive a
functional integral representation of these states using
Feynman-Kac formula. In the following we will furthermore explain how this
representation is related to the Gibbs and DLR measures described in
Subsection~\ref{subsec:2.3}. We will not
only link
in the following the finite volume Gibbs measures with the quantum
mechanical states described in this subsection but we will also explain how one can
derive from the main part of the paper also the existence of the
thermodynamic limits of these quantum states.

The sum in the definition of $H_{\gamma, \lambda',
  \Lambda}$ in \eqref{260611_1} is well-defined, for example, in the sense
of quadratic forms and $H_{\gamma, \lambda',
  \Lambda}$ is selfadjoint and $e^{-\beta H_{\gamma, \lambda',
  \Lambda}}$ is trace-class, see Section~6.3.1 in \cite{BRII} for more
details. According to the Feynman-Kac formula,  for $q \in
Q^\Lambda_n$ we get the following representation
\[
e^{-\beta \mathcal{H}_{\gamma, \lambda',
  \Lambda}} G(q) = \int_{\Lambda^n} k_{\bar{q},\Lambda}(q;q_1', \ldots q'_n)
G(q'_1, \ldots q'_n)  dq_1' \ldots dq_n' ,
\]
where the kernel $k_{\bar{q},\Lambda}$ is defined as
\[
k_{\bar{q},\Lambda}(q;q') :=
\frac{1}{(2\pi \beta)^{nd/2}}\int_{\Q_\Lambda(q:q')}
e^{-H_{\beta,\gamma,\lambda'}(\q|\bar{q})} \otimes_{k=1}^n
W^\beta_{q_k,q'_k}(d\q),
\]
with $q=(q_1, \ldots ,q_n)$, $q=(q'_1, \ldots ,q'_n)$ and  $\otimes_{k=1}^n W^\beta_{q_k,q'_k}$ denotes the
$n$-fold product measure. The pre-factor can be absorbed in the
chemical potential, the effective chemical potential obtained in this way is
denoted by $\lambda$; analogously as one usually absorbs the kinetic part of the Hamiltonian for non-quantum systems into an effective chemical potential.
For finite rank operators with continuous kernel one can easily
derive \tobi{following for example the proof of} the Lemma after Theorem~XI.31 in \cite{RSIII} that
\begin{equation}\label{eq220811}
\omega_{\Lambda,\bar{q}} (A) =
\sum_{n=0}^\infty \frac{1}{n!} \int_{\Lambda^n}
a(q';q)K_{\bar{q},\Lambda}(q;q') dq_1 \ldots dq_n,dq_1' \ldots dq_n' ,
\end{equation}
where $a$ denotes the integral kernel of $A$ in
$\mathcal{F}(L^{2}(\Lambda))$ and the kernel
$K_{\bar{q},\Lambda}(q;q')$ is defined as
\[
K_{\bar{q},\Lambda}(q;q') := \frac{1}{Z_{\beta,\gamma,\lambda}(\Lambda|\bar{q})}
\int_{\Q_\Lambda(q;q')}
e^{-H_{\beta,\gamma,\lambda}(\q|\bar{q})} \otimes_{k=1}^n
W^\beta_{q_k,q'_k}(d\q),
\]
 with
$Z_{\beta,\gamma,\lambda}(\Lambda| \bar{q}):= \int_{Q^\Lambda}
k_{\bar{q},\Lambda}(q;q) \nu_\Lambda(dq)$ denotes the normalization constant.
Similarly, one can derive for multiplication operators $AG(q):=
a(q) G(q)$ that
\[
\omega_{\Lambda,\bar{q}} (A) =
\sum_{n=0}^\infty \frac{1}{n!} \int_{\Lambda^n}
a(q)K_{\bar{q},\Lambda}(q:q) dq_1 \ldots dq_n ,
\]
when $a$ is continuous \tobi{and bounded}. One can interpret \tobi{the latter }representations as a special case of
the representation \eqref{eq220811} for finite rank operators, if one
\tobbi{ represents the  multiplication operator via} the generalized kernel $a(q')\delta_{q'}(dq)$.
In the same way, one can obtain explicit
functional integral representations for tensor products of finite rank operators
with continuous kernel and multiplication operators for continuous
functions. Hence \tobi{from \eqref{eq220811}}, one obtains the following representation
for the projection $\omega_{\Delta,\Lambda,\bar{q}}$ \tobi{on the algebra of all finite rank operators with continuous kernel}:
\[
\omega_{\Delta, \Lambda,\bar{q}} (A) =
\sum_{n=0}^\infty \frac{1}{n!} \int_{\Delta^n}\int_{\Delta^n}
a(q;q') K_{\bar{q},\Delta, \Lambda}(q';q) dq_1 \ldots dq_n,dq_1' \ldots dq_n' ,
\]
where for \tobi{$q= (q_1, \ldots , q_n),q'=(q_1' , \ldots , q_n') \in Q^\Delta$} and $\bar q \in Q^{\La^c}$
\begin{eqnarray*}
K_{\bar{q},\Delta, \Lambda}(q';q) :=
\frac{1}{Z_{\beta,\gamma,\lambda,\Lambda}(\bar{q})} \int_{\Q_\Lambda(q:q')}
\int_{\Q^{\Lambda \setminus \Delta}_\Lambda}e^{-H_{\beta,\gamma,\lambda}(\q,\q' |\bar{q})} \nu_{\beta,
\Lambda \setminus \Delta}(d\q) \otimes_{k=1}^n
W^\beta_{q_k,q'_k}(d\q').
\end{eqnarray*}
\tobi{Furthermore, for the multiplication operators given by continuous bounded functions} we obtain the
representation
\[
\omega_{\Delta, \Lambda,\bar{q}} (A) =
\frac{1}{Z_{\beta,\gamma,\lambda,\Lambda}(\bar{q})}
\int_{\Q_{\Lambda }}a(\q_\Delta)e^{-H_{\beta,\gamma,\lambda}(\q|\bar{q})}
\nu_{\beta,\Lambda}(d\q),
\]
where $\q_\Delta$ denotes the sub-configuration formed from the loops
starting in $\Delta$.  \tobi{By linearity of $\omega_{\Delta, \Lambda,\bar{q}}$, we obtain a functional integration representation on the algebra generated by the finite rank operators with continuous kernel and by  multiplication operators for continuous bounded functions.}

The \tobi{state $\omega_{\Delta, \Lambda,\bar{q}}$ restricted to the commutative sub-algebra of operators $A$ which are multiplication operators for a continuous bounded function $a$}  is nothing else than the projections of the
Euclidean Gibbs measures in the volume $\Lambda$, defined in \eqref{2.a12}, on $\Delta$ with boundary
condition $\bar q$
\[
\omega_{\Delta, \Lambda,\bar{q}} (A)  = \int_{\Q^{\Delta}}a(\q_\Delta) \mu_{\beta,\ga,\la,\La}(d\q|\bar
 q).
\]
Hence the quantum state restricted to the algebra of multiplication
operators is completely described by the Euclidean Gibbs
measures. Furthermore, the existence of the thermodynamic limit of the
state restricted to the algebra of multiplications
operators in the
weak-$*$ topology is equivalent to the existence of  the thermodynamic limit
of the Euclidean Gibbs measure on cylinder functions\tobi{, in the sense that the starting points of the loops are restricted to a bounded set.}

\tobi{More surprising, however, is that} we can also characterize the full quantum state \tobi{from the Euclidean Gibbs measure} using the
following explicit
description of the kernel $K_{\bar{q},\Delta, \Lambda}$
\begin{eqnarray}
K_{\bar{q},\Delta, \Lambda}(q';q) &=&
\int_{\Q_\Lambda(q;q')}
\frac{1}{Z_{\beta,\gamma,\lambda,\Lambda}(\bar{q})}
\int_{\Q_\Lambda^{\Lambda \setminus \Delta}} e^{-H_{\beta,\gamma,\lambda}(\q' | \q ,\bar{q})}
e^{-H_{\beta,\gamma,\lambda}(\q|\bar{q})} \nu_{\beta,
\Lambda \setminus \Delta}(d\q) \otimes_{k=1}^n
W^\beta_{q_k,q'_k}(d\q')  \nonumber \\
&=&  \int_{\Q^{\Lambda }_\Lambda, \q_\Delta = \emptyset} \left( \int_{\Q_\Lambda(q;q')}
 e^{-H_{\beta,\gamma,\lambda}(\q' | \q ,\bar{q})}
\otimes_{k=1}^n
W^\beta_{q_k,q'_k}(d\q')  \right) \mu_{\beta,\ga,\la,\La}(d\q|\bar q) . \label{eq:092020_1}
\end{eqnarray}
As a consequence of the Pirogov-Sinai scheme \tobii{derived}{estabished} in the main body
of the paper, \tobi{we can derive the convergence of the Euclidean Gibbs measure in the thermodynamics limit $\Lambda \uparrow
\mathbb{R}^d$ in a sense strong enough that also 
 $K_{\bar{q},\Delta,
  \Lambda}$ converges in the thermodynamics limit}, see Subsection~\ref{sec:quantum} for more details. \tobi{The
limiting objects can be expressed in terms of the thermodynamics limits of Euclidean
Gibbs measures (we obtain at least two distinct Euclidean Gibbs measure in the limit)
}

\begin{lemma}
\tobi{Assume that for a given boundary condition $\mu_{\beta,\ga,\la,\La}(d\q|\bar q)$ converges in the thermodynamics limit $\Lambda \uparrow
\mathbb{R}^d$ to a measure $\mu_{\bar q}$ such that \eqref{eq:092020_1} converges to 
$\mu_{\bar q}$
\begin{equation} \label{eq220811_3}
K_{\bar{q},\Delta,\mathbb{R}^d}(q';q) := \int_{\Q(q;q')}
\int_{\Q , \q_\Delta = \emptyset}e^{-H_{\beta,\gamma,\lambda}(\q' | \q,\bar{q})}
\mu_{\bar q }(d\q) \otimes_{k=1}^n
W^\beta_{q_k,q'_k}(d\q').
\end{equation}
Denote by $\rho_{\bar{q}, \Delta, \mathbb{R}^d}$ the trace class operator associated to $K_{\bar{q},\Delta,\mathbb{R}^d}(q';q) $. Then the state $\omega_{\Delta, \Lambda,\bar{q}}$ converges in the weak-* topology with
respect to the quasi-local algebra generated by the algebras $B(\mathcal{F}(L^{2}(\Delta)))$ to a locally normal state $\omega_{\bar{q}}$. Explicitly, for all $A \in B(\mathcal{F}(L^{2}(\Delta)))$ we have that
\[
\lim_{\Lambda \uparrow \mathbb{R}^d}\omega_{\Delta, \Lambda,\bar{q}}
(A) =  \mathrm{Tr}(A\rho_{\bar{q}, \Delta, \mathbb{R}^d}),
\]
where, if $A$ has a continuous kernel $a$, we have additionally that 
\[
= \sum_{n=0}^\infty \frac{1}{n!} \int_{\Delta^n}\int_{\Delta^n}
a(q;q') K_{\bar{q},\Delta, \mathbb{R}^d}(q';q) dq_1 \ldots dq_n,dq_1' \ldots dq_n' .
\]}
\end{lemma}

 The representations \eqref{eq220811_3} gives an explicit
characterization of the quantum state via the Gibbs measures defined
in Subsection~\ref{subsec:2.3} and identifies them as locally normal states.

\begin{proof}
\tobi{Then by dominated convergence, using \eqref{2.a11}, for all bounded operators $A$ which are either {finite rank}
operators with a continuous kernel or multiplication operators with a bounded continuous functions we obtain that
\[
\lim_{\Lambda \uparrow \mathbb{R}^d}\omega_{\Delta, \Lambda,\bar{q}}
(A) = \sum_{n=0}^\infty \frac{1}{n!} \int_{\Delta^n}\int_{\Delta^n}
a(q;q') K_{\bar{q},\Delta, \mathbb{R}^d}(q';q) dq_1 \ldots dq_n,dq_1' \ldots dq_n' .
\]
As the set of finite rank operator is irreducible and
$K_{\bar{q},\Delta,\mathbb{R}^d}$ is
continuous we can identify the limit $\omega_{\bar{q},\Delta,\mathbb{R}^d}$
of $\omega_{\Delta, \Lambda,\bar{q}}$ as a normal state, using again the Lemma after Theorem~XI.31 in \cite{RSIII}. The associated trace
class operator $\rho_{\bar{q},\Delta,\mathbb{R}^d}$ is uniquely determined by the kernel
$K_{\bar{q},\Delta,\mathbb{R}^d}$. As states have norm
one in the uniform topology, continuous extension in uniform topology gives immediately that
\begin{equation}\label{eq220811_2}
\lim_{\Lambda \uparrow \mathbb{R}^d}\omega_{\Delta,
  \Lambda,\bar{q}}(A) = \mathrm{Tr}(A\rho_{\bar{q},\Delta,\mathbb{R}^d})
\end{equation}
holds for all compact operators $A$ on $\mathcal{F}(L^{2}(\Delta))$. Furthermore, according
to Proposition~2.6.15 in \cite{BRI}, we see that the convergence in \eqref{eq220811_2} holds even for all
bounded linear operators $A$ on $\mathcal{F}(L^{2}(\Delta))$. 
}
\end{proof}

\vskip1cm

\subsection{Mean-field model}
Existence of phase transitions is actually trivial
at the level of mean field approximation of the model.
The mean field approximation of the LMP model is described
via the mean field free energy density
  \begin{eqnarray}
    \label{2.a14}
&&  f_{\beta,\la}(x)= e_\la(x ) -\frac{s(x)}{\beta},\qquad
s(x)=-x (\ln x-1),
  \end{eqnarray}
where $e_\la(x )$ is the energy density defined in
\eqref{2.a4}, $s(\cdot)$  the entropy density and $x\ge 0$
the particles density.

For $\beta \leq \beta_c:=(3/2)^{3/2}$,
$f''_{\beta,\la}(x)\geq 0$ for all $\la$ and $x$ and
$f_{\beta,\la}$ is a convex function. For any $\beta >
\beta_c$ there is an interval of $\la$'s in which the
function $f_{\beta,\la}(\cdot)$ has two local minima and a
unique value, $\la=\la_\beta$, where  these local minima are equal (thus
being both global minima). The two corresponding minimizers are
denoted by $\rho_\beta^- < \rho_\beta^+$, otherwise the
minimizer is unique. Thus phase transitions at the mean
field level (i.e.\ non uniqueness of the minimizers) occur
at $\beta>\beta_c$ and $\la=\la_\beta$; the order parameter
of the transition is the density and $\rho_\beta^{\pm}$ are
the equilibrium densities in the $\pm$ phases.

\vskip1cm

\subsection{Phase transitions}

As in \cite{LMP}, we restrict hereafter to $\beta \in
(\beta_c,\beta_0)$, where $\beta_0>\beta_c$ is defined in
\eqref{dA.6}.  In \cite{LMP} it is proved that for any
$\ga>0$ small enough, there is $\la_{\text{\rm
class}}(\beta,\ga)$, and two distinct DLR measures at
$\beta,\ga,\la_{\text{\rm class}}(\beta,\ga)$.  Here we
will prove the quantum extension of the result.
First, for the Gibbs measures on loop configurations we will show the following

\vskip.5cm

 \begin{thm}

 \label{thm2.1}
For any $d\ge 2$ and any $\beta \in (\beta_c,\beta_0)$ and
any $\ga>0$ small enough, there is $\la(\beta,\ga)$ and two
distinct DLR measures at $\beta,\ga, \la(\beta,\ga)$.

 \end{thm}

\vskip.5cm
This yields also the existence of phase transition for the quantum states

\vskip.5cm

 \begin{thm}

 \label{thm2.1q}
For any $d\ge 2$ and any $\beta \in (\beta_c,\beta_0)$ and
any $\ga>0$ small enough, there exists a $\la(\beta,\ga)$ such that
one can find two boundary conditions as defined in Section~\ref{sec:quantum}
leading to two distinct limiting
quantum states at $\beta,\ga, \la(\beta,\ga)$.
 \end{thm}

\vskip.5cm

As in the classical case, we will prove  Theorem
\ref{thm2.1} by constructing the two distinct DLR measures
via a thermodynamic limit procedure, after ``imposing $+$
and $-$ boundary conditions''. The persistent memory of the
boundary conditions follows from Peierls estimates obtained
by extending the Pirogov-Sinai methods.  The proof yields a
detailed control on the structure of the typical
configurations, showing that in the two distinct DLR
measures of Theorem \ref{thm2.1}, the coarse grained image
of the particle configurations $\q(0)$ have approximately a
homogeneous density; moreover the values of the density are
close to the corresponding mean field values
$\rho_\beta^{\pm}$. We will also show that the typical
loops are ``short'' so that the loops remain confined
around their starting points and the energy of the loops is
for small $\ga$,  typically very close to $\beta$ times the
classical energy of the starting points of the loops. Theorem~\ref{thm2.1q}
follows then almost directly from Theorem~\ref{thm2.1}. Two
distinct DLR measures corresponds to distinct limiting quantum
states. Theorem~\ref{thm2.1} will be shown for a rather rich class of
boundary conditions.
The special choice of the boundary condition in Theorem~\ref{thm2.1q} is made
to enable a straight forward physical interpretation of the quantum
states.

\vskip3cm

\setcounter{equation}{0}

\section{{\bf Contours}}
       \label{sec:a3}

In this section, we introduce the important notion of
contours. Contours are designed to indicate location and
nature of ``large deviations'' from equilibrium which
may occur in a configuration on a coarse grained level.
Throughout the sequel, the inverse
temperature $\beta$ is kept fixed inside the interval
$(\beta_c,\beta_0)$ and frequently dropped from the notation;
$\la$ instead may vary in the interval $\la_\beta\pm 1$, but
it will be eventually fixed to a value $\la(\beta,\ga)$, the
exact value at which the phase transition occurs, identified in Section~\ref{sec:b4}.

\vskip0.5cm

$\bullet$\; For any $\ell \in \{ 2^{n} :  n \in \mathbb
Z\}$, we set
  \begin{equation}
     \label{a3.1}
C_0^{(\ell)}:= \Big\{ r \in \mathbb R^d: 0\leq r_i< \ell, \;
1\le i \le d\Big\}
     \end{equation}
and we denote by $\mathcal D^{(\ell)}$ the partition of $\mathbb
R^d$ made up out of cubes which are translates of $C_0^{(\ell)}$
by vectors whose coordinates are  integer multiples of
$\ell$.  We will denote
by $C^{(\ell)}_r$, $r\in \mathbb R^d$, the cube of
$\mathcal D^{(\ell)}$ which contains $r$.

\noindent Before we introduce the coarse graining procedure itself, we
will describe an {\it abstract contour model}, which means a model on
configurations and on contours as entities in itself. The coarse
graining procedure will yield such an abstract contour model as the
image of the LMP model.

$\bullet$\; Given
$\ell_-<\ell_+$ both in $\{ 2^{n} :  n \in \mathbb Z\}$,   we will
consider pairs of functions which will be indicators whether a cube in
the coarse graining procedure is in a particular phase or is passed by
a ``long'' loop. We will choose later $1 \ll \ell_- \ll \ell_+$. The
first function $\eta$ belongs to
$L^{\infty}(\mathbb R^d,\{0,\pm 1\})$ and is $\mathcal
D^{(\ell_-)}$ measurable. Namely $\eta(r)$ is constant on
each cube of $\mathcal D^{(\ell_-)}$ with possible values $0, \pm 1$. The second function is
denoted by $\si\in L^{\infty}(\mathbb R^d,\{0, 1\})$ and is $\mathcal D^{(\ell_+)}$ measurable. Given $\eta$ and $\si$ as above we then define a third function $\Theta$
by setting
 \begin{equation}
      \label{a3.5a}
\Theta(\eta,\si;r) :=\begin{cases} \pm 1
 &\hskip1cm\text{\parbox[t]{10cm}{if  $\eta(r')\equiv \pm 1$
 and  $\si(r')=1$  
 for all
 $r'\in C^{(\ell_{+})}_r \cup \delta_{\text{\rm out}}^{\ell_{+}}
 [C^{(\ell_{+})}_r]$ }}
\\
\\
0 &\hskip1cm\text{otherwise}\end{cases},
     \end{equation}
where we use the following notation frequently used later on: for $A$ a $\mathcal
D^{(\ell)}$-measurable set, we denote by $\delta_{\text{\rm
    out}}^{\ell}[A]$ the union of all cubes in $\mathcal D^{(\ell)}$
which are in $A^{c}$ and are connected to $A$, (two
sets are considered as being connected if their closures have non empty
intersection); furthermore we define $\delta_{\text{\rm in}}^{\ell}[A]:=\delta_{\text{\rm
out}}^{\ell}[A^c]$. Notice that $\Theta$ as defined in \eqref{a3.5a} is a $\mathcal
D^{(\ell_+)}$ measurable function.

\notes{ \blue{
         \begin{tabular}{|l|} \hline \\
according to \\ the refereee\\ this \\ definition \\ is not \\ the same \\ as in the \\ paper II
\\ but I do not\\  see any \\  difference \\
 \hline
         \end{tabular}}
         .}

      \vskip0.5cm

$\bullet$\; {\it Contours.} Let $\eta$ and $\si$ be two
functions of the type considered above,
then the configuration of contours $\{\Ga_i\}$ associated to the pair $(\eta,\si)$ is the set of all
triples $(\text{sp}(\Ga_i),\eta_{\Ga_i},\si_{\Ga_i})$,
where $\text{sp}(\Ga_i)$ are the maximal connected components of $\{ r \in \mathbb{R}^d : \Theta(\eta,\si;r)=0 \}$, the supports of the contours,
 and
$\eta_{\Ga_i}$ and $ \si_{\Ga_i}$ are the restrictions of
$\eta$ and $\si$ to
$\text{sp}(\Ga_i)$.
All contours considered in this paper will have finite support and hence we will include in the definition that the volume of the support of the contours  $|\text{sp}(\Ga_i)| < \infty$. {If we say that $\Ga$ is a contour we mean that there is a triple $(\text{sp}(\Ga_i),\eta_{\Ga_i},\si_{\Ga_i})$ such that $\eta_{\Ga_i},\si_{\Ga_i}$ can be extended in such a way to functions $\eta,\si$ on all of $\mathbb{R}^d$ such that $\Ga$ is the contour associated to $(\eta,\si)$.}
It follows directly from the definition that:

\vskip.5cm

\begin{lemma}
  \label{lemma3a.1}
If $\Ga$ is one of the contours associated to $(\eta,\si)$ then, on
each one of the maximal connected components of
$\delta_{\text{\rm out}}^{\ell_+}[\text{sp}(\Ga)]\cup
\delta_{\text{\rm in}}^{\ell_+}[\text{sp}(\Ga)]$, the functions $\si=1$
and $\eta$ is constant and non zero.

  \end{lemma}

Given a contour $\Ga$, we call
 \begin{equation}
      \label{a4.2}
K := \delta_{\text{\rm
in}}^{\ell_{+}/2}[\text{sp}(\Ga)],\qquad
K^{\pm}:=\Big\{r\in K: \eta_{\Ga}(r)=\pm 1 \Big\} .
     \end{equation}
Notice that, Lemma \ref{lemma3a.1} implies $K=K^{+}\cup K^{-}$,
with $K^{+}$ and $K^{-}$ mutually disconnected.
As $|\text{sp}(\Ga_i)| < \infty$ , for any contour $\Ga$ there is a unique unbounded component of sp$(\Ga)^c$. We say that $\Ga$ is a plus [minus] contour if $K^+$ [$K^-$] is
connected to the unbounded component of sp$(\Ga)^c$.  Let
us proceed by supposing that $\Ga$ is a plus contour, the
definitions for minus contours are completely analogous. We call int$^-_i(\Ga)$ the maximal connected
components of sp$(\Ga)^c$ which are connected to $K^-$ and
set
 \begin{equation}
      \label{a4.3}
\text{{\rm int}$^-(\Ga) := \bigcup_i$\;{\rm
int}$^-_i(\Ga)$},\qquad A^- := \delta_{\text{\rm
in}}^{\ell_{+}}[ \text{\rm int}^-(\Ga)]
     \end{equation}
noticing that Lemma \ref{lemma3a.1} implies $\eta=-1$ on
$A^-$. By int$_i^+(\Ga)$ we then denote the remaining
maximal connected components of sp$(\Ga)^c$ with the exception of the unbounded one, int$^+(\Ga)$ is defined as the union of all int$_i^+(\Ga)$.

Let $\La$ be a bounded region,
 \begin{eqnarray}
&& \mathcal C_\La^{+} = \Big\{ \{\Ga_i\}: \;\text{ all
$\Ga_i$ are $+$ contours, sp$(\Ga_i) \subset
\La$,}\nonumber\\&& \hskip2cm \text{dist$($ sp$(\Ga_i),$
sp$(\Ga_j))\ge \ell_{+}$, for all $j\ne i$\;}\Big\}
      \label{a4.8p}
  \end{eqnarray}
with $ \mathcal C_\La^{-}$ defined analogously.

Given weights ${\hat W}_{\ga,\la}^{\pm}(\Ga_i;\q)$, we can  define an abstract partition function:
let $\La$ be a bounded $\mathcal
D^{(\ell_{+})}$-measurable region, let $\mathcal A
\subset \mathcal C_\La^{\pm}$, then we define
 \begin{equation}
      \label{a4.13}
{\hat Z}^{\pm}_{\ga,\la}(\La;\mathcal A |\bar\q):=
\sum_{\{\Ga_i\}\in \mathcal A } \int_{ \Q^\La_{\pm}}
\nu_{\beta,\La}(d\q)e^{-H_{\ga,\la}(\q|\bar\q_{\La^c})}
  \prod_{i}{\hat W}_{\ga,\la}^{\pm}(\Ga_i;\q)
 \end{equation}
and we simply write ${\hat
Z}^{\pm}_{\ga,\la}(\La|\bar\q) $ for ${\hat
Z}^{\pm}_{\ga,\la}(\La; \mathcal C_\La^{\pm} |\bar\q)$.

\vskip3cm

\setcounter{equation}{0}

\section{{\bf Course graining and contour models}}
       \label{sec:a4}

\vskip.5cm

$\bullet$\; {\it Phase indicator.}  We will associate
to each loop configuration $\q$ three functions
$\eta(\q;r)$, $\si(\q;r)$ and $\Theta(\q;r)$, where for fixed $\q$ the resulting functions (as functions in $r$) will be used to define contours along the lines followed above; the function $\Theta(\q;r)$ being ``the indicator of  phases'' of the configuration $\q$. The functions depend on the parameters $\beta$, $\ga$, $\ell_-$, $\ell_+$ and $\zeta$: usually the dependence on these parameters will not be indicated, however, occasionally we will add superscripts to stress the dependence
on these parameters.

The function $\si(\q;r)$ is an indicator of the existence
or absence of long loops; more concretely, a loop $\om(t), 0\le t \le
\beta$, is defined as being ``long'' if
 \begin{equation}
      \label{a3.3}
\sup_{0\le t \le \beta} |\om(t)-\om(0)| > \ga^{-1/2}
     \end{equation}
and as being ``short'' otherwise. Furthermore, let us denote by
$\q^<$ and $\q^>$ the subconfigurations of loops obtained from $\q$ by
selecting only its short, long loops respectively.
We set $\si(\q;r)=0$, if there is at least one
long loop $\q_i\in \q$ and a time $t\in [0,\beta]$ such  that
$\q_i(t) \in C_r^{(\ell_+)}$; otherwise, we set $\si(\q;r) =1$.
In other words, $\{r : \si(\q;r)=0\}$ is the minimal
$\mathcal D^{(\ell_+)}$-region which covers  all the long loops
$\q^>$, while on its complement all loops are short.

The function $\eta(\q;r)$ indicates the regions  where the
empirical density is close to the liquid (plus) and vapor (minus)
densities of the mean field model, or where it deviates from both; that is, given $\ell_-$ and a
positive ``accuracy'' parameter $\zeta$, we set
      \begin{equation}
       \label{a3.2}
\eta(\q;r) :=\begin{cases} \pm 1
 &\hskip1cm\text{if $ \Big| \ell_-^{-d} |\q^<(0)\cap C^{(\ell_-)}_r|
- \rho^{\pm}_{\beta}\Big|\le \zeta$}
\\
\\ 0 &\hskip1cm\text{otherwise}\end{cases}
     \end{equation}
(note that the definition of $\eta(\q;r)$ depends also on
$\beta$).

Finally, the phase indicator $\Theta(\q;r)$ is defined via
\eqref{a3.5a} substituting for $\eta$ and $\si$ the random variables
$\eta(\q;r)$ and $\si(\q;r)$, namely
 \begin{equation}
      \label{a3.5}
\Theta(\q;r):=\Theta\Big( \eta(\q;\cdot),\si(\q;\cdot);r\Big).
     \end{equation}

The function $\Theta(\q;r)$ is  ``a phase indicator'' in
the sense that we consider a configuration $\q$ as being in the liquid, in the vapor phase respectively, if $\Theta(\q;r)=1$, or equal to $-1$ respectively. In the complimentary region, namely  $\{ r :\Theta(\q;r)=0\}$, we consider the system as being in neither phase and hence our aim is to show that the existence of such a region is a ``large deviation from equilibrium ''.

\vskip.5cm

$\bullet$\; {\it The contours of $\q$} are defined as  the
contours $\{\Ga_i\}$ associated to the pair $(\eta(\q;\cdot),\si(\q;\cdot))$.
We denote by $\und \Ga$ the map which associates $\{\Ga_i\}$ to
$\q$, writing $\{\Ga_i\}=\und \Ga(\q)$.

%
%
%
%

\vskip1cm

\subsection {Choice of parameters}
\label{subsec:a4.1}

The main aim of the article is to show that for small but non-zero
$\ga$ there is a phase transition  for a suitably chosen chemical
potential which we will achieve by showing that contours are rare and
small. This effect appears on a particular scale which is captured by
the correct choice
of the $\ga$ dependence of
$\zeta$ and $\ell_{\pm}$.
We will consider
     \begin{equation}
      \label{a4.1}
\zeta:= \ga^a,\quad \ell_{\pm,\ga}:= \ga^{-(1\pm \alpha)}
     \end{equation}
where $1 \gg \alpha \gg a >0$, the more precise requirements
will be fixed later, see below \eqref{b6.31bb}.
Unless specified otherwise, contours are defined relative
to the parameters as given in \eqref{a4.1}, that is, the functions  $\eta$, $\si$
and $\Theta$ without superscripts are defined for the choice
\eqref{a4.1} of parameters.

\vskip1cm

\subsection {Partition functions with constraint, dilute partition functions}
\label{subsec:a4.3aa}

A special role is played in the sequel by the so-called plus and
minus, ``dilute'' partition functions, whose definition
includes not only a restriction on the chosen boundary
conditions  but also additionally a restriction on the contours admissible.
Throughout the paper we use the  notation
 \begin{equation}
      \label{a4.6}
 \q_A := \{\q_i\in \q: \q_i(0)\in A\}.
     \end{equation}

\vskip.5cm

$\bullet$\; Analogously to the definition  of $\und \Ga(\q)$
in the previous section, we define for any $\mathcal
D^{(\ell_{\red{+},\ga})}$-measurable region $\La$ the maps
${\und \Ga}_{\La}^{\pm}$ in the following way: denote by ${\und
\Ga}_{\La}^{\pm}(\q)$ the configuration of contours $\{\Ga_i\}$ associated to the pair of random variables
$(\eta_{\La,\pm},\si_{\La,+})$, where
 \begin{equation}
      \label{a3.5p}
\si_{\La,+}(r):=\si(\q_\La;r),\qquad \eta_{\La,\pm}(r)
:=\begin{cases} \eta(\q_\La;r) & \text{ if  $ r\in \La$}
\\
\pm 1 &\text{otherwise.}\end{cases}
     \end{equation}
For any subset $\mathcal A$ of configurations of contours (contained in the range of ${\und
\Ga}_{\La}^{\pm}$) we define the partition function with
constraint $\mathcal A$ as follows
      \begin{equation}
      \label{a4.5}
 Z_{\ga,\la}\Big(\La;\mathcal A| \bar \q\Big) := \int_{{\und \Ga}_{\La}^{\pm}(\q)\in \mathcal A}
  \nu_{\beta,\La}(d\q)  e^{-H_{\ga,\la}(\q|\bar \q_{\La^c})}.
     \end{equation}

\vskip.5cm

$\bullet$\; A configuration $\bar\q$\notes{{$\bar q$ bad \\ notation?}} is called a $\pm$ boundary
condition for $\La$, if ${\und \Ga}_{\La^c}^\pm({\bar
\q})\sqsubset \Lambda^c$.
where, for any subset $B\subset \La$ (including $\La$
itself) we say that $\{\Ga_i \} \sqsubset B$ if and only if for all $i$ holds that
$\mathrm{sp}(\Ga_i) \subset B$. For short we denote by
      \begin{eqnarray}
\{{\und \Ga}_{\La}^{\pm} \sqsubset B\}:= \Big\{ \q : {\und
  \Ga}_{\La}^{\pm}(\q) \sqsubset B \Big\} .
      \label{a4.14a}
     \end{eqnarray}
The configuration $\bar \q$ is a $+$-boundary condition iff in a layer
of width $\ell_{+,\ga}$ all $\mathcal{D}^{(\ell_{-,\ga})}$-cubes have a
density near to $\rho^+_\beta$ in the sense of \eqref{a3.2} and non long
loops start or enter this layer. Hence, in particular, all
configurations of short loops for which $\eta$ is identical $1$ fulfill
this condition.


\vskip.5cm

$\bullet$\; Let $\q$ be a $\pm$ boundary condition for $\La$, then we define the $\pm$ dilute partition function $Z_{\ga,\la}^{\pm}(\La|\bar\q)$ in $\La$ with b.c.\ $\bar\q$ as
 \begin{equation}
      \label{a4.7}
Z_{\ga,\la}^{\pm}(\La|\bar\q):= Z_{\ga,\la}\Big(\La ;
\{{\und \Ga}_{\La}^{\pm} \sqsubset \La\} \big|\bar\q\Big) .
     \end{equation}
Furthermore, we may also write in order to stress the role of the boundary condition (see for instance \eqref{a4.11} below)
$Z_{\ga,\la}^{\pm}(\La;\mathcal A|\bar\q)$ for
$Z_{\ga,\la}\Big(\La;\mathcal A| \bar \q\Big)$ if $\bar\q$
is a $\pm$ b.c.\ and all contours $\{ \Ga_i \} \in \mathcal A$ fulfill $\{ \Ga_i \} \sqsubset \La$.
%
We need to introduce further notations. Call
 \begin{equation}
      \label{a4.7d}
A :=\delta_{\text{\rm in}}^{\ell_{+,\ga}}[\La],\qquad
\La_0:=\La\setminus A .
     \end{equation}
We write for configurations of points
 \begin{equation}
      \label{a4.7dd}
q\subset B \text{ iff } q_i\in B,\text{\; for all $q_i\in q $}
     \end{equation}
and,in a slight abuse of notations,  we write analogously for configurations of loops
 \begin{equation}
      \label{a4.7ddd}
\q\subset B \text{ iff } \q(t) \subset B, \text{ for all }0 \le t\le
\beta .
     \end{equation}

The constraint in the partition function
\eqref{a4.7} consists out of two conditions, as
  \begin{equation}
      \label{a4.7e}
\{{\und \Ga}_{\La}^{\pm}(\q) \sqsubset \La\} = \{ \q :
\eta(\q_\La;r)=1 \text{\; for all $r\in A$}\} \cap
\{\q: \si(\q_\La;r)=1 \text{\; for all $r\in A$}\} .
     \end{equation}
The first condition (the one referring to $\eta$) is local and involves only
loops starting from $A$; the second one is instead global,
as it requires that any long loop $\q_i$  has to stay
completely inside $\La_0$.

Let us further remark  about \eqref{a4.7} that loops of
$\bar\q$ which originate from outside $\delta_{\text{\rm
out}}^{\ell_{+,\ga}}[\La]\cup \La$ do not interact (for $\ga$ small
enough) with the loops $\q$ in the integrator in \eqref{a4.5}: in fact in order to be within
interaction range, a loop should be long and enter into
$\delta_{\text{\rm out}}^{\ell_{+,\ga}}[\La]$, which is excluded by the
condition that the contours of ${\und
\Ga}_{\La^c}^{\pm}(\q)$ are inside $\La^c$. Thus
$Z_{\ga,\la}^+(\La|\bar\q)$ is insensitive to changes or
removals of loops of $\bar \q_{\La^c}$ which originate in
the complement of $\delta_{\text{\rm
out}}^{\ell_{+,\ga}}[\La]\cup \La$, provided $\bar\q$ remains a
$\pm$ boundary condition.


\vskip1cm

\subsection {Partition function as a gas of contours}
\label{subsec:a4.3bb}
\label{subsec:a4.2}

The general strategy is first to write the partition function \eqref{a4.7} as the
partition function of ``a gas of contours'' and then to
prove that this gas has low density. Each contour will
contribute to the partition function with a statistical
weight. The weight of a contour is the ratio of the
probability of its presence over the probability of its absence.
The weight of a plus contour $\Ga=(\mathrm{sp}(\Ga),\eta_\Ga,\si_\Ga)$ with b.c.\ $\q$ is defined as
 \begin{equation}
      \label{a4.10}
W_{\ga,\la}^{+}(\Ga;\q):= \frac{N_{\ga,\la}(\q)} {
D_{\ga,\la}(\q)} ,
     \end{equation}
where, shorthanding below $\Delta^-:=$ int$^-(\Ga)$, we define
 \begin{eqnarray}
N_{\ga,\la}(\q) &:=&\int_{\eta=\eta_\Ga, \si = \si_\Ga}
\nu_{\beta,\text{\rm sp}(\Ga)\setminus K^+}(d\q')\;
e^{-H_{\ga,\la}(\q'|\q_{K^+})}
 Z^-_{\ga,\la}\Big( \Delta^-;  \{\und \Ga_{\Delta^-}^{-}
  \sqsubset \Delta^-\setminus A^-\} \big| \q'\Big)  \nonumber \\
      \label{a4.11}
     \end{eqnarray}
 \begin{eqnarray}
D_{\ga,\la}(\q) &:=&\int_{\eta= \si =1} \nu_{\beta,\text{\rm
sp}(\Ga)\setminus K^+}(d\q')\;
e^{-H_{\ga,\la}(\q'|\q_{K^+})}
Z^+_{\ga,\la}\Big( \Delta^-;  \{\und \Ga_{\Delta^-}^{+}
\sqsubset \Delta^-\setminus A^-\} \big| \q'\Big).\nonumber \\
      \label{a4.12}
     \end{eqnarray}

By a simple inductive argument, nowadays standard in
Pirogov-Sinai theory, one can show that
 \begin{equation}
      \label{a4.8}
Z_{\ga,\la}^{\pm}(\La|\bar\q)=   \sum_{\{\Ga_i\} \in
\mathcal C_\La^{\pm} }\int_{ \Q^\La_{\pm} }
\nu_{\beta,\La}(d\q)e^{-H_{\ga,\la}(\q|\bar\q_{\La^c})}
  \prod_{i}W_{\ga,\la}^{\pm}(\Ga_i;\q) ,
 \end{equation}
where $\Q^\La_{\pm}$ are the plus (minus) restricted
ensembles in $\La$,  namely
 \begin{equation}
      \label{a4.8a}
\Q^\La_{\pm} :=   \Big\{\q\in \Q^\La:
{\und\Ga}_{\La}^{\pm}(\q)=\emptyset\Big\}
   \end{equation}
and $W_{\ga,\la}^{+}(\Ga;\q)$ are the weights defined as in
\eqref{a4.10}.

In the classical nearest neighbor Ising model, the analogous expression to \eqref{a4.8} becomes
much simpler. In this case $\Q^\La_{\pm}$ is just a single  configuration,
namely the ground state configuration with all spins equal to $+1$
(resp.\ $-1$). The contours are just surfaces (in the dual
lattice) and one sees easily that their weights are small (equal to the
exponential of minus $\beta$ times the perimeter of the
contour, if the strength of the spin-spin couplings is set equal to $1$).
Thus for the nearest neighbor Ising model, \eqref{a4.8} becomes the partition function of a non
interacting gas of contours (except for the compatibility
condition). For $\beta$ large, the statistical weights of
the contours are small enough such that cluster expansion
techniques become applicable. Whereas, in our case, the representation \eqref{a4.8} is
not as simple. The gas of contours is non-trivially coupled
to a ``restricted ensemble of loop configurations'',
$\Q^\La_{\pm}$ through the weights $W_{\ga,\la}^{\pm}(\Ga;\q)$. In our case both the contour and the restricted ensemble have a non-trivial internal structure. A possible approach is to fix the
configuration of contours $\{\Ga_i\}$ in \eqref{a4.8} and to integrate
over $\nu_{\beta,\La}$. As result, we would obtain an effective gas of
contours alone, but the contours would interact with each other in a non trivial way.

However, we will not follow this approach and we will study instead the
full combined system of loop configurations from the restricted
ensembles $\Q^\La_{\pm}$ coupled with a gas of contours.
The strategy to derive a bound for $W_{\ga,\la}^{+}(\Ga;\q)$ is
based on two points: first, one proves that the two partition
functions $Z_{\gamma,\lambda}^\pm$ which appear in \eqref{a4.11} and in
\eqref{a4.12} have approximately the same value; secondly,
one proves that the integral on the r.h.s.\ of \eqref{a4.11} is much
smaller than its analogue on the r.h.s.\ of \eqref{a4.12}.
While the second point is quite robust, as it follows from an
analysis of the cost of deviations from equilibrium (in the sense of
contours) via a non-local free energy functional. The cost is of the order of the surface of the contour.  The first point is
much more delicate. The equality in the bulk is established by a special choice of the
chemical potential $\la$, as a function of the inverse
temperature $\beta$ and of the Kac parameter $\ga$. Exactly at this value of the chemical potential the phase transition appears.

Technically, the most challenging point is to show that the
quotient of the two partition functions (which by the choice of the
parameters is of
surface order) is smaller than the exponential of the cost of the deviations
from equilibrium which is also of surface order, cf. Section~\ref{sec:a6}.

\vskip1cm

  \subsection {Abstract contour models}
  \label{subsec:a4.4}

In order to show that large contours are rare and small, we will use
Zahradn{\'i}k's version of the Pirogov-Sinai
theory. Let us outline the underlying idea before we enter into a more
detailed explanation.  First, one introduces an abstract
contour model, that is a model on configurations of loops and contours
where the weights of the contours do not coincide a priori with the
weights \eqref{a4.10} but are given by \eqref{a4.19} below. For short,
we will call these new weights ${\hat W}_{\ga,\la}^{\pm}(\Ga_i;\q)$
{\it abstract weights}.
Second, one has to show a bound for the abstract weights small enough to be able to conclude
that the abstract weights coincide with the original
weights of the LMP-model.

The abstract weights ${\hat W}_{\ga,\la}^{\pm}(\Ga_i;\q)$ will be
defined inductively using definition \eqref{a4.13} for the abstract
partition function ${\hat Z}^{\pm}_{\ga,\la}(\La;\mathcal A
|\bar\q)$.

For any contour $\Ga$ it holds $|\ssp(\Ga)|\ge \ell_{+,\ga}^d 3^d$ and
hence there are no contours in any $\La$ with
$|\La|=\ell_{+,\ga}^d$. For such $\La$ holds that ${\hat
  Z}^{\pm}_{\ga,\la}(\La;\mathcal A |\bar\q)={
  Z}^{\pm}_{\ga,\la}(\La;\mathcal A |\bar\q)$ independent on the
choice of ${\hat W}_{\ga,\la}^{\pm}(\Ga_i;\q)$, because in this case
neither ${\hat W}_{\ga,\la}^{\pm}(\Ga_i;\q)$ nor ${
  W}_{\ga,\la}^{\pm}(\Ga_i;\q)$ appear in the definition \eqref{a4.8}
and \eqref{a4.13}.

Analogously to equations \eqref{a4.11}--\eqref{a4.12} we define:

      \begin{eqnarray}
{\hat N}^{+}_{\ga,\la}(\Ga,\q_{K^+}) &:=&   \int_{\eta=\eta_\Ga, \si
= \si_\Ga} \nu_{\beta,\text{\rm sp}(\Ga)\setminus K^+}(d\q')\;
e^{-H_{\ga,\la}(\q'|\q_{K^+})}
  {\hat Z}_{\ga,\la}^-\Big(\text{int$^-(\Ga)$};
  \mathcal C_{\text{int$^-(\Ga)$}\setminus  A^-
  }^- \big| \q'\Big) \nonumber \\
      \label{a4.14}
     \end{eqnarray}
 \begin{eqnarray}
{\hat D}^{+}_{\ga,\la}(\Ga,\q_{K^+}) &:=&  \int_{\eta= \si = 1}
\nu_{\beta,\text{\rm sp}(\Ga)\setminus K^+}(d\q')\;
e^{-H_{\ga,\la}(\q'|\q_{K^+})}
  {\hat Z}_{\ga,\la}^+\Big(\text{int$^-(\Ga)$};
  \mathcal C_{\text{int$^-(\Ga)$}\setminus  A^-
  }^+ \big| \q'\Big) \nonumber \\
      \label{a4.15}
     \end{eqnarray}
and analogously ${\hat N}^{-}_{\ga,\la}$ and
${\hat D}^{-}_{\ga,\la}$.  Note that
 ${\hat N}^{+}_{\ga,\la}(\Ga,\q_{K^+})$ and ${\hat D}^{+}_{\ga,\la}(\Ga,\q_{K^+})$ are
functions of the weights ${\hat W}_{\ga,\la}^{\pm}(\Ga';\q)$ via
\eqref{a4.13} only for $\Ga'$ with $ |\ssp (\Ga')|< |\text{int$(\Ga)$}|$ .
In particular, if $|\text{int$(\Ga)$}|=\ell_{+,\ga}^d$ then ${\hat N}^{+}_{\ga,\la}(\Ga,\q)=
{ N}^{+}_{\ga,\la}(\Ga,\q)$ and ${\hat D}^{+}_{\ga,\la}(\Ga,\q)=
{ D}^{+}_{\ga,\la}(\Ga,\q)$.
The abstract weights are defined inductively in the volume of the contour:
let $\Ga$ be a contour, assume that ${\hat W}_{\ga,\la}^{\pm}(\Ga';\q)$
is defined for
any $\Ga'$ such that  $ |\ssp (\Ga')|< |\text{int$(\Ga)$}|$, then we define:

 \begin{equation}
      \label{a4.19}
{\hat W}_{\ga,\la}^{\pm}(\Ga;\q):= \min
\Big\{e^{-c_f\ga^{-1} N_\Ga} \;\;;\;\;\frac{{\hat
N}^{\pm}_{\ga,\la}(\Ga,\q_{K^+})}{{\hat
D}^{\pm}_{\ga,\la}(\Ga,\q_{K^+})}\Big\} ,
     \end{equation}
where $ \ell_{+,\ga}^d N_\Ga= \big|\text{sp}(\Ga)\big|$ and $0<c_f<c_1$ for $c_1$ as in Proposition~\ref{propa7.1}. The constant $c_1$  depends only on $\beta$ and the stability constant $b$.
 We call  ${\hat W}_{\ga,\la}^{\pm}(\Ga;\q)$ also {\em cutoff weights}. This inductive procedure is given explicitly for example in \cite{errico}.

     \vskip .5cm
Let us suppose that for all $\Ga$ and $\q$ we can show the following bound for ${\hat W}_{\ga,\la}^{\pm}(\Ga;\q)$
 \begin{equation}
      \label{a4.20a}
{\hat W}_{\ga,\la}^{\pm}(\Ga;\q)< e^{-c_f\ga^{-1} N_\Ga}, \qquad \text{where }
 \ell_{+,\ga}^d N_\Ga= \big|\text{sp}(\Ga)\big|.
     \end{equation}
Then \eqref{a4.19} yields that
\begin{equation}
      \label{a4.16a}
{\hat W}_{\ga,\la}^{\pm}(\Ga;\q)= \frac{{\hat
N}^{+}_{\ga,\la}(\Ga,\q_{K^+})} {{\hat D}^{+}_{\ga,\la}(\Ga,\q_{K^+})}
     \end{equation}
%
and by induction in the volume of the contour one can prove the following theorem (for details see \cite{errico}, paragraph 10.5):
  \begin{thm}
  \label{thma4.1}
Suppose that for all $\Ga$ and $\q$ as above,
 \begin{equation}
      \label{a4.16}
{\hat W}_{\ga,\la}^{\pm}(\Ga;\q)= \frac{{\hat
N}^{\pm}_{\ga,\la}(\Ga,\q_{K^+})} {{\hat D}^{\pm}_{\ga,\la}(\Ga,\q_{K^+})}
     \end{equation}
then one has for all $\La$ and $\bar\q$.
 \begin{equation}
      \label{a4.17}
{\hat W}_{\ga,\la}^{\pm}(\Ga;\q)=
W_{\ga,\la}^{\pm}(\Ga;\q);\qquad {\hat
Z}^{\pm}_{\ga,\la}(\La|\bar\q)=
Z^{\pm}_{\ga,\la}(\La|\bar\q).
     \end{equation}
  \end{thm}
This implies immediately:
  \begin{thm}
  \label{thma4.2}
Suppose that for some $\la(\beta,\ga)$, the cutoff weights
${\hat W}_{\ga,\la}^{\pm}(\Ga;\q)$ defined by \eqref{a4.19}
satisfy the following bound for all $\Ga$ and $\q$
 \begin{equation}
      \label{a4.20}
{\hat W}_{\ga,\la}^{\pm}(\Ga;\q)< e^{-c_f\ga^{-1} N_\Ga} .
     \end{equation}
Then \eqref{a4.17} holds.

  \end{thm}

 \vskip .5cm
Thus the main task is to prove that the cutoff weights defined in
\eqref{a4.19} are (at least for $\ga$ small enough)
strictly smaller than $e^{-c_f \ga^{-1} N_\Ga}$.
We prove in the following that there exists $\la(\beta,\ga)$ such that
the cutoff weights satisfy \eqref{a4.20}, indeed we will prove:
 \begin{equation}
      \label{a4.18}
 {\hat W}_{\ga,\la(\beta,\ga)}^{\pm}(\Ga;\q) < e^{-2c_f\ga^{-1} N_\Ga}
     \end{equation}
and Theorem \ref{thma4.2} will yield directly that also  $
W_{\ga,\la(\beta,\ga)}^{\pm}(\Ga;\q) < e^{-2c_f\ga^{-1} N_\Ga}$. Such
a bound on the probability of contours is sufficient for an effective
control of the contours via a Peierls' type estimate. Further
properties, in particular Theorem~\ref{thm2.1} and \ref{thm2.1q} can
then be rather easily derived, cf.  Section~\ref{sec:11}.

\vskip1cm

\setcounter{equation}{0}

\section{{\bf Choice of chemical potential}}
       \label{sec:b4}

In this and the next section, we will outline the proof of
\eqref{a4.18}. The intermediate results are stated as theorems which
are proofed in the sucessive sections and in the appendices.
One essential part is the correct choice of
the chemical potential $\la(\beta,\ga)$, which, according
to the Pirogov-Sinai technique, has to be chosen such that the pressures in the two
restricted ensembles become equal. In this section, we will
prove that such a chemical potential indeed exists and we will point out
in the next section where specifically this choice is used.

We consider the $\pm$ restricted ensembles
$\Q_{\pm}^\La$  defined in \eqref{a4.8a} for $\La$ a
bounded $\mathcal D^{(\ell_{+,\ga})}$-measurable region

Since the thermodynamic
pressure, as we will show later, is independent of the boundary conditions,
it is sufficient to choose particular boundary conditions:
 \begin{equation}
      \label{a5.1}
\hat Z^{\pm}_{\ga,\la}(\La|\rho_\beta^{\pm}\text{\bf
1}_{\La^c}) : = \sum_{\{\Ga_i\}\in \mathcal C^{\pm}_\La
}\int_{\Q^\La_{\pm}}
\nu_{\beta, \La}(d\q)e^{-H_{\ga,\la}(\q|\rho_\beta^{\pm}\text{\bf
1}_{\La^c})}
  \prod_{i}{\hat W}_{\ga,\la}^{+}(\Ga_i;\q) ,
 \end{equation}
where we denote by
 \begin{eqnarray*}
&& H_{\ga,\la}\big(\q|\rho_\beta^{\pm}\text{\bf
1}_{\La^c}\big) := \int_0^\beta
h_{\ga,\la}\big(\q(t)|\rho_\beta^{\pm}\text{\bf
1}_{\La^c}\big)dt
 \end{eqnarray*}
and $h_{\ga,\la}\big(\q(t)|\rho_\beta^{\pm}\text{\bf
1}_{\La^c}\big)$ is defined as in \eqref{230512_1}.

For our purpose it is sufficient  to prove the existence of
the thermodynamic limit of the pressure for cubes
$\La_n$ of side length $2^n\ell_{+,\ga}$ and for the aforementioned
special boundary conditions (however the existence result for the
thermodynamic limit is valid in much wider generality).
In Section \ref{sec:a5} we will prove the following
two theorems

     \vskip.5cm

    \begin{thm} [Existence of pressure]
    \label{thm5.1}
The limits
 \begin{equation}
      \label{a5.4}
\lim_{n\to \infty} \frac {\log {\hat
Z}^{\pm}_{\ga,\la}(\La_n|\rho_\beta^{\pm}\text{\bf
1}_{\La^c})}{\beta |\La_n|} = : {\hat P}^{\pm}_{\ga,\la}
 \end{equation}
 exist and are both continuous function of $\la$  in the interval
$|\la-\la_\beta|\le 1$.

     \end{thm}

\vskip1cm

    \begin{thm}
    \label{thm5.2}
There is a constant $c>0$ and for all $\ga$ small enough
there is $\la(\beta,\ga)\in [\la_\beta-
c\ga^{1/2},\la_\beta+c\ga^{1/2}]$ such that
 \begin{equation}
      \label{a5.24}
{\hat P}^+_{\ga,\la(\beta,\ga)}={\hat
P}^-_{\ga,\la(\beta,\ga)} .
 \end{equation}

     \end{thm}

\vskip2cm

\setcounter{equation}{0}

\section{{\bf Scheme of proof}}
       \label{sec:a6}

Having found the ``correct'' chemical potential, hereafter
denoted for short by $\la_\ga :=\la(\beta,\ga)$, we  will outline the proof
that the cutoff weights in \eqref{a4.19} with $\la =
\la_\ga$ are strictly smaller than
$e^{-c_f \ga^{-1} N_\Ga}$ (for $\ga$ small enough), confer also the remarks at the end of Subsection~\ref{subsec:a4.2}. When either temperature and/or chemical potential are missing as
subscripts, they are meant to have the above values,
for instance, $H_\ga$ is just a shorthand for $H_{\beta,\ga,\la_\ga}$.
We write also for short $\nu_\La$ for $\nu_{\beta, \La}$.

For the sake of definiteness, we suppose that $\Ga$ is a
plus contour and we
want to show that
 \begin{eqnarray}
{\hat N}^{+}_{\ga}(\Ga,\q_{K^+})\le  e^{-2c_f \ga^{-1} N_\Ga}
{\hat D}^{+}_{\ga}(\Ga, \q_{K^+})
      \label{a6.1}
     \end{eqnarray}
where   l.h.s.\ and   r.h.s.\  are defined respectively in
\eqref{a4.14} and \eqref{a4.15}.


%

We will prove \eqref{a6.1} via a chain of inequalities
for  ${\hat N}^{+}_{\ga}$, recovering ${\hat D}^{+}_{\ga}$ at the
final step. In Subsection~\ref{subsec:a4.4} we indicated how this bound implies the main
theorems of the paper.

\vskip.5cm

\subsection { Contribution of long loops}

Long loops  produce one of the two gain term, denoted by
$g_{1,\ga}$ (where $g$ stands for ``gain'').  Let be
    \begin{eqnarray}
g_{1,\ga}:= e^{-c_1 \ga^{-1}N^{(\si)}_\Ga},\quad
N^{(\si)}_\Ga: = \frac{\big |\big\{r\in
\text{sp}(\Ga):\si_\Ga(r)= 0\big\}\big
|}{\ell_{+,\ga}^d}
      \label{b6.2}
     \end{eqnarray}
and let be
\begin{eqnarray}
 \text{sp}_{\pm}(\Ga) := \Big\{ r\in  \text{sp}(\Ga):
 \Theta(\eta_\Ga,1;r)=\pm 1\Big\} .
      \label{b6.3}
     \end{eqnarray}
In words $ \text{sp}_{\pm}(\Ga)$ are those subregions of sp$(\Ga)$
which belong to the contour only due to long loops.
In Appendix \ref{sec:a7} we will prove that for a suitable
value of $c_1>0$
    \begin{eqnarray}
{\hat N}^{+}_{\ga}(\Ga,\q_{K^+}) &\leq&
g_{1,\ga}\int_{\eta=\eta_\Ga,\si=1}
 \nu_{ \text{sp}(\Ga)\setminus[K^+\,\cup  \,\text{sp}_-(\Ga)]}(d\q)\;
 e^{-H_\ga(\q|\q_{K^+})}\nonumber
\\ && \hskip1cm \times {\hat Z}^-_\ga \big(\Delta_-\,;\, \mathcal
C_{\text{\rm int}^-(\Ga)\setminus A^-}^-\big| \;\q\big)
      \label{b6.4}
     \end{eqnarray}
where $\Delta_-=\text{\rm int}^-(\Ga) \cup  \text{sp}_-(\Ga)$
and ${\hat Z}^-_\ga \big(\Delta_-\,;\, \mathcal
C_{\text{\rm int}^-(\Ga)\setminus A^-}^-\big| \;\q\big)$ is
the partition function with an extra constraint expressed by the
second of its arguments, namely the sum in  \eqref{a4.13}
is restricted to all $\{\Ga_j\} \in \mathcal
C_{\Delta^-}^-$ such that $ \text{\rm sp}(\{\Ga_j\})\cap [
A^- \cup\, \text{sp}_-(\Ga)] =\emptyset$, see
\eqref{a4.8p}.

\vskip.5cm

\subsection { Contours with only short loops}

We rewrite \eqref{b6.4} in terms of the contours  generated
just by the density fluctuations. Denote by
$\{$sp$(\Ga_i)\}$ the configuration of the maximal connected components of
 $\{r\in \text{sp}(\Ga):\Theta(\eta_\Ga,1;r)=0\}$
and, correspondingly, we denote by
      \begin{eqnarray}
 K_i^+ = \delta_{\text {\rm in}}^{\ell_{+,\ga}/2}[ \text{\rm
 sp$(\Ga_i)$}] \cap \{r\in \ssp(\Ga):\eta_\Ga=1\}, \qquad A_i^- =
 \delta_{\text {\rm in}}^{\ell_{+,\ga}}[ \text{\rm
int}^-(\Ga_i) ] .
      \label{b6.6}
      \end{eqnarray}
Since for any $i$, $A_i^- \subset \{A^- \cup
\text{sp}_-(\Ga)\} $, one has that
   \[
\mathcal C_{\Delta_-\setminus \{ A^- \cup  \,
\text{sp}_-(\Ga)\}
  }^-
 \; \subset \;
\mathcal C_{\Delta_-\setminus \cup_i \, A_i^-
  }^- ,
   \]
where $\Delta_-=  \text{\rm
int}^-(\Ga) \cup \text{sp}_-(\Ga)$  as in \eqref{b6.4}.
Hence one can rewrite the right hand side of \eqref{b6.4} as
    \begin{eqnarray}
{\hat N}^{+}_{\ga}(\Ga,\q_{K^+}) &\leq&
g_{1,\ga}\int_{\eta=\si=1}
 \nu_{[\cup_i K_i^+\cup\, \text{sp}_+(\Ga)]\setminus K^+}(d\q)\;
 e^{-H_\ga(\q|\q_{K^+})}  \prod_i  M_{\ga}^{(i)}(\q,\q_{K^+})
      \label{b6.7} ,
     \end{eqnarray}
where we define
    \begin{eqnarray}
 M_{\ga}^{(i)}(\q,\q_{K^+}) &:=&  \int_{\si=1;\, \eta=\eta_\Ga
  \text{ on sp$(\Ga_i)$, $\eta=-1$ on
 $A^-_i$}}
 \nu_{A_i^-\,\cup\, \text{sp}(\Ga_i) \setminus K_i^+}(d\q')\;
 e^{-H_\ga(\q'|\q_{K^+},\q)}
  \nonumber
\\ && \hskip2cm\times
 {\hat Z}^-_\ga \Big( \text{\rm int}^-(\Ga_i)  \setminus A_i^-\;\Big|
 \;\q'\Big)
      \label{b6.8}
     \end{eqnarray}
The right hand side of \eqref{b6.7} is the expression for $\hat{N}_\ga^+$ relative to a contour
$\Ga$ which has only short loops.

\vskip.5cm

\subsection {Main estimate}

There exists a constant $c>0$ such that for
    \begin{eqnarray}
g^{(i)}_{2,\ga}:= \dis{e^{-c \ga^{-(1-\alpha)d +2a}
N_{\Ga_i}}},
      \label{b6.2b}
     \end{eqnarray}
(the parameters $a$ and $\alpha$ are as \eqref{a4.1})
holds
     \begin{eqnarray}
&& M_{\ga}^{(i)}(\q,\q_{K^+}) \leq \;
 g_{2,\ga}^{(i)} \,{\hat
Z}^+_\ga \Big( \text{\rm int}^-(\Ga_i) \cup
\text{sp}(\Ga_i)\setminus K_i^+ \,;\, \mathcal C_{\text{\rm
int}^-(\Ga_i)\setminus A_i^-}^+ \;\Big| \q_{K_i^+}\Big) .\nonumber\\
       \label{b6.26}
    \end{eqnarray}
Postponing the proof of \eqref{b6.26}, we can conclude the proof in
the next subsection.

\vskip.5cm

\subsection {Conclusion of the proof}

Calling $\dis{g_{2,\ga} := \prod_i g_{2,\ga}^{(i)}}$ and
shorthanding $\Delta= \text{sp}(\Ga) \cup \text{\rm
int}^-(\Ga) \setminus K^+$, we get
    \begin{eqnarray}
&&{\hat N}^{+}_{\ga}(\Ga,\q_{K^+}) \leq  g_{1,\ga}g_{2,\ga}
\,{\hat Z}^+_\ga \Big( \Delta
\,;\, \mathcal C_{\Delta
\setminus \mathcal B}^+ \;\Big| \q_{K^+}\Big),
 \nonumber
     \end{eqnarray}
where
    \begin{eqnarray}
&& \mathcal B=  \text{sp}_+(\Ga) \; \cup\;
  \{\cup_i [\text{sp}(\Ga_i)\cup A_i^-]\} .
      \label{b6.28}
     \end{eqnarray}
In order to reconstruct ${\hat D}^{+}_{\ga}( \q_{K^+})$,
cf. equation \eqref{a4.15}, we need to replace $\mathcal B$ by
sp$(\Ga) \cup A^-$.  As we have
    \begin{eqnarray}
&& \text{sp}(\Ga)= \text{sp}_-(\Ga) \cup \text{sp}_+(\Ga)
\cup_i \text{sp}(\Ga_i)
      \label{b6.29}
     \end{eqnarray}
    \begin{eqnarray}
&& A^- \subset \bigsqcup_i\;\Big( A_i^-  \cup
\delta_{\text{\rm out}}^{\ell_{+,\ga}}[
\text{sp}_-(\Ga)]\Big)
      \label{b6.30}
     \end{eqnarray}
we need to extend the constraint of no  contours in the partition
function to a larger region, the new part is contained in sp$_-(\Ga)
\cup\delta_{\text{\rm out}}^{\ell_{+,\ga}}[
\text{sp}_-(\Ga)]$.  The number of $\mathcal
D^{(\ell_{+,\ga})}$ cubes in such a region is bounded by
$3^d | \text{sp}_-(\Ga)| \ell_{+,\ga}^{-d}$ (recalling that
the number of cubes contiguous to a given one is $3^d-1$).
Then, Lemma \ref{lemmaF.2} shows that
    \begin{eqnarray}
&&{\hat N}^{+}_{\ga}(\Ga,\q_{K^+})\leq   g_{1,\ga}g_{2,\ga}\;
e^{3^d |\text{sp}_-(\Ga)|\ell_{+,\ga}^{-d}  e^{-c_f
\ga^{-1}/2} }\;\,{\hat D}^{+}_{\ga}(\Ga,\q_{K^+}) .
      \label{b6.31aa}
     \end{eqnarray}
The new factor is smaller than each of the gain terms for $\ga$ small enough.
Hence \eqref{a6.1} holds for any $c_f$ such that $c_f < c_1$
for $c_1$ as in \eqref{b6.2}, which only depends on $\beta$ and the stability constant $b$.

\vskip1cm

\centerline{{\bf Proof of main estimate}}

It remains to show \eqref{b6.26}. In Section \ref{sec:a9}
we will prove that with a small cost we can replace in the
energy $\la(\beta,\ga)$ by the mean field value
$\la_\beta$, because by stability one can restrict (with large
probability) to configurations whose  density in
$\mathcal{D}^{(\ell_{-,\ga})}$-cubes is bounded by
$e^{2-\beta b}$.  Recalling from Theorem \ref{thm5.2} that
 $|\la_\beta -\la(\beta,\ga)|\leq c \ga^{1/2}$, we then deduce
 directly that there is a
constant $c_2$ so that
    \begin{eqnarray}
M_{\ga}^{(i)}(\q,\q_{K^+}) &\leq& b_{2,\ga}^{(i)}
\int_{\si=1;\, \eta=\eta_\Ga \text{ on sp$(\Ga_i)$,
$\eta=-1$ on
 $A^-_i$}}
 \nu_{A_i^-\,\cup\, \text{sp}(\Ga_i) \setminus K_i^+}(d\q')\;
 e^{-H_{\beta,\ga,\la_\beta}(\q'|\q_{K^+},\q)}
  \nonumber
\\ && \hskip2cm\times
 {\hat Z}^-_\ga \Big( \text{\rm int}^-(\Ga_i) \setminus A_i^-\;\Big|
 \;\q'\Big),  \qquad  b_{2,\ga}^{(i)} = e^{c_2 \ga^{1/2} |\text{sp}(\Ga_i)|}
%
      \label{b6.9}
     \end{eqnarray}
where the loss term is denoted by $b_{n,\ga}^{(i)}$.

\vskip.5cm

\subsection {Separating corridors}
  \label{subsec:6.5}

In Section \ref{sec:a8} we will prove a ``small deviations
result'', namely that ``well inside'' regions where
$\eta\equiv 1$ (or $\eta\equiv -1$), the density is in
fact`` much closer to'' $\rho_\beta^{\pm}$ than what is
implied directly by $\eta$ being 1 (or $-1$). Therefore, we introduce
two corridors $C_i^+$ and $C_i^-$ of width
$\ell_{+,\ga}/6$,
    \begin{eqnarray}
&& C_i^- = \delta_{\text{\rm
in}}^{\ell_{+,\ga}/6}[\text{\rm sp}(\Ga_i)] \;\cap\;
\{\eta_{\Ga_i}=-1\},\quad C_i^+ = \Big( \delta_{\text{\rm
in}}^{5\ell_{+,\ga}/6}[\text{\rm sp}(\Ga_i)] \;\cap\;
\{\eta_{\Ga_i}=1\}\Big) \setminus B_i^+
  \nonumber \\      \label{b6.10}
\\ &&
B_i^+ = \Big( \delta_{\text{\rm
in}}^{4\ell_{+,\ga}/6}[\text{\rm sp}(\Ga_i)] \;\cap\;
\{\eta_{\Ga_i}=1\}\Big)  \nonumber
     \end{eqnarray}
Note that $C_i^-$ is at the minus boundary of sp$(\Ga_i)$ and
$C_i^+$ is separated from $K_i^+$ by $\ell_{+,\ga}/6$ (and hence
also separated by the same distance from the inner part of the support
sp$(\Ga) \setminus \delta_{\text{\rm
in}}^{\ell_{+,\ga}}[\text{sp}(\Ga)]$ of the spatial support
of $\Ga_i$). Important point for what follows is that
these corridors are well inside the regions where $\eta =
\pm 1$, respectively.
We can  decompose sp$(\Ga_i) \setminus K_i$ in mutually disjoint sets
    \begin{eqnarray}
&& \text{\rm sp}(\Ga_i) \setminus K_i^+ = (B_i^+\setminus K_i^+) \cup C_i^+ \cup T_i
\cup C_i^-
    \label{b6.11}
     \end{eqnarray}
where $T_i$ is define via the identity
\eqref{b6.11}.
There exists  $c_3>0$ such that for
    \begin{eqnarray}
b_{3,\ga}^{(i)} := \exp\Big\{c_3 \gamma^{1/2} (|C_i^-|
+|C_i^+|)\Big\}
      \label{b6.12}
     \end{eqnarray}
holds the following, where
$F_{\beta,\la}(\cdot)$ denotes the functional defined as in \eqref{aB.1},
 \begin{eqnarray}
 M_{\ga}^{(i)}(\q,\q_{K^+}) &\leq &\;
 b_{2,\ga}^{(i)}\, b_{3,\ga}^{(i)} \;\exp\big\{-\beta \ga^{-d}
 [F_{\beta,\la_\beta}(\rho_\beta^- \text{\bf
1}_{\ga C_i^-})+F_{\beta,\la_\beta}(\rho_\beta^+ \text{\bf
1}_{\ga C_i^+})]\big\}
 \nonumber
\\ &&  \times
 \int_{\eta=\si=1}
 \nu_{ B_i^+\setminus K_i^+}(d\q')\;
 e^{-H_{\beta,\ga,\la_{\beta}}(\q'|\q_{K_i^+},\rho_\beta^+\text {\bf 1}_{C_i^+})}
 \nonumber
\\ &&  \times
  \int_{\eta=\eta_\Ga,\si=1}
 \nu_{ T_i}(d\q_{T_i})\;
 e^{-H_{\beta,\ga,\la_{\beta}}(\q_{T_i}|\rho_\beta^-\text {\bf 1}_{C_i^-},
 \rho_\beta^+\text {\bf 1}_{C_i^+})} 
 \nonumber
\\ && \times
{\hat Z}^-_\ga \Big( \text{\rm int}^-(\Ga_i)\,;\,\mathcal
C_{\text{\rm int}^-(\Ga_i)\setminus A_i^-}^- \;\Big|
 \;\rho_\beta^-\text {\bf 1}_{C_i^-}\Big)
      \label{b6.13}
     \end{eqnarray}
The proof of this step is based on the Lebowitz-Penrose coarse
graining techniques and properties of the non local
functional.  We are going to use these techniques again and thus
postpone comments on this point.

\vskip.5cm

\subsection{A variational problem}\label{subsec:6.6t}

The  gain term $g_{2,\ga}^{(i)}$  derives from the integral
over $\q_{T_i}$ in \eqref{b6.13}. Following the classical
Lebowitz-Penrose approach, we first bound the integral
in terms of a minimization problem for the non local
functional $F_{\beta,\la_\beta}(\rho)$ (see Appendix
\ref{app:aC}) and then derive lower bounds for such a
problem exploiting the imposed deviations from
equilibrium present in the integral. Details are given in
Section \ref{sec:a9} and in several appendices. We obtain in this way
 \begin{eqnarray}
&& \int_{\eta=\eta_\Ga,\si=1}
 \nu_{ T_i}(d\q_{T_i})\;
 e^{-H_{\beta,\ga,\la_{\beta}}(\q_{T_i}|\rho_\beta^-\text {\bf 1}_{C_i^-},
 \rho_\beta^+\text {\bf 1}_{C_i^+})} \nonumber
 \\&&\hskip2cm \;\leq \; (g_{2,\ga}^{(i)})^2\,
\exp\Big\{ - \beta \ga^{-d}\Big(
f_{\beta,\la_\beta}(\rho_\beta^+)|\ga T_i| +I_{\ga T_i, \ga
C_i^-}^-+I_{\ga T_i,\ga C_i^+}^+\Big)\Big\}
  \label{b6.14b}
     \end{eqnarray}
with $I^{\pm}_{A,B}$ defined in \eqref{dA.9}. The terms
$\beta I_{ \text{\rm int}^-(\Ga_i),C_i^-}^{\pm}$ \notes{$T_i$ ??}
are the
surface corrections to the pressure in the limit $\ga=0$.

\vskip.5cm

\subsection {Ratio of partition functions}

Using the three identities
   \begin{eqnarray}
   && I_{\ga T_i,\ga C_i^-}^-+ F_{\beta,\la_\beta}(\rho_\beta^- \text{\bf 1}_{\ga
   C_i^-}) + I_{\ga A^-_i,\ga C_i^-}^-
    =   f_{\beta,\la_\beta}(\rho_\beta^+)|\ga C_i^-|
  \nonumber      \label{b6.15}\\
   &&   f_{\beta,\la_\beta}(\rho_\beta^+)|\ga C_i^+|  =
   I_{\ga T_i,\ga C_i^+}^++ F_{\beta,\la_\beta}(\rho_\beta^+ \text{\bf 1}_{\ga
   C_i^+}) +
   I_{\ga B^+_i,\ga C_i^+}^+
  \nonumber
\\
   &&   f_{\beta,\la_\beta}(\rho_\beta^+)|\ga (C_i^- \cup T_i \cup C_i^+)|  =
   I_{\ga B_i^+,\ga C_i^+}^++ F_{\beta,\la_\beta}(\rho_\beta^+ \text{\bf 1}_{\ga
   (C_i^- \cup T_i \cup  C_i^+)}) +
   I_{\ga A^-_i,\ga C_i^-1}^+
  \nonumber
     \end{eqnarray}
\notes{$I_{\ga B_i^+,\ga C_i^+}^+$ change to $I_{\ga C_i^+,\ga B_i^+}^+$}

proved in Proposition \ref{propdB.1}, we get from
\eqref{b6.13} and \eqref{b6.14b} that
     \begin{eqnarray}
&& M_{\ga}^{(i)}(\q,\q_{K^+}) \leq \;(g_{2,\ga}^{(i)})^2
 b_{2,\ga}^{(i)}\, b_{3,\ga}^{(i)}
 \int_{\eta=\si=1}
 \nu_{ B_i^+\setminus K_i^+}(d\q')\;
 e^{-H_{\beta,\ga,\la_{\beta}}(\q'|\q_{K_i^+},\rho_\beta^+\text {\bf 1}_{C_i^+})}
 \nonumber
\\ &&  \hskip2cm \times
\exp\Big\{-\beta \ga^{-d}\Big(
F_{\beta,\la_\beta}(\rho_\beta^+ \text{\bf 1}_{\ga
[C_i^+\cup T_i\cup C_i^-]}) -
 I_{\ga A_i^-,\ga C_i^-}^- +
 I_{\ga A_i^-,\ga C_i^-}^+%
%
%
%
\Big)\Big\} \nonumber
\\ &&  \hskip2cm \times
\;\,{\hat Z}^-_\ga \Big( \text{\rm int}^-(\Ga_i)\,;\,
\mathcal C_{\text{\rm int}^-(\Ga_i)\setminus A_i^-}^-
\;\Big|
 \;\rho_\beta^-\text {\bf 1}_{C_i^-}\Big) .
       \label{b6.16}
\end{eqnarray}

The most delicate part of the proof is the following
bound, derived in  Section \ref{sec:10}
 \begin{eqnarray}
&&  \frac{ \exp\{\beta \ga^{-d} I_{\ga A_i^-,\ga C_i^-}^-\}
\;{\hat Z}^-_\ga \Big( \text{\rm int}^-(\Ga_i)\,;\,
\mathcal C_{\text{\rm int}^-(\Ga_i)\setminus A_i^-}^-
\;\Big|
 \;\rho_\beta^-\text {\bf 1}_{C_i^-}\Big)}
 { \exp\{\beta \ga^{-d} I_{\ga A_i^-,\ga C_i^-}^+\}\; {\hat Z}^+_\ga \Big(
 \text{\rm int}^-(\Ga_i)\,;\,
\mathcal C_{\text{\rm int}^-(\Ga_i)\setminus
A_i^-}^+\;\Big|
 \;\rho_\beta^+\text {\bf 1}_{C_i^-}\Big)}\;
\leq \;  b_{4,\ga}^{(i)}   \nonumber\\
     \label{b6.17}
     \\&&
 \text{for } \qquad b_{4,\ga}^{(i)}:=  \exp \{ c_4 \ga^{a'}|A_i^-|\}
     \end{eqnarray}
with $c_4$ a positive constant and $a'$ \notes{can one choose $a'$ arbitrary} is such that $1\gg a'\gg \alpha\gg
a>0$, for more details see below \eqref{b6.31bb}.

At this point the choice of the chemical
potential becomes important.  In fact, \eqref{b6.17} is an
estimate on the finite volume corrections to the pressure:
the log of the partition functions are to first order given
by $\beta P^{\pm}_{\beta, \la(\beta,\ga)} | \text{\rm
int}^-(\Ga_i)|$. Hence according to \eqref{a5.24} these contributions
cancel between numerator and denominator.  As said before,
the terms $\beta I_{ A_i^-,C_i^-}^{\pm}$ \notes{index changed}
are the surface corrections in the limit $\ga=0$ of the pressure, and
$b_{4,\ga}^{(i)}$ bounds its finite $\ga$ correction.
Hidden in this description is an estimate of exponential
decay of correlations which allows to localize around and
close to the surface, the corrections to the pressure
(which are then computed by taking the limit $\ga=0$
and then estimating its $\ga>0$ correction).
So far we have shown that
     \begin{eqnarray}
&& M_{\ga}^{(i)}(\q,\q_{K^+}) \leq \; ( g_{2,\ga}^{(i)})^2
b_{2,\ga}^{(i)}\cdots b_{4,\ga}^{(i)}
 \int_{\eta=\si=1}
 \nu_{ B_i^+\setminus K_i^+}(d\q')\;
 e^{-H_{\beta,\ga,\la_{\beta}}(\q'|\q_{K_i^+},\rho_\beta^+\text {\bf 1}_{C_i^+})}
 \nonumber
\\ && \hskip2cm \times
e^{-\beta \ga^{-d} F_{\beta,\la_\beta}(\rho_\beta^+
\text{\bf 1}_{\ga [C_i^+\cup T_i\cup C_i^-]})} \,{\hat
Z}^+_\ga \Big( \text{\rm int}^-(\Ga_i)\,;\,\mathcal
C_{\text{\rm int}^-(\Ga_i)\setminus A_i^-}^+\;\Big|
 \;\rho_\beta^+\text {\bf 1}_{C_i^-}\Big)   \nonumber\\
       \label{b6.19}
    \end{eqnarray}

\vskip.5cm

\subsection {Reconstruction of a plus partition function}

In order to reconstruct $\hat{D}^+_\ga$ we will reverse the
previous steps. Call $F_{\ga}=F_{\beta,\la(\beta,\ga)}$. Then,  as in
\eqref{b6.9}, one can replace $F_\ga$ by $F_{\beta,\la_\beta}$ in the
sense that  there exists a $c_5>0$ such that with
    \begin{eqnarray}
&& b_{5,\ga}^{(i)}: = e^{c_5(|C_i^+|+ |T_i|+ |C_i^-|)
\ga^{1/2}}
       \label{b6.20}
   \end{eqnarray}
 the following inequality holds
     \begin{eqnarray}
&& M_{\ga}^{(i)}(\q,\q_{K^+}) \leq \;
 ( g_{2,\ga}^{(i)})^2  b_{2,\ga}^{(i)}\cdots b_{5,\ga}^{(i)}
 \int_{\eta=\si=1}
 \nu_{ B_i^+\setminus K_i^+}(d\q')\;
 e^{-H_{\beta,\ga,\la(\beta,\ga)}(\q'|\q_{K_i^+},\rho_\beta^+\text {\bf 1}_{C_i^+})}
 \nonumber
\\ && \hskip2cm \times
e^{-\beta \ga^{-d} F_{\ga}(\rho_\beta^+ \text{\bf 1}_{\ga
[C_i^+\cup T_i\cup C_i^-]})} \,{\hat Z}^+_\ga \Big(
\text{\rm int}^-(\Ga_i)\,;\, \mathcal C_{\text{\rm
int}^-(\Ga_i)\setminus A_i^-}^+\;\Big|
 \;\rho_\beta^+\text {\bf 1}_{C_i^-}\Big) .  \nonumber\\
       \label{b6.21}
    \end{eqnarray}
Next, we want to replace the free
energy functional by a partition function using the Lebowitz-Penrose
coarse graining explained in Appendix~\ref{app:aC}. To do so, we
rewrite the r.h.s.\ of
\eqref{b6.21} as
     \begin{eqnarray}
&&  ( g_{2,\ga}^{(i)})^2  b_{2,\ga}^{(i)}\cdots
b_{5,\ga}^{(i)}
 \int_{\eta=\si=1}
 \nu_{ B_i^+\setminus K_i^+}(d\q')\;
 \int_{\eta=\si=1}
 \nu_{ A_i^-}(d\q'')\;
  e^{-H_{\beta,\ga,\la(\beta,\ga)}(\q'|\q_{K_i^+},\rho_\beta^+\text {\bf 1}_{C_i^+})}
 \nonumber
\\ && \hskip2cm \times
e^{-\beta \ga^{-d} F_{\ga}(\rho_\beta^+ \text{\bf 1}_{\ga
[C_i^+\cup T_i\cup C_i^-]})}
e^{-H_{\beta,\ga,\la(\beta,\ga)}(\q''|\rho_\beta^+\text
{\bf 1}_{C_i^-})} \,{\hat Z}^+_\ga \Big( \text{\rm
int}^-(\Ga_i) \setminus A^-_i\,\Big|
 \; \q_{A^-_i}''\Big) .   \nonumber\\
    \end{eqnarray}
Moreover, adding subscripts to remind the regions where the
loops start from, we obtain that
    \begin{eqnarray}
&&H_{\beta,\ga,\la(\beta,\ga)}(\q'_{B^+_i \setminus K^+_i}
|\q_{K_i^+},\rho_\beta^+\text {\bf 1}_{C_i^+}) +\beta
\ga^{-d} F_{\ga}(\rho_\beta^+ \text{\bf 1}_{\ga
[C_i^+\cup T_i\cup C_i^-]})
+H_{\beta,\ga,\la(\beta,\ga)}(\q''_{A^-_i}|\rho_\beta^+\text {\bf 1}_{C_i^-})\nonumber \\
&=&H_{\beta,\ga,\la(\beta,\ga)}(\q'_{B^+_i \setminus
K^+_i}|\q_{K_i^+}) +\beta \ga^{-d} F_{\ga}(\rho_\beta^+
\text{\bf 1}_{\ga [C_i^+\cup T_i\cup C_i^-]}|
\q'_{B^+_i \setminus K^+_i},\q''_{A^-_i})
+H_{\beta,\ga,\la(\beta,\ga)}(\q''_{A^-_i}), \nonumber
     \end{eqnarray}
Then \eqref{Ku132} implies that
    \begin{eqnarray}
&& -\beta \ga^{-d} F_{\ga}(\rho_\beta^+ \text{\bf 1}_{\ga
[C_i^+\cup T_i\cup C_i^-]}| \q'_{B^+_i \setminus
K^+_i},\q''_{A^-_i})
 \leq \log Z_{ \ga}( T_i \cup
C_i^-\cup C_i^+ |  \q'_{B^+_i \setminus K^+_i},\q''_{A^-_i})\nonumber\\
&& \hskip3.5cm  + c_6\ga^{1/2}( |T_i|+ |C_i^-|+ |C_i^+|)
        \label{b6.23}
    \end{eqnarray}
hence we proved that with $b_{6,\ga}^{(i)}:=e^{c_6 \ga^{1/2} \text{sp}(\Ga_i)}$ holds
     \begin{eqnarray}
&& M_{\ga}^{(i)}(\q,\q_{K^+}) \leq \;  ( g_{2,\ga}^{(i)})^2
b_{2,\ga}^{(i)}\cdots b_{6,\ga}^{(i)}\,{\hat Z}^+_\ga \Big(
\text{\rm int}^-(\Ga_i) \cup \text{sp}(\Ga_i)\setminus
K_i^+ \,;\, \mathcal C_{\text{\rm
int}^-(\Ga_i)\setminus A_i^-}^+ \;\Big| \q_{K_i^+}\Big) . \nonumber\\
    \end{eqnarray}

Summarizing, for $\gamma$ small enough there exists a
constant $c_7=c_2+c_3+c_5 + c_6$ such that
    \begin{eqnarray}
 ( g_{2,\ga}^{(i)})^2  b_{2,\ga}^{(i)}\cdots
b_{6,\ga}^{(i)} \leq e^{- \Big( 2c \ga^{-(1-\alpha)d +2a} - c_7
  \ell_{+,\ga}^d {\gamma^{1/2}} -
c_4
\ell_{+,\ga}^d {\gamma^{a'}}\Big) N_{\Ga_i}}
%
      \label{b6.31bb}
     \end{eqnarray}
By the choice of $a'$ holds that $d\alpha +2a< -d\alpha +
a'$. Recalling that $d\alpha + 2a <1/2$,  we obtain (\ref{b6.26}).
\notes{more precise formulation}

\vskip3cm

\setcounter{equation}{0}

\section{{\bf Equality of pressures}}
       \label{sec:a5}

In this section we will prove Theorems \ref{thm5.1} and
\ref{thm5.2}.

 \vskip1cm

\subsection{  Proof of Theorem \ref{thm5.1}.}

For notational simplicity we only refer to the $+$ case.
Let
 \begin{equation}
      \label{a5.17}
D_{\ga,\la}(n) := \frac {\log {\hat
Z}^{+}_{\ga,\la}(\La_n|\rho_\beta ^{+}\text{\bf
1}_{\La_n^c})}{\beta |\La_n|} - \frac {\log {\hat
Z}^{+}_{\ga,\la}(\La_{n-1}|\rho_\beta ^{+}\text{\bf
1}_{\La_{n-1}^c})}{\beta |\La_{n-1}|} .
 \end{equation}
We claim that there is a constant $c$ such that
 \begin{equation}
      \label{a5.18}
\sup_{|\la-\la_\beta|\le 1}\big|D_{\ga,\la}(n)\big| \leq
c2^{-n} .
 \end{equation}
Since for any $n$
 \begin{equation}
      \label{a5.19}
  \frac {\log {\hat Z}^{+}_{\ga,\la}(\La_n|\rho_\beta ^{+}\text{\bf
1}_{\La_n^c})}{\beta |\La_n|}
 \end{equation}
is a continuous function of $\la$, existence of the limit
\eqref{a5.4} and its continuity then follow from
\eqref{a5.18}--\eqref{a5.19}, thus it is sufficient to prove
\eqref{a5.18}.

\vskip.5cm

{\it Lower bound}.  Decomposing $\La_n$ into cubes
$\La_{n-1}(i)$ of side $2^{n-1}$.
We call $\La^0$ the union over all $i$ of
$\La_{n-1}(i)\setminus \delta_{\text{\rm
in}}^{\ell_{+,\ga}}[\La_{n-1}(i)]$ and $\La_{n-1}^0:=\La_{n-1} \setminus \delta_{\text{\rm
in}}^{\ell_{+,\ga}}[\La_{n-1}]$. We have
 \begin{equation}
      \label{a5.20}
{\hat Z}^{+}_{\ga,\la}(\La_n|\rho_\beta ^{+}\text{\bf
1}_{\La_n^c}) \ge
 \sum_{\{\Ga_i\}\in
\mathcal C^+_{\La^0} } \int_{\Q_+^{\La_n}}
\nu_\La(d\q)e^{-H_{\ga,\la}(\q|\rho_\beta ^{+}\text{\bf
1}_{\La_n^c})}
  \prod_{i}{\hat W}_{\ga,\la}^{+}(\Ga_i;\q) .
  \end{equation}
Exploiting that the configurations are in the restricted
ensemble, we can gain control of the interaction energy.
Referring to Lemma \ref{lemmaa5.3} for details, one obtains in this way
 \begin{equation}
      \label{a5.21}
\log {\hat Z}^{+}_{\ga,\la}(\La_n|\rho_\beta ^{+}\text{\bf
1}_{\La_n^c}) \ge 2^d \log {\hat
Z}^{+}_{\ga,\la}\Big(\La_{n-1}; \mathcal
C^+_{\La_{n-1}^0}\;\big|\rho_\beta ^{\pm}\text{\bf
1}_{\La_{n-1}^c} \Big) - c (2^n \ell_{+,\ga})^{d-1}
\ga^{-1} .
 \end{equation}%

As the weights are small (by definition) we can bound the
cost of droping their restriction to $\La_{n-1}^0$ in \eqref{a5.21}.
Referring to Lemma  \ref{lemmaa5.1} for details, we get that
 \begin{equation}
      \label{a5.22}
 \log {\hat
Z}^{+}_{\ga,\la}\Big(\La_{n-1};\mathcal
C^+_{\La_{n-1}^0}\;\big|\rho_\beta ^{+}\text{\bf
1}_{\La_{n-1}^c}\Big) \ge  \log {\hat
Z}^{+}_{\ga,\la}\big(\La_{n-1} |\rho_\beta ^{+}\text{\bf
1}_{\La_{n-1}^c}\big) -  2d (2^{n-1} )^{d-1}
e^{-c_f\ga^{-1}/2} .
 \end{equation}
Summarizing, the inequalities \eqref{a5.21} and \eqref{a5.22} show
that there is a constant
$c>0$ such that
 \begin{equation}
      \label{a5.23}
D_{\ga,\la}(n) \ge - c 2^{-n}.
 \end{equation}

       \vskip.5cm

{\it Upper bound}.  Using    Lemma  \ref{lemmaa5.1},
instead of \eqref{a5.20} we get
 \begin{eqnarray*}
&& {\hat Z}^{+}_{\ga,\la}(\La_n|\rho_\beta ^{+}\text{\bf
1}_{\La_{n}^c}) \le \; e^{2^d2d(2^{n-1} )^{d-1}
e^{-c_f\ga^{-1}/2}}\\&&\hskip3cm \times
 \sum_{\{\Ga_i\}\in
\mathcal C^+_{\La^0} } \int_{\Q^{\La_n}_+}
\nu_\La(d\q)e^{-H_{\ga}(\q|\rho_\beta ^{+}\text{\bf
1}_{\La_{n}^c})}
  \prod_{i}{\hat W}_{\ga,\la}^{+}(\Ga_i;\q)
 \end{eqnarray*}
and, instead of  \eqref{a5.21}, using Lemma  \ref{lemmaa5.1} and
\ref{lemmaa5.3} we obtain
 \begin{eqnarray*}
\log {\hat Z}^{+}_{\ga,\la}(\La_n|\rho_\beta ^{+}\text{\bf
1}_{\La_{n}^c}) &\leq & 2^d \Big[\log {\hat
Z}^{+}_{\ga,\la}\Big(\La_{n-1}; \mathcal
C^+_{\La_{n-1}^0}\;\big|\rho_\beta ^{+}\text{\bf
1}_{\La_{n-1}^c} \Big)
\\&&+\;2d(2^{n-1} )^{d-1} e^{-c_f\ga^{-1}/2}\Big] + c (2^n \ell_{+,\ga})^{d-1}
\ga^{-1} .
 \end{eqnarray*}
Hence, one concludes that $D_{\ga,\la}(n)\le c 2^{-n}$, which together with
\eqref{a5.23} proves \eqref{a5.18} and thus the theorem. \qed

\vskip1cm

\subsection{  Proof of Theorem \ref{thm5.2}}

In the traditional approach to the Pirogov-Sinai technique, the statement is
proved using the implicit function theorem after
establishing bounds on the derivatives w.r.t.\ $\la$; such
an approach yields local uniqueness as well. To avoid bounds on
derivatives, which are here not so straightforward, we will
use a different, weaker approach, showing that the
difference ${\hat P}^+_{\ga,\la}-{\hat P}^-_{\ga,\la}$ is a
continuous function of $\la$ which undergoes a change of
sign. Continuity in $\la$ has been proved in Theorem
\ref{thm5.1}. The change of sign property follows from the analogous
change of sign property in the mean field limit together by showing
that the mean field for $\ga$ is close enough to ${\hat P}^+_{\ga,\la}-{\hat
P}^-_{\ga,\la}$ for small enough $\ga$.

We restrict $\la$ to
the interval $|\la-\la_\beta| \le \ga^{2a}$, recalling that
$\zeta=\ga^a$ and $\ga^{2a}\gg \ga^{1/2}$, and we want to show that ${\hat P}^+_{\ga,\la}-{\hat
P}^-_{\ga,\la}$ has a change of sign for $\la$ in such an interval.

Using  Lemma \ref{lemmaa5.1}, we can drop, with a ``small
error'', the sum over $\{\Ga_i\}$ in \eqref{a5.1},
retaining only the term without contours.  We then use
Proposition \ref{propaB.1} to get
    \begin{eqnarray}
&&  {\hat P}^{\pm}_{\ga,\la} \leq -\lim_{n\to \infty}
\inf_{\rho \in \mathcal A^{\pm}_n} \frac { F_{\beta,\la,
\ga\La_n }(\rho|\rho_\beta ^{\pm}\text{\bf
1}_{\La_{n}^c})}{\ga^d |\La_n|} + \frac c \beta\, \ga^{1/2},
      \label{a5.25}
    \end{eqnarray}
where $\mathcal A^{\pm}_n$ is the set of all $\rho \in
L^\infty(\ga\La_n,\mathbb R_+)$ such that the averages in
the cubes of $\mathcal D^{(\ga\ell_{-,\ga})}$ are close to
$\rho^{\pm}_{\beta}$ within $\zeta=\ga^a$.   By
inequality \eqref{aB.6} we see that
    \begin{eqnarray}
&&  {\hat P}^{\pm}_{\ga,\la} \leq  \frac c \beta\,
\ga^{1/2} - \lim_{n\to \infty} \inf_{\rho \in \mathcal
A^{\pm}_n} \frac{1}{\ga^d|\La_n|}\int_{\ga \La_n}
f_{\beta,\la}(j*[\rho_{\La_n}+ \rho_\beta^{\pm}\text{\bf
1}_{\La_n^c}]) dr .
      \label{a5.26}
    \end{eqnarray}
By the bound \eqref{d0.4}, for $\rho \in \mathcal A^{\pm}_n$, there
is $c'>0$ so that
    \begin{eqnarray}
&& j*[\rho_{\La_n}+ \rho_\beta^{\pm}\text{\bf 1}_{\La_n^c}]
\in [\rho_\beta^{\pm} -\zeta- c' \ga\ell_{-,\ga},
\rho_\beta^{\pm} +\zeta + c' \ga\ell_{-,\ga}]
      \label{a5.27}
    \end{eqnarray}
 Then, by \eqref{a5.26}, we derive that
    \begin{eqnarray*}
&&  {\hat P}^{\pm}_{\ga,\la} \leq  \frac c \beta\,
\ga^{1/2} -
 \inf_{|s- \rho_\beta^{\pm}|\le \zeta + c' \ga\ell_{-,\ga}}
f_{\beta,\la}(s) .
    \end{eqnarray*}
Recalling that $f_{\beta,\la_\beta}(\cdot)$ is a double
well function with two minimizers $\rho_\beta^{\pm}$, we
deduce that for $\la-\la_\beta$ small enough,
$f_{\beta,\la}(\cdot)$ is still double well with local
minima at $\rho_{\beta,\la}^{\pm}$.  Moreover,
$|\rho_{\beta,\la}^{\pm}- \rho_{\beta}^{\pm}|\le
c|\la-\la_\beta| \le c\ga^{2a}$. For $\ga$ small, it hold
$|\rho_{\beta,\la}^{\pm}- \rho_\beta^{\pm}|\le \zeta + c'
\ga\ell_{-,\ga}$ and
    \begin{eqnarray}
&&  {\hat P}^{\pm}_{\ga,\la} \leq  \frac c \beta\,
\ga^{1/2} - f_{\beta,\la}(\rho_{\beta,\la}^{\pm}) .
      \label{a5.28}
    \end{eqnarray}

For the lower bound we use \eqref{aC.7} to get that
    \begin{eqnarray*}
&&  {\hat P}^{\pm}_{\ga,\la} \geq -\lim_{n\to \infty}
\frac { F_{\beta,\la, \ga\La_n }(\rho^{(n)}|\rho_\beta
^{\pm}\text{\bf 1}_{\La_{n}^c})}{\ga^d |\La_n|} - \frac c
\beta\, \ga^{1/2}, \qquad \rho^{(n)} \in \mathcal A^{\pm}_n .
    \end{eqnarray*}
By the previous argument, for $\ga$ small enough
$\rho^{(n)}(r) = \rho_{\beta,\la}^{\pm} \text{\bf
1}_{\La_n} \in A^{\pm}_n $, so that
    \begin{eqnarray}
&&  {\hat P}^{\pm}_{\ga,\la} \geq -
f_{\beta,\la}(\rho^{\pm}_{\beta,\la}) -  \frac c \beta\,
\ga^{1/2} .
      \label{a5.29}
    \end{eqnarray}
Summarizing, the inequalities \eqref{a5.28}--\eqref{a5.29} yield that
    \begin{eqnarray*}
&& \Big|[ {\hat P}^{+}_{\ga,\la}- {\hat P}^{-}_{\ga,\la}]-
[f_{\beta,\la}(\rho^{-}_{\beta,\la}) -
f_{\beta,\la}(\rho^{+}_{\beta,\la})]\Big|  \leq \frac{2
c}{\beta}\, \ga^{1/2} ,
    \end{eqnarray*}
which completes the proof of Theorem \ref{thm5.2}, because iit holds that
   \[
f_{\beta,\la_\beta}(\rho^-_{\beta})=f_{\beta,\la_\beta}(\rho^+_{\beta}),\qquad
\frac{d}{d\la}[f_{\beta,\la}(\rho^-_{\beta,\la})-f_{\beta,\la}(\rho^+_{\beta,\la})]
\Big|_{\la=\la_\beta} =  \rho^+_{\beta}-\rho^-_{\beta} > 0 .
   \]
 \qed

\vskip3cm

\setcounter{equation}{0}

\section{{\bf Small deviations}}
       \label{sec:a8}

In this section we will prove \eqref{b6.13}.  Let $\La$ be
a bounded $\mathcal D^{(\ell_{+,\ga})}$-measurable region,
which satisfies the following ``fatness property'':
  \begin{equation}
     \label{a8.0}
\Big| \big\{r: \text{dist}(r,\La)\leq 4 \ga^{-1}\big\}\Big|
\le 3^d |\La| ,
    \end{equation}
where $3^d-1$ is the number of neighbours on $\mathbb Z^d$. Let then
   \begin{equation}
     \label{a8.1}
Z_{\ga,\la_\beta}(\La;   \Q^\La_{\pm} \, |\bar\q) =
\int_{\Q^\La_{\pm}}\nu_\La(d\q)\,
e^{-H_{\beta,\ga,\la_\beta}(\q|\bar \q)}
    \end{equation}
and suppose $\si(\bar\q)\equiv 1$, $\bar\q(0) \subset
\La^c$ and that Av$^{(\ell_{-,\ga})}(\bar\q(0);\cdot)\le
2\rho_{\beta^+}$, conditions which are fulfilled in the
case of \eqref{b6.13}.

Let finally $\Delta$ be a $\mathcal
D^{(\ell_{+,\ga})}$-measurable subset of $\La$ and suppose
   \begin{equation}
     \label{a8.2}
\text{\rm dist}(\Delta,\La^c) \ge 10 \ga^{-1}+ \ga^{-1}
\ell,\quad \ell>0 .
    \end{equation}
In the proof of  \eqref{b6.13}, we take
for $\La$ a maximal connected component either of
    \[
  A_i^- \;\cup\;
 \Big( \delta_{\text{\rm in}}^{\ell_{+,\ga}}[\text{sp}(\Ga_i)]
 \cap \{r\in  \text{sp}(\Ga_i): \eta_{\Ga_i} =-1\}\Big)
    \]
or else of
   \[
 \Big(\delta_{\text{\rm in}}^{\ell_{+,\ga}}[\text{sp}(\Ga_i)] \cap
 \{r\in  \text{sp}(\Ga_i): \eta_{\Ga_i} =1\}\Big) \setminus K_i^+ .
    \]
In either case \eqref{a8.0} is satisfied.

In the proof of  \eqref{b6.13} one chooses $\Delta$ as either $C_i^{-}$
or $C_i^{+}$. Since $\ga^{-1}\ell > \ell_{+,\ga}/100$, then
\eqref{b6.13} follows from \eqref{a8.3} below.

\vskip1cm

    \begin{prop}
    \label{propa8.1}
There are $\om>0$ and $c>0$ such that for any $\La$ and
$\Delta$ as above one has
%
%
    \begin{eqnarray}
Z_{\ga,\la_\beta}(\La; \Q^\La_{\pm} \,|\bar\q) &\leq &
e^{c( \ga^{1/2} |\La| + e^{- \om \ell}|\Delta|)}
e^{-\beta\ga^{-d} F_{\beta,\la_\beta}(\rho_\beta^{\pm}
\text{\bf 1}_{\ga\Delta})}\nonumber \\&& \hskip-2cm\times
Z_{\ga,\la_\beta}\Big(\La\setminus \Delta; \Q^{\La\setminus
\Delta}_{\pm} \, |\bar\q_{\La^c}, \rho_\beta^{\pm}
\text{\bf 1}_{\Delta}\Big) .
     \label{a8.3}
    \end{eqnarray}

    \end{prop}

\vskip.5cm

{\bf Proof.}  For the sake of definiteness we restrict ourselves to
the plus case.  By using \eqref{aC.6} and using the assumption
\eqref{a8.0}, there is $c>0$ such that
    \begin{eqnarray}
&& \hskip-1cm  \log Z_{\ga,\la_\beta}(\La;\Q^{\La}_{\pm} \,
|\bar\q) \le - \inf_{\rho \in \mathcal A^*} \beta \ga^{-d}
F_{\beta,\la_\beta}(\rho|\bar\rho) + c\ga^{1/2} |\La|
     \label{a8.4}
    \end{eqnarray}
where
   \begin{eqnarray}
&& \mathcal A^*:= \Big\{ \rho \in L^\infty(\ga \La,\mathbb
R_+): \eta^{(\zeta,\ga\ell_{-,\ga})}(\rho; r)= 1 \Big\}
     \label{a8.5}
    \end{eqnarray}
and $\eta^{(\zeta,\ell)}(\rho;r)$ is defined as in
\eqref{a3.2} with $ |q\cap C_r^{(\ell)}|$ replaced by
$\int_{C_r^{(\ell)}}
   \rho(r') \, dr $.
 For $\eps>0$, let $\rho_\eps\in  \mathcal A^*$  be such that
   \begin{eqnarray}
&& F_{\beta,\la_\beta}(\rho_\eps|\bar\rho) \leq \; \eps+
\inf_{\rho \in \mathcal A^*}
F_{\beta,\la_\beta}(\rho|\bar\rho) .
     \label{a8.6}
    \end{eqnarray}
Calling
    \begin{eqnarray}
 D= \delta_{\text{\rm in}}^{10\ga^{-1}}[\La],\quad  \mathcal B^*=
  \Big\{ \rho \in L^\infty(\ga (\La\setminus D),\mathbb R_+):
\eta^{(\zeta,\ga\ell_{-,\ga})}(\rho;r)= 1 \Big\}
     \label{a8.7}
    \end{eqnarray}
then one gets that
   \begin{eqnarray}
&&   \inf_{\rho \in \mathcal A^*}
F_{\beta,\la_\beta}(\rho|\bar\rho) \geq - \eps
+F_{\beta,\la_\beta}(\rho_\eps\text{\bf 1}_D|\bar\rho) +
\inf_{\rho \in \mathcal B^*} F_{\beta,\la_\beta}(\rho|
\rho_\eps\text{\bf 1}_D).
     \label{a8.8}
    \end{eqnarray}
In Theorem~\ref{thmdB.1} it is proved that the inf on the
r.h.s.\ is actually a minimum, the minimizer is unique, say
$\hat\rho_\eps$, and there are $\omega >0$ and $c_\omega$
such that for all $r \in \Lambda \setminus D$:
   \begin{eqnarray}
&& | \hat \rho_\eps(r) -\rho_\beta^+| \leq c_\om e^{-\om
\text{dist}(r,D)} .
     \label{a8.9}
    \end{eqnarray}
Thus, calling $\tilde \rho_\eps=\rho_\eps\text{\bf
1}_D+\hat\rho_\eps$,
   \begin{eqnarray*}
 &&  \inf_{\rho \in \mathcal A^*}
F_{\beta,\la_\beta}(\rho|\bar\rho)+  \eps \geq
 F_{\beta,\la_\beta}(\tilde\rho_\eps |\bar\rho)
 \geq
F_{\beta,\la_\beta}(\hat \rho_\eps\text{\bf 1}_{\ga
\Delta}) + F_{\beta,\la_\beta}(\tilde\rho_\eps \text{\bf
1}_{\ga
(\La\setminus \Delta)}|\bar\rho+\hat \rho_\eps\text{\bf 1}_{\ga\Delta}) \nonumber \\
&&\hskip1cm \geq -c e^{-\om \ell}|\ga \Delta| +
F_{\beta,\la_\beta}( \rho_\beta^+\text{\bf 1}_{\ga
\Delta})+ F_{\beta,\la_\beta}\Big(\tilde\rho_\eps \text{\bf
1}_{\ga (\La\setminus \Delta)}\Big |\bar\rho +
 \rho_{\beta}^+\text{\bf 1}_{\ga \Delta}\Big )
    \end{eqnarray*}
and choosing $\eps <c e^{-\om \ell}|\ga \Delta|$,
   \begin{eqnarray}
 &&\hskip-1cm   \inf_{\rho \in \mathcal A^*}
F_{\beta,\la_\beta}(\rho|\bar\rho)  \geq  -2c e^{-\om
\ell}|\ga \Delta| + F_{\beta,\la_\beta}(
\rho_\beta^+\text{\bf 1}_{\ga \Delta})+
F_{\beta,\la_\beta}\Big(\tilde\rho_\eps \text{\bf 1}_{\ga
(\La\setminus \Delta)}\Big |\bar\rho +
 \rho_{\beta}^+\text{\bf 1}_{\ga \Delta}\Big ) .
     \label{a8.10}
    \end{eqnarray}
Inequality \eqref{aC.7} yields that
    \begin{eqnarray}
&& \hskip-1cm \exp\Big\{-\beta \ga^{-d} F_{\beta,\la_\beta
}\Big(\tilde\rho_\eps \text{\bf 1}_{\ga (\La\setminus
\Delta)} \Big|\bar\rho +
 \rho_{\beta}^+\text{\bf 1}_{\ga \Delta}\Big) \Big\} \nonumber
 \\&& \hskip2cm
\leq  e^{ c\ga^{1/2} |\La|}\;
Z_{\ga,\la_\beta}\Big(\La\setminus \Delta;\Q^{\La}_{\pm}
\,\Big|\bar\q, \rho_{\beta}^+\text{\bf 1}_{ \Delta}\Big) .
     \label{a8.11}
    \end{eqnarray}
Summarizing, the inequalities \eqref{a8.4},  \eqref{a8.10} and  \eqref{a8.11} then
yield \eqref{a8.3}.  \qed

\vskip3cm

\setcounter{equation}{0}

\section{{\bf Large deviations}}
       \label{sec:a9}

In this section, we will prove the statements about the ``large deviations'' from
equilibrium to which we refered to in Section
\ref{sec:a6}. We will repeatedly use Lemma \ref{lemma9a.1}
below

Denote in the following by
    \begin{equation}
         \label{9a.1}
N:=e^{-\beta b} \ell_{-,\ga}^d   e^{2}
    \end{equation}
and denote by
    \begin{equation}
         \label{9a.1d}
n(r): = |\q(0) \cap C^{(\ell_{-,\ga})}_r|,
    \end{equation}
for a given configuration $\q$.

 \vskip1cm

 \begin{lemma}
 \label{lemma9a.1}
Let $\Delta$ be a bounded $\mathcal
D^{(\ell_{+,\ga})}$-measurable region, $\eta^*$ a $\{0,\pm
1\}$-valued function on $\Delta$, $\bar \q$ any finite loop
configuration. Then for $N$ as in \eqref{9a.1} holds
    \begin{eqnarray}
&& \frac{Z_{\ga,\beta,\la(\beta,\ga)} \Big(\Delta; \{
\eta=\eta^*,\si=1\}\big|
\bar\q\Big)}{Z_{\ga,\beta,\la(\beta,\ga)} \Big(\Delta; \{
\eta=\eta^*,\si=1, n(r)\leq N, r\in \Delta \}\big|
\bar\q\Big)}  \leq \; 3^{
 |\Delta|/ \ell_{-,\ga}^d}. 
      \label{9a.2}
     \end{eqnarray}

 \end{lemma}

\vskip1cm

{\bf Proof.} Denote by $X$ be the set of the centers of the
cubes of $\mathcal D^{(\ell_{-,\ga})}$ which are in
$\Delta$. Then one can write
    \begin{eqnarray}
&& Z_{\ga,\beta,\la(\beta,\ga)} \Big(\Delta; \{
\eta=\eta^*,\si=1\}\big| \bar\q\Big) = \sum_{ Y\subset X}
Z_{\ga,\beta,\la(\beta,\ga)} \Big(\Delta; \{
\eta=\eta^*,\si=1, \nonumber \\&&  \hskip3cm
 n(x)\leq N, x\in X\setminus Y, n(y)>N, y\in Y
\}\big| \bar\q\Big) .
      \label{9a.3}
     \end{eqnarray}
Next, we will show
    \begin{eqnarray}
&&
 \frac{Z_{\ga,\beta,\la(\beta,\ga)} \Big(\Delta; \{ \eta=\eta^*,\si=1,
 n(x)\leq N, x\in X\setminus Y, n(y)>N, y\in Y
\}\big| \bar\q\Big) }
 {  Z_{\ga,\beta,\la(\beta,\ga)}
\Big(\Delta; \{ \eta=\eta^*,\si=1,
 n(x)\leq N, x\in X\setminus Y, n(y)=N, y\in Y
\}\big| \bar\q\Big)}
 \nonumber
\\&& \hskip3cm
 \leq  \Big(\sum_{p>0} \frac{N!}{(N+p)!} [e^{-\beta b
}\ell_{-,\ga}^d]^p\Big)^{|Y|} =: C(Y) .
      \label{9a.4}
     \end{eqnarray}
To derive \eqref{9a.4}, we first write
   \[
\nu_{\beta,\Delta}(d\q) = \prod_{x\in X}
\nu_{\beta,{C^{(\ell_{-,\ga})}_x}}(d\q_{C^{(\ell_{-,\ga})}_x}) .
  \]
Then in each cube $C^{(\ell_{-,\ga})}_y$, $y\in Y$, which
has $N+p$ particles, $p>0$, we pick $p$   particles  and
denote by $\q'$ the collection of all this particles from all
the cubes with $y\in Y$. Calling $\q$ the the configuration of the remaining particles,
by \eqref{2.a11} one obtains
    \begin{equation}
            \label{9a.6}
H_{\ga}(\q+\q'| \bar\q) - H_{\ga}(\q| \bar\q)=
 H_{\ga}(\q'|\q, \bar\q)
 \ge b |\q'| .
    \end{equation}
This yields   \eqref{9a.4}, because
$\eta(\q(0);x)=\eta(n(x))$ is as a function of the number
$n(x)$ constant if
$n(x)\geq N$ due to defintion \eqref{a3.2}.

Recalling the Stirling formula
      \begin{equation}
    \label{3e.13}
n! = n^{n+1/2} e^{-n} \sqrt{2\pi}\Big( 1 +
0\left(\frac{1}{\sqrt n}\right)\Big)
     \end{equation}
we can write
      \begin{equation*}
(N+p)!\ge  N^{(N+p)+1/2} e^{-(N+p)} \sqrt{2\pi}\Big( 1 +
0\left(\frac{1}{\sqrt {N+p}}\right)\Big) \ge \frac12\;\;N!
N^p e^{-p}
     \end{equation*}
for $N$ large enough, that is, recalling \eqref{9a.1}, for
$\ga$ small enough.  Then definition \eqref{9a.1} implies
      \begin{equation*}
\frac{[e^{-\beta b }\ell_{-,\ga}^d]^{p}}{(N+p)!} \leq
\frac{2}{N!} \left(\frac{[e^{-\beta b }\ell_{-,\ga}^d] e
}{N} \right)^p \leq \frac{2}{N!}e^{-p},
     \end{equation*}
which yields, recalling the definition of $C(Y)$ in \eqref{9a.4},
    \begin{eqnarray}
&& C(Y) \leq 2^{|Y|}.
      \label{9a.7}
     \end{eqnarray}
Summarizing, using \eqref{9a.3}-\eqref{9a.4} one can extimate
    \begin{eqnarray}
&& Z_{\ga,\beta,\la(\beta,\ga)} \Big(\Delta; \{
\eta=\eta^*,\si=1\}\big| \bar\q\Big) \leq 3^{|X|}
Z_{\ga,\beta,\la(\beta,\ga)} \Big(\Delta; \{
\eta=\eta^*,\si=1, \nonumber
\\&&  \hskip3cm
 n(x)\leq N, x\in X
\}\big| \bar\q\Big),
      \label{9a.8}
     \end{eqnarray}
which proves the lemma.  \qed

\vskip1cm

{\it Proof of \eqref{b6.9}}.  Applying Lemma \ref{lemma9a.1} for
$\Delta = A_i^-\,\cup\, \text{sp}(\Ga_i) \setminus
K_i^+$, one obtains
    \begin{eqnarray*}
M_{\ga}^{(i)}(\q,\q_{K^+}) &\leq&
3^{|\Delta|/\ell_{-,\ga}^d} \int_{\si=1;\, \eta=\eta_\Ga
\text{ on sp$(\Ga_i)$, $\eta=-1$ on
 $A^-_i$; $n(r)\leq N, r\in \Delta$}}
 \nu_{\Delta}(d\q')\;
  \nonumber
\\ && \hskip2cm\times  e^{-H_{\beta,\ga,\la(\beta,\ga)}(\q'|\q_{K^+},\q)}
 {\hat Z}^-_\ga \Big( \text{\rm int}_i^-(\Ga) \setminus A_i^-\;\Big|
 \;\q'\Big)\nonumber\\
     \end{eqnarray*}
The bound \eqref{b6.9} follows from
  \[
 H_{\beta,\ga,\la_\beta}(\q'|\q_{K^+},\q)-H_{\beta,\ga,\la(\beta,\ga)}(\q'|\q_{K^+},\q)
 = |\q'|\big(\la(\beta,\ga)-\la_\beta\big),
  \]
that by Theorem~\ref{thm5.2} holds
$|\la_\beta-\la(\beta,\ga)|\leq c \ga^{1/2}$, that by \eqref{dA.5a} holds $|\q'|
\leq [e^{-\beta b } e^2] |\Delta|$ and from dDropping the
condition $n(\cdot)\le N$. \qed

\vskip1cm

{\it Proof of \eqref{b6.14b}.}  To make the notation lighter we
drop the suffix $i$ and write  $T$, $C^{\pm}$ instead of $T_i$,
$C_i^{\pm}$. Due to \eqref{9a.2}, we can bound the left hand side of \eqref{b6.14b} by
 \begin{eqnarray*}
&& \leq 3^{
 |T|/ \ell_{-,\ga}^d}
\int_{\eta=\eta_\Ga,\si=1, n(r)\leq N, r \in T}
 \nu_{ T}(d\q_{T})\;
 e^{-H_{\beta,\ga,\la_{\beta}}(\q_{T}|\rho_\beta^-\text {\bf 1}_{C^-},
 \rho_\beta^+\text {\bf 1}_{C^+})}
     \end{eqnarray*}
According to \eqref{Ku133}, there is a constant   $c>0$, such that the
integral on the right hand side can be bounded by the expression
    \begin{eqnarray}
&&  \leq - \inf_{\rho \in
\mathcal A^*} \beta \ga^{-d} F_{\beta,\la_\beta
}(\rho|\bar\rho) + c\ga^{1/2}  |T|
     \label{9a.9}
    \end{eqnarray}
where
   \begin{eqnarray}
&&  \hskip-1cm \mathcal A^*= \Big\{ \rho \in L^\infty(\ga
T,\mathbb R_+): \rho(r)\le X_{{0}}=e^{-\beta b+2},\;
\eta^{(\zeta,\ga \ell_{-,\ga})}(\rho;r)=\eta_\Ga(\ga r),
r\in \ga T\Big\}
      \label{9a.10}
    \end{eqnarray}
and
   \begin{eqnarray}
&&   \bar\rho = (\rho_\beta^-\text {\bf 1}_{\ga C^-},
 \rho_\beta^+\text {\bf 1}_{\ga C^+}) .
     \label{9a.11}
    \end{eqnarray}
A bound for the minimization problem in \eqref{9a.9} is
proved in Theorem \ref{thmD.1}. Taking
$\La=\ga T$ and $\ell=\ga \ell_{-,\ga}$ and for $\ga$
small enough, this bound implies
   \begin{eqnarray}
&&   N_0+ N_{\pm} \geq  3^{-d}N_T , 
     \label{9a.12}
    \end{eqnarray}
where $N_0$ and $N_{\pm}$ and the families $\{C_i^0, i \leq N_0\}$ and
$\{C_i^{\pm},i \leq N_\pm\}$ are defined in
\eqref{D.8b}-\eqref{D.8c}, while $N_T$ is the number of
cubes from $\mathcal D^{(\ell_{+,\ga})}$ which are contained
in $T$.  In fact if $C$ is a $\mathcal
D^{(\ga\ell_{+,\ga})}$ cube in $\gamma T$ and $C^*$ is the union
of $C$ and its neighboring cubes, it cannot happen, by the
definition of contour, that $\eta^{(\zeta,\ga
\ell_{-,\ga})}(\rho;r)$ is constantly equal to 1 or to $-1$
in $C^*$, hence $C^*$ must contain at least one element of
$\{C^0_i\}$ or of $\{C_j^{\pm}\}$.
Since the number of the cubes contiguous to a given cube is
$3^d-1$, one obtains \eqref{9a.12}.

 Thus, the bound \eqref{D.5} yields, where we call the constant in \eqref{D.5}
 $c'$ instead of $c$,
    \begin{eqnarray}
&&  \text{l.h.s.\ of\ \eqref{b6.14b}} \leq -   \beta
\ga^{-d}\Big\{ [c' \zeta^2 (\ga \ell_{-,\ga})^d ]3^{-d}N_T
\nonumber
\\&&\hskip3cm + f_{\beta,\la_\beta}(\rho_\beta^+) |\ga T|+I_{\ga
T,\ga C^-}^- +I_{\ga T,\ga C^+}^+ \Big\} + c\ga^{1/2}  |T|.
     \label{9a.13}
    \end{eqnarray}

\vskip3cm

\section{{\bf Finite volume corrections to the pressure}}
       \label{sec:10}

We need to prove that the absolute value of the difference
between
 \begin{eqnarray}
&&  \log  \Big\{  e^{\beta \ga^{-d} I_{\ga \La,\ga
\La^c}^+} {\hat Z}^+_\ga \Big(
 \La \,;\, \mathcal C^+_{\La\setminus A}\;\Big|
 \;\rho_\beta^+\text {\bf 1}_{\La^c}\Big)\Big\}
     \label{10.1}
     \end{eqnarray}
(shorthanding $\La=\text{\rm int}^-(\Ga_i)$ \notes{index changed} and $ A=
A_i^-$) and the corresponding term relative to the minus
ensemble,  is bounded by $c |A|\ga^{a'}$, with $c$ a
constant and $a'$ as in \eqref{b6.17}.

\vskip1cm

\subsection{ Main result}

This will be achieved by proving that
 \begin{eqnarray}
&&  \eqref{10.1} = {\beta} {\hat
P}^+_{\ga,\la(\beta,\ga)}|\La^0| \tobias{-}{+} \beta
f_{\beta,\la_\beta}(\rho^+_\beta) |\La\setminus \La^0| +
\int_{\La\cup \delta_{\text{\rm
out}}^{\ell_{+,\ga}}[\La]} L_{\ga,\La}(r)\,dr
     \label{10.2}
     \end{eqnarray}
where
 \begin{eqnarray}
&& \hskip-1cm\La^0:= \Delta_{10}
     \label{10.2a}
     \end{eqnarray}
having called $\Delta_k=
\Delta_{k-1}\setminus \delta_{\text{\rm
in}}^{\ell_{+,\ga}}[\Delta_k]$, $\Delta_0=\La$, $10$ is not
optimal.  The remainder term $L_{\ga,\La}(r)$ is ``small''
in the following sense. There are $\om_0$, $c$ and $c'$
positive so that
   \begin{eqnarray}
&&  \big|L_{\ga,\La}(r)\big| \leq c e^{- \om_0 \ga
\text{\rm dist}(r,\La^c )},\;\; r \in \La^0;
   \label{10.3a}
   \\
&& \big| \int_{\La\cup \delta_{\text{\rm
out}}^{\ell_{+,\ga}}[\La]\setminus \La^0} L_{\ga,\La}(r)
\;dr\big| \leq c' \ga^{a'}|A|
     \label{10.3b}
     \end{eqnarray}
After proving the analogous bounds in the minus case, we
exploit the identities ${\hat
P}^+_{\ga,\la(\beta,\ga)}={\hat P}^-_{\ga,\la(\beta,\ga)}$
and
$f_{\beta,\la_\beta}(\rho^+_\beta)=f_{\beta,\la_\beta}(\rho^-_\beta)$,
to conclude that
 \begin{eqnarray}
&&  \text{l.h.s.\ of \eqref{b6.17} } \leq  \int_{\La^0} 2 c
e^{- \om_0 \ga\text{\rm dist}(r,\La^c)}\,dr + 2c'
\ga^{a'}|A|
     \label{10.4}
     \end{eqnarray}
which  proves \eqref{b6.17}.

\vskip.5cm

\subsection{Strategy of proof}

\eqref{10.2}-\eqref{10.3b} is an estimate of the  ``finite
volume corrections to the pressure'', a result usually
proved using cluster expansion techniques to exponentiate
the partition function.  One then obtains a sum (or an
integral) of quasi local terms, which differ from the
pressure by a term exponentially small with the distance
from the boundary of the domain. The terms which are close
to the boundaries however, giving a contribution which is
proportional to the surface times the range of the
interaction, are a priori larger than the gain term from
the energy of the contour. The special choice of the
boundary conditions, given by the constant value
$\rho_\beta^+$, allows to compute such dangerous terms to
leading order as $\ga\to 0$, giving $\dis{e^{-\beta
\ga^{-d} I_{\ga \La,\ga \La^c}^+}}$, which exactly cancels
with the prefactor in \eqref{10.1}. The next order
correction in $\ga$ gives rise to the term $ c'
\ga^{a'}|A|$ in \eqref{10.3b}, which covers also the
exponentially small corrections to the bulk terms and it is
infinitesimal w.r.t.\ the gain term from the contour as
$\ga\to 0$.

In order to prove \eqref{10.2}-\eqref{10.3b},  we will
choose among the several
 Dobrushin-Shlosman versions of cluster expansion, \cite{DS}, the one based
on an interpolation of the Hamiltonian.
\tobias{$L_{\ga,\La}(r)$}{$\pi^+_{\ga,\La}(r)$} will then be
the expectation of a local function translated by $r$; the
expectation is with respect to \tobias{a Gibbs measure for
a convex combination of Hamiltonians}{convex combinations of
Gibbs measures with the interpolated Hamiltonians}.
Exponential decay of correlations for such measures is then
responsible for \eqref{10.3a}, while \eqref{10.3b} will be
proved using a Lebowitz-Penrose argument.

\vskip1cm

\subsection{An Hamiltonian representing contours}

By using cluster expansion we can regard the contribution
of the contours to the partition function as an additional
term in the Hamiltonian:
     \begin{equation}
      \label{10a.5}
\sum_{ \{\Ga_i\} \subset \mathcal C^+_{\La\setminus A}}
\;\; \prod_{i} {\hat W}_{\ga,\la}^{+}(\Ga_i;\q) =
e^{-K^+_{\La\setminus A}(\q_{\La\setminus A})}
      \end{equation}
Referring to Appendix \ref{app:F} for more details, we
mention here that the Hamiltonian $K^+_{\La\setminus
A}(\q_{\La\setminus A})$ is expressed in terms of
potentials $U_\Delta^+$:
     \begin{equation}
      \label{10a.6}
K^+_{\La\setminus A}(\q_{\La\setminus A})= \sum_{\Delta\in
\La\setminus A} U^+_\Delta(\q_\Delta)
      \end{equation}
with $U^+_\Delta$ as in \eqref{gg.8} and \eqref{gg.9}.  We
then call
     \begin{equation}
      \label{10a.7}
H^+_{\ga}(\q  |\rho_\beta^+\text {\bf 1}_{\La^c} )=
H_{\beta,\ga,\la(\beta,\ga)}(\q |\rho_\beta^+\text {\bf
1}_{\La^c} ) + K^+_{\La\setminus A}(\q_{\La\setminus A})
      \end{equation}
and have
 \begin{eqnarray}
&& \hat Z^+_\ga \Big(
 \La \,;\, \mathcal C^+_{\La\setminus A}\;\Big|
 \;\rho_\beta^+\text {\bf 1}_{\La^c}\Big) = \int_{\Q^\La_+}
 \nu_\La(d\q) e^{-H^+_{\ga}(\q  |\rho_\beta^+\text {\bf 1}_{\La^c}
 )}
     \label{10a.8}
     \end{eqnarray}

\vskip1cm

\subsection{ Interpolating Hamiltonians}

Given a   a reference Hamiltonian $H_0(\q)$, we set
     \begin{equation}
      \label{10a.9}
H^+_{\ga,u} = uH^+_{\ga}+ (1-u)H_0,\qquad u\in [0,1]
      \end{equation}
with $H^+_{\ga}$ defined  in \eqref{10a.7}. $H^+_{\ga,u}$
is the interpolating Hamiltonian and $u$ the interpolating
parameter. Then
 \begin{eqnarray}
&& \hskip-2cm \log {\hat Z}^+_\ga \Big(
 \La \,;\,
  \mathcal C^+_{\La\setminus A}\;\Big|
 \;\rho_\beta^+\text {\bf 1}_{\La^c}\Big)   =  \log   {\hat Z}^{+,0}_\ga \Big(
 \La \, \Big|
 \;\rho_\beta^+\text {\bf 1}_{\La^c}\Big)    \nonumber\\
\hskip2cm  &&- \int_0^1 du~  \mu^+_{\ga ,u,\La} \left[
H^+_{\ga}(\q|\rho_\beta^+ \text{\bf 1}_{\La^c}) -
 H_{0}(\q )   \right]
      \label{10.5}
     \end{eqnarray}
where ${\hat Z}^{+,0}_\ga$ is given by the r.h.s.\ of
\eqref{10a.8} with $H^+_{\ga}$ replaced by $H_0$
\tobias{and $\mu^+_{\ga,u,\La}$ is defined as it is described right before
\eqref{Ku211}.}{}

 $H_0$ will be the Hamiltonian of a free gas, for which the representation \eqref{10.2}
 is easily derived.  The difference
$H^+_{\ga}(\q|\rho_\beta^+ \text{\bf 1}_{\La^c}) - H_{0}(\q
)$ in \eqref{10.5} will be a sum of local potentials, and
the crucial point will be a proof that the measure
$\mu^+_{\ga ,u,\La}$ is well approximated by its
thermodynamic limit and that this is translational
invariant.

The typical example where the interpolation method works is
at high temperatures.  In that case, the reference
Hamiltonian $H_0$ is taken equal to zero. The interpolating
Hamiltonian is then the same original Hamiltonian, with
$\beta \to u\beta$, i.e.\ at higher temperatures, and the
whole scheme works. We are in a sense not too far from this
case, as we are working in the restricted ensemble with a
single well situation, but the extension requires some
care.

To choose $H_0$ (that will be called $H_{0,\La}$) we first
approximate
      \begin{eqnarray*}
      H^+_{\ga}(\q|\rho_\beta^+\text{\bf 1}_{\La^c})\longrightarrow
\beta h_{\la_\beta}(\q(0)|\rho_\beta^+\text{\bf
1}_{\La^c}),\qquad \q = \q_\La
       \end{eqnarray*}
and then expand the latter to first order in
$j_\ga*(\q(0)-\rho_\beta^+\text{\bf 1}_{\La})$,
      \begin{eqnarray}
 h_{\tobias{}{\beta,}\la_\beta}(\q(0)|\rho_\beta^+\text{\bf
1}_{\La^c})& \approx &
h_{\la_\beta}(\rho_\beta^+\text{\bf
1}_{\La}|\rho_\beta^+\text{\bf 1}_{\La^c})+
e'_{\la_\beta}(\rho_\beta^+)\int j_\ga * (\q(0)-
\rho_\beta^+\text{\bf 1}_{\La})\;dr \nonumber
\\&=:&\beta^{-1} H_{0,\La}(\q)
         \label{10.6a}
       \end{eqnarray}

Using the identity
      \begin{eqnarray*}
&&     h_{\la_\beta}(\rho_\beta^+\text{\bf
1}_{\La}|\rho_\beta^+\text{\bf 1}_{\La^c})=
e_{\la_\beta}(\rho_\beta^+) |\La|+ \ga^{-d}
I^+_{\ga\La,\ga\La^c}
      \end{eqnarray*}
proved in \tobias{\eqref{B.18a}}{ Appendix~\ref{app:dA}} and
       \begin{eqnarray*}
\int j_\ga * (\q(0)- \rho_\beta^+\text{\bf 1}_{\La})\;dr =
|\q(0)| -\rho_\beta^+|\La|
      \end{eqnarray*}
we can rewrite
      \begin{eqnarray}
&& H_{0,\La}(\q ) =- \beta \la_{\mathrm{eff}} |\q| +\beta
E_\La
 \nonumber \\
         \label{10.7}\\
&& \la_{\mathrm{eff}}:=  -
e'_{\la_\beta}(\rho_\beta^+),\qquad E_\La =
[e_{\la_\beta}(\rho^+_{\beta})+ \la_{\mathrm{eff}}
\rho^+_{\beta} ]|\La| + \ga^{-d} I^+_{\ga\La,\ga\La^c}
 \nonumber
       \end{eqnarray}
The Hamiltonian $ H_{0,\La}$ has only a one body potential
and it describes an ideal gas, which is thus easy to study.
The measure $\mu^+_{\ga ,u,\La}$ is Gibbs on $\Q^\La_+$
with Hamiltonian
 \begin{eqnarray}
\label{Ku211}
&&    H^+_{ \ga,u}(\q { |\rho_\beta^+\text {\bf
1}_{\La^c}})= u H^+_{ \ga} (\q | \rho_\beta^+\text{\bf
1}_{\La^c}) +(1-u) (-\beta\la_{\mathrm{eff}}|\q|)
        \label{10.8}
     \end{eqnarray}
having dropped from $ H_{0,\La}$ in \eqref{10.8} the
constant term $\beta E_\La$ present in \eqref{10.7}.

\vskip1cm

\subsection{Pressure of reference Hamiltonian}
We are going to prove that
 \begin{eqnarray}
      \label{10a.18a}
&&  \log  \Big\{  e^{\beta \ga^{-d} I_{\ga \La,\ga
\La^c}^+} {\hat Z}^{+,0}_\ga \Big\}=  p^{+,0}_\ga|\La|
\\&& |p^{+,0}_\ga \tobias{+}{-}\beta f_{\beta,\la_\beta}(\rho_\beta^+)|\le
e^{-c\ga^{-1}}
     \label{10a.18b}
     \end{eqnarray}
Calling  $C$ a cube of side  $\ell_{-,\ga}$
 \begin{eqnarray*}
&&   {\hat Z}^{+,0}_\ga =\int_{\Q_+^\La} \nu_\La(d\q) e^{-
H_{0,\La}(\q )} =\left(\int_{\eta=1} \nu_C^<(d\q) e^{\beta
\lambda_{\text{eff}} |\q|} \right)^{|\La|/\ell_{-,\ga}^d}
e^{-\beta E_\La}
     \end{eqnarray*}
Then, recalling the definition of $E_\La$ in \eqref{10.7},
we get \eqref{10a.18a} with
 \begin{eqnarray}
&& e^{p^{+,0}_\ga \ell^{d}_{-,\ga}}=   \left(\int_{\eta=1}
\nu_C^<(d\q) e^{\beta \lambda_{\text{eff}} |\q|}\right)e^{
-\beta( e_{\la_\beta}(\rho_\beta^+)+ \la_{\text{eff}}\
\rho^+_\beta)\ell^{d}_{-,\ga}}
        \label{10a.19}
     \end{eqnarray}

It remains to prove \eqref{10a.18b}. We have
 \begin{eqnarray}
&&
 \int \nu_C(d\q) e^{\beta \lambda_{\text{eff}} |\q|}=
\exp\Big\{e^{\beta \la_{\mathrm{eff}}}\ell_{-,\ga}^d\Big\}
= e^{ \rho_\beta^+ \ell_{-,\ga}^d}
        \label{10a.20}
     \end{eqnarray}
because $\la_{\mathrm{eff}} =
 (\log \rho_\beta^+)/\beta$,   by the mean field equation. Recalling the definition of
\tobias{$p_\ga^{+,0}$}{$E_\La$} in \tobias{\eqref{10a.19}}{\eqref{10.7}}, we then get \tobias{}{from \eqref{10.7}}
      \begin{eqnarray*}
&& e^{p^{+,0}_\ga \ell^{d}_{-,\ga}}=\;
e^{(\rho_\beta^+-\beta e_{\la_\beta}(\rho^+_{\beta})+
\tobias{\beta}{}\la_{\mathrm{eff}} \rho^+_{\beta} ) )\ell^{d}_{-,\ga}}\;
R\nonumber
\\&& R=
 \frac{\dis{\int_{\eta=1} \nu_C^<(d\q) e^{\beta
\lambda_{\text{eff}} |\q|}}}{\dis{ \int \nu_C(d\q) e^{\beta
\lambda_{\text{eff}} |\q|}}}
     \end{eqnarray*}
Since $ \rho_\beta^+ - \beta \la_{\mathrm{eff}}
\rho_\beta^+ = s( \rho_\beta^+)$, we have
     \begin{eqnarray}
&&  e^{p^{+,0}_\ga \ell^{d}_{-,\ga}} = \; e^{\tobias{-}{}\beta
f_{\beta,\la_\beta}(\rho_\beta^+) \ell_{-,\ga}^d} \;e^{\log
(1-(1-R))}
        \label{10.21}
     \end{eqnarray}
and \eqref{10a.18b} follows once we show that $1-R$ is
bounded by $ e^{-c\ga^{-1}}$, $c>0$.

There are two contributions to $1-R$: one comes from
density deviations, $\eta\ne 1$. By \eqref{10a.20} in fact
the average number $N$ of particles is
$\rho_\beta^+\ell_{-,\ga}^d$, thus the probability that
$|N-\rho_\beta^+\ell_{-,\ga}^d| \ge \ga^a \ell_{-,\ga}^d$
is a large deviation estimate (for an ideal gas) and it is
bounded by $\exp\{-c'\ga^{2a} \ell_{-,\ga}^d\}$. The second
contribution to $1-R$ is due to the occurrence of a long
loop, this is bounded by $\eps
\rho_{\beta}^+\ell_{-,\ga}^d$, where $\eps$ is the
probability that a Brownian bridge is longer than
$\ga^{-1/2}$, and $ \rho_{\beta}^+\ell_{-,\ga}^d$ \tobias{is the mean number
of particles}{}.  Hence
$1-R \le e^{-c\ga^{-1}}$ and
\eqref{10a.18a}-\eqref{10a.18b} are proved.

\vskip1cm

\subsection{Energy difference}

We have proved so far that
 \begin{eqnarray}
&& \hskip-1cm \eqref{10.1} =  p^{+,0}_\ga|\La| - \int_0^1
\mu^+_{\ga ,u,\La}\left[ H^+_{ \ga }(\q|\rho_\beta^+
\text{\bf 1}_{\La^c}) -
 H_{0,\La}(\q )   \right]\, du
     \label{10.13}
     \end{eqnarray}
with $ p^{+,0}_\ga$ satisfying \eqref{10a.18b}. By the help
of \tobias{\eqref{10.7}}{\eqref{10.6a}}, we can write more explicitly the energy
difference:
 \begin{eqnarray}
   \label{10.14}
 \hskip-1.5cm
H_{\beta,\ga,\la(\beta,\ga)}(\q|\rho_\beta^+ \text{\bf
1}_{\La^c}) -
 H_{0,\La}(\q ) &=& \nonumber\\
 &&\hspace{-3cm}=\int  \phi (\q,r)\text{\bf 1}_{\La^0}  + \psi_\La(\q,r)\text{\bf
 1}_{(\La^0)^c}
 +   v_\La(\q,r)\text{\bf 1}_{\La\setminus A} \,dr
     \end{eqnarray}
     where
      \begin{eqnarray}
&& { \phi(\q,\cdot) =
 \int_0^\beta~
e_{\la(\beta,\ga)}(j_\ga*\q(t) ) -
e_{\la_\beta}(\rho^+_{\beta}) + \la_{\mathrm{eff}}
j_\ga*[\q(0)- \rho^+_{\beta}
] \,dt} \nonumber \\&&
 \psi_\La(\q;\cdot) =
 \int_0^\beta~
\{e_{\la(\beta,\ga)}(j_\ga*[\q(t)  +\rho^+_{\beta}\text{\bf
1}_{\La^c}])-
 e_{\la(\beta,\ga)}(j_\ga*[\rho^+_{\beta}\text{\bf 1}_{\La^c}]) -
 \text{\bf 1}_{\La\setminus
 \La^0}e_{\la_\beta}(\rho^+_{\beta})\}dt
 \nonumber \\&& \hspace{5cm}
 -
\beta\ga^{-d}I^+_{\ga\La,\ga\La^c}+\int_0^\beta~\la_{\mathrm{eff}}
j_\ga*[\q(0)- \rho^+_{\beta}\text{\bf 1}_{\La}] \, dt
\nonumber
\\&&  v_\La(\q,r)=\;\; \sum_{\Delta\ni r, \Delta \subset \La\setminus A}\;\;
\frac{1}{|\Delta|} U^+_\Delta(\q_{\Delta}) \label{10.19}
     \end{eqnarray}
Notice that {$\psi_\La(\q;r)=\phi(\q,r)$ if $r \in \La^0$
and }
 $\psi_\La(\q;r)=0$ if $r \in \Big( \La\cup \delta_{\text{\rm
out}}^{\ell_{+,\ga}}[\La]\Big)^c$.  According to
Lemma~\ref{lemmaa5.2}, $\phi$ and $\psi_\La$ are uniformly
bounded.

\vskip1cm

\subsection{Decay of correlations}
  \label{subsec:10.6}

In \cite{BKMP2} , Theorem 5.2, it is proved that for $\ga$
small enough the system of conditional probabilities
$\{\mu^+_{\ga,u,C_i}(\cdot|\q_{C_i^c}), C_i\in \mathcal
D^{(\ell_{-,\ga})}\}$ satisfies the Dobrushin uniqueness
condition, uniformly in $u\in [0,1]$.  In particular, it is
shown that for each $u\in [0,1]$ there exists a unique
measure $\mu^+_{\ga,u}$ which is DLR w.r.t.\ the above
system of conditional probabilities; moreover
$\mu^+_{\ga,u}$ is invariant w.r.t.\ translations by
integer multiples of $\ell_{+,\ga}$ and there are {$c$} and
{$c_0$} positive so that for any bounded measurable
function $f$  which depends only on $\q_\Delta$ and any
$u\in [0,1]$,
    \begin{eqnarray}
&& \Big| \;\mu^+_{\ga ,u,\La} [ f ]- \mu^+_{\ga ,u} [ f ]
\Big| \le {c}|\Delta| e^{-{c_0}\ga \,\text{dist}(\Delta,
\La^c)}\|f\|_\infty
     \label{10.16}
     \end{eqnarray}

\vskip1cm

\subsection{The thermodynamic pressure}

Take $\La=\La_n$ in \eqref{10.13}, $\La_n$ a cube of side
$2^n\ell_{+,\ga}$, i.e.\ one of those used in
Theorem~\ref{thm5.1} to define ${\hat
P}^+_{\ga,\la(\beta,\ga)}$. By letting $n\to \infty$, and
using the   results of Subsection \ref{subsec:10.6}, we can
then identify the thermodynamic pressure as
    \begin{eqnarray}
&& \hspace{-1cm}\beta{\hat P}^+_{\ga,\la(\beta,\ga)}= p^{+,0}_\ga -
 \int_0^1 \tobias{\frac{1}{|C|}\int_C \mu^+_{\ga ,u}\!\!\left[
  \phi(\q,r)+ \!\!\!\!\!\!\sum_{\stackrel{\scriptstyle \Delta
\in \mathcal{D}^{(\ell_{+,\ga})}}
{\hspace{-0.7cm}r \in \Delta}}\;
\!\!\frac{1}{|\Delta|} U^+_\Delta(\q_{\Delta}) \right]dr~du}
{\mu^+_{\ga ,u}\left[ \sum_{\Delta\ni 0} U^+_\Delta(\q_{\Delta}) \right]}
     \label{10a.25}
     \end{eqnarray}
having used that $\psi_\La$  and $U_\Delta$ are uniformly
bounded and that $U_\Delta$ decays exponentially, see
\eqref{gg.9}. $C$ denotes the cube from $\mathcal{D}^{(\ell_{+,\ga})}$
which contains $0$.

\vskip1cm

\subsection{Bounds on remainders}

We use \eqref{10a.25} to express $p^{+,0}_\ga$ in terms of
the other quantities appearing in \eqref{10a.25} itself.
We insert such an expression in \eqref{10.13} and get
 \begin{eqnarray*}
&& \hskip-1cm \eqref{10.1} =\beta {\hat
P}^+_{\ga,\la(\beta,\ga)}|\La^0| { \tobias{-}{+} \beta
f_{\beta,\la_\beta}(\rho_\beta^+)
 |\La \setminus \La^0|} +
\tobias{\int L_1(r) +L_2(r)+L_3(r)+L_4(r)\, dr}{L_1 +L_2+L_3+L_4}
    \end{eqnarray*}
    where the remainders $L_1,\ldots,L_4$ are
  \begin{eqnarray}
&&\tobias{L_1(r)= \left(p^{+,0}_\ga + \beta
f_{\beta,\la_\beta}(\rho_\beta^+)\right)
 \text{\bf 1}_{\La \setminus \La^0}(r)}{L_1= \left(p^{+,0}_\ga - \beta
f_{\beta,\la_\beta}(\rho_\beta^+)\right)
 |\La \setminus \La^0|} \label{10.19a} \\
&&
\tobias{L_2(r)= - \int_0^1 \mu^+_{\ga
,u,\La}\left[\phi(\cdot;r)+v_\La(\cdot;r)\right] -
\mu^+_{\ga ,u}\left[\phi(\cdot;r)+v_\La(\cdot;r)\right]
 \,du\, \text{\bf 1}_{\La^0}(r)}
{L_2= - \int_{\La^0}\int_0^1 \mu^+_{\ga
,u,\La}\left[\phi(\cdot;r)+v_\La(\cdot;r)\right] -
\mu^+_{\ga ,u}\left[\phi(\cdot;r)+v_\La(\cdot;r)\right]
 \,du~dr} \nonumber\\&&
\tobias{L_3(r)=   \int_0^1  \mu^+_{\ga
,u}\big[\sum_{\Delta\ni r, \Delta \not \subset
\La\setminus A}\frac{1}{|\Delta|}U_\Delta \big] \,du\,
\text{\bf 1}_{\La^0}(r)
  - \int_0^1  \mu^+_{\ga
,u}\big[v_\La(\cdot;r)\big]
  \,du \,\text{\bf 1}_{\La\setminus[A\cup\La^0]}(r)}
{L_3=   \int_{\La^0}\int_0^1  \mu^+_{\ga
,u}\big[\sum_{\Delta\ni r, \Delta \not \subset
\La\setminus A}\frac{1}{|\Delta|}U^+_\Delta \big] \,du~dr
  - \int_{\La\setminus[A\cup\La^0]}\int_0^1  \mu^+_{\ga
,u}\big[v_\La(\cdot;r)\big]
  \,du~dr} \nonumber\\&&
 \tobias{ L_4(r)=
 \hskip0.5cm  
 -\int_0^1 \mu^+_{\ga ,u,\La}\left[
\psi_\La(\cdot,r) \right]  \,du\,
 \text{\bf 1}_{(\La^0)^c}(r)}
{ L_4=
 \hskip0.5cm  
 -\int_{(\La^0)^c}\int_0^1 \mu^+_{\ga ,u,\La}\left[
\psi_\La(\cdot,r) \right]  \,du\,dr
 +\beta\ga^{-d}I^+_{\ga\La,\ga\La^c}}
\nonumber
     \end{eqnarray}
By \eqref{10a.18b}, $L_1$ satisfies the bound
\eqref{10.3b}.  By Subsection \ref{subsec:10.6} and
\eqref{gg.9}, $L_2$ satisfies \eqref{10.3a}.  Using again
\eqref{gg.9}, we prove that  $L_3$ satisfies
\eqref{10.3a} and \eqref{10.3b}, and it only remains to bound  $L_4$.

By Corollary \ref{corolaA.1},
   \begin{eqnarray}
&&   \big| \int_{(\La^0)^c} \psi_\La(\q,r) -
\psi^{(\ga^{-1/2})}_\La(\q(0),r) dr\big| \le c \ga^{1/2}
|A|
     \label{10a.30}
     \end{eqnarray}
where $\psi^{(\ga^{-1/2})}_\La(\q(0),r)$ is defined as in
the second one of the equations \eqref{10.19} (i.e.\ the
one relative to $\psi_\La(\q,r)$), with $\q(t)$ on the
r.h.s.\ replaced by $\q(0)$ and $j_\ga$ by
$j^{(\ga^{-1/2})}_\ga$, see \eqref{d0.2}.  Letting
   \begin{eqnarray}
&&  \tobias{ L_5(r):=  -\int_0^1 \mu^+_{\ga
,u,\La}\left[
\psi^{(\ga^{-1/2})}_\La(\q(0),r)\right] du
\, \text{\bf 1}_{(\La^0)^c}(r)}
{ L_5:=  -\int_0^1 \mu^+_{\ga
,u,\La}\left[\int_{(\La^0)^c}
\psi^{(\ga^{-1/2})}_\La(\q(0),r) \,dr\right] \,du
 + \beta\ga^{-d}I^+_{\ga\La,\ga\La^c}}
    \label{10a.31}
     \end{eqnarray}
we have
   \begin{eqnarray}
&& \tobias{\int |  L_4(r)   |\, dr \le  \int |  L_5(r)   |\, dr
+ c \ga^{1/2} |A|}
 {|  L_4   | \le  |  L_5   |+ c \ga^{1/2} |A|}
\label{10a.32}
     \end{eqnarray}
and it only remains to prove the bound
   \begin{eqnarray}
&&  \tobias{\int |  L_5(r)  |\, dr \le c \ga^{a'} |A|}
{|  L_5  | \le c \ga^{a'} |A|}
   \label{10a.33}
     \end{eqnarray}

For any   $\q(0)=\q_\La(0) $, 
 let
   \begin{eqnarray}
&&  \mathcal B:= \Big\{ r \notin  \La^0 :
|j_\ga^{(\ga^{-1/2})} *[\q(0)+\rho_\beta^+\text{\bf
1}_{\La^c}](r) -\rho_\beta^+| \ge   \ga^{a'} \Big\}
     \label{10a.34}
     \end{eqnarray}
Recalling that $|\la_{\beta,\ga}-\la_\beta|\leq
c\ga^{1/2}$, there is a constant $k_1$ such that
   \begin{eqnarray*}
&&  \big|
\int_{(\La^0)^c}\psi^{(\ga^{-1/2})}_\La(\q(0),r)\, dr
 \big| \le k_1 \ga^{2a'}
 \big|(\La\cup \delta_{\text{\rm
out}}^{(\ell_{+,\ga})}[\La])\setminus \La^0\big|, \quad r
\notin \mathcal B
     \end{eqnarray*}
We distinguish in the expectation in \eqref{10a.31} whether
$|\mathcal B|$ is either smaller or larger than $|A|
\ga^{a'}$, getting
        \begin{eqnarray*}
&&  \tobias{\int |L_5(r)|\,dr}{L_5}
\le k_1\ga^{2a'} \big|(\La\cup \delta_{\text{\rm
out}}^{(\ell_{+,\ga})}[\La])\setminus \La^0\big| + \|
\psi^{(\ga^{-1/2})}_\La\|_\infty |A| \ga^{a'}\\&& \hskip2cm
+  \mu^+_{\ga ,u,\La}\left[\text{\bf 1}_{\q(0) : |\mathcal
B | \ge |A| \ga^{a'}} \int_{(\La^0)^c}
\psi^{(\ga^{-1/2})}_\La(\q(0),r) \,dr\right] \,du
     \end{eqnarray*}
There are therefore constants $k_2,k_3$ so that
        \begin{eqnarray}
&&  \tobias{\int |L_5(r)|\,dr}{L_5}
\le  k_2\ga^{ a'}  |A| + k_3|A| \sup_{u\in [0,1]}
\mu^+_{\ga ,u,\La}\left[\{\q(0) : |\mathcal B | \ge |A|
\ga^{a'}\} \right]
    \label{10a.35}
     \end{eqnarray}
Call $\Delta= \La^0\setminus \delta_{\text{\rm
in}}^{\ell_{+,\ga}}[\La^0]$, then, by the DLR property,
        \begin{eqnarray}
&&   \mu^+_{\ga ,u,\La}\left[\{|\mathcal B | \ge |A|
\ga^{a'}\} \right]=
 \mu^+_{\ga ,u,\La}\left[ \mu^+_{\ga ,u,\La\setminus \Delta}
 \left[\{ |\mathcal B | \ge |A| \ga^{a'}\}\big|
 \q_\Delta, \rho_\beta^+\text{\bf 1}_{\La^c} \right]\right]
    \label{10a.36}
     \end{eqnarray}
where $\mu^+_{\ga ,u,\La\setminus
\Delta}[\;\cdot\;|\q_\Delta, \rho_\beta^+\text{\bf
1}_{\La^c}]$ is the finite volume Gibbs measure in
$\La\setminus \Delta$ in the restricted ensemble of
configurations
\tobias{$\Q_+^{\La\setminus \Delta}$}{$\Q_{\La\setminus \Delta}$}
with interaction
determined by the Hamiltonian $H_{\ga,u}^+$.  The
contribution to the corresponding Gibbs factor coming from
$K^+_{\La\setminus A}$ can be bounded using \eqref{gg.9}.
We then get
        \begin{eqnarray}
&&    \mu^+_{\ga ,u,\La\setminus \Delta}
 \left[\{ |\mathcal B | \ge |A| \ga^{a'}\}\big|
 \q_\Delta, \rho_\beta^+\text{\bf 1}_{\La^c} \right]
  \le \exp\big\{c |A| e^{-c_f \ga^{-1}/2}\big\} \nonumber \\&& \hskip2cm \times
 p^+_{ \ga ;u;\La\setminus \Delta} \left[\{ |\mathcal B | \ge |A| \ga^{a'}\}\big|
 \q_\Delta, \rho_\beta^+\text{\bf 1}_{\La^c} \right]
    \label{10a.37}
     \end{eqnarray}
where $p$ on the r.h.s.\ denotes the Gibbs measure on
$\Q_+^{\La\setminus \Delta}$ with Hamiltonian
       \begin{eqnarray}
\tobias{uH_{\beta,\ga,\la(\beta,\ga)}
(\q_{\La\setminus \Delta}|\rho_\beta^+\text{\bf
1}_{\La^c},\q_\Delta) + (1-u) (-\beta \la_{\text{eff}})
|\q_{\La\setminus \Delta}(0)|}
{uH^+_{\ga}(\q_{\La\setminus \Delta}|\rho_\beta^+\text{\bf
1}_{\La^c},\q_\Delta) + (1-u) (-\la_{\text{eff}})
|\q_{\La\setminus \Delta}(0)|}
    \label{10a.38}
     \end{eqnarray}
By an error bounded by $e^{c_2 \ga^{1/2} |\La\setminus
\Delta|}$, see \eqref{b6.9}, we can replace
$\la(\beta,\ga)$ by $\la_\beta$, we call $p^+_{
\ga,\la_\beta ;u;\La\setminus \Delta}$ the corresponding
Gibbs measure. By the argument used in Subsection
\ref{subsec:6.5}
        \begin{eqnarray}
&&
 p^+_{ \ga,\la_\beta ;u;\La\setminus \Delta} \left[\{ |\mathcal B | \ge |A| \ga^{a'}\}\big|
 \q_\Delta, \rho_\beta^+\text{\bf 1}_{\La^c} \right]\nonumber
  \\&&\hskip2cm  \le e^{c_3 \ga^{1/2}|A|}
 p^+_{ \ga,\la_\beta
;u;\La\setminus \La^0}\left[\{ |\mathcal B | \ge |A|
\ga^{a'}\}\big|
  \rho_\beta^+\text{\bf 1}_{(\La\setminus \La^0)^c} \right]
    \label{10a.39}
     \end{eqnarray}
To bound the r.h.s.\ we perform the continuum approximation
using the analogue of Proposition \ref{propaB.1}.  Due to
the presence of the reference Hamiltonian, the free energy
functional is modified into a $u$-dependent functional.
Call $D=\ga (\La\setminus \La^0)$ and define
    \begin{equation}
     \label{10a.40}
F^0_{\beta}(\rho_D) = \int_D f^0_{\beta}(\rho_D)\,dr,\qquad
f^0_{\beta}(x)= -\la_{\text{eff}} x - \frac {s(x)}{\beta}
     \end{equation}
        \begin{eqnarray}
&& B = \Big\{ \rho\in L^{\infty}(D,\mathbb R_+):
\eta^{(\ga^a,\ell_-)}(\rho;r)=1, r\in D;\nonumber
  \\&&\hskip2cm   |\{ r : |j^{(\ga^{1/2})}*(\rho+\rho_\beta^+\text{\bf 1}_{D^c})(r)-\rho_\beta^+| >
\ga^{a'}\}| > \ga^{a'}(\ga^d|A|) \Big\}
    \label{10a.41}
     \end{eqnarray}
Then, by the analogue of Proposition \ref{propaB.1},
        \begin{eqnarray}
&&
 p^+_{ \ga,\la_\beta
;u;\La\setminus \La^0}\left[\{ |\mathcal B | \ge |A|
\ga^{a'}\}\big|
  \rho_\beta^+\text{\bf 1}_{(\La\setminus \La^0)^c} \right] \le \exp\Big\{ c  \ga^{1/2}|A|
  \nonumber
  \\&&\hskip2cm  - \ga^{-d} \inf_{\rho_D \in B}
\Big( u \mathcal
F_{\beta,\la_\beta}(\rho_D|\rho_\beta^+\text{\bf 1}_{D^c} )
+(1-u)\mathcal F^0_{\beta}(\rho_D) \Big)\Big\}
    \label{10a.42}
     \end{eqnarray}
where the functionals on the r.h.s.\ are the excess free
energies, namely
 $\mathcal F_{\beta,\la_\beta}(\rho_D|\rho_\beta^+\text{\bf 1}_{D^c} )=
F_{\beta,\la_\beta}(\rho_D|\rho_\beta^+\text{\bf 1}_{D^c} )
- F_{\beta,\la_\beta}(\rho_\beta^+\text{\bf
1}_{D}|\rho_\beta^+\text{\bf 1}_{D^c} )$ and similarly,
$\mathcal F^0_{\beta}(\rho_D)= F^0_{\beta}(\rho_D)-
F^0_{\beta}(\rho_\beta^+\text{\bf 1}_{D})$.

\tobias{Using as in \eqref{D.8} that $\rho_\beta^+$ is a minimum of
$f_{\beta,\la_\beta}$ we obtain for $\rho_D \in \mathcal{B}$}
{By \eqref{aBa.5}, if $\rho_D \in \mathcal B$,}
        \begin{eqnarray}
&&
 F_{\beta,\la_\beta}(\rho_D|\rho_\beta^+\text{\bf 1}_{D^c} )
 \ge  F_{\beta,\la_\beta}(\rho_\beta^+\text{\bf 1}_{D} |\rho_\beta^+\text{\bf 1}_{D^c} )
 + c \ga^{3a'} \ga^d|A|
    \label{10a.43}
     \end{eqnarray}
Since
        \begin{eqnarray}
&& F^0_{\beta}(\rho_D)
 \ge  F^0_{\beta}(\rho_\beta^+\text{\bf 1}_{D} )
 + c' \ga^{3a'} \ga^d|A|
    \label{10a.44}
     \end{eqnarray}
from \eqref{10a.42} we get
        \begin{eqnarray}
&&
 p^+_{ \ga,\la_\beta
;u;\La\setminus \La^0}\left[\{ |\mathcal B | \ge |A|
\ga^{a'}\}\big|
  \rho_\beta^+\text{\bf 1}_{(\La\setminus \La^0)^c} \right] \le
   \exp\Big\{ c  \ga^{1/2}|A|  - c''\ga^{3a'}|A|  \Big\}
    \label{10a.45}
     \end{eqnarray}
and \eqref{10a.33} is proved.

\vskip3cm

\section{{\bf Conclusions}}
       \label{sec:11}

So far we have proved that given $\beta\in
(\beta_c,\beta_0)$ there exists $\ga_\beta>0$ such that for all
$\ga\le \ga_\beta$ there exists a $\la(\ga,\beta)$ with the property
that the
weights $W_{\ga,\la(\beta,\ga)}^{\pm}(\Ga;\q)$, as defined in
\eqref{a4.10}, satisfy the bound
        \begin{eqnarray}
&&W_{\ga,\la(\beta,\ga)}^{\pm}(\Ga;\q) \le e^{-2c_f
\ga^{-1} N_\Ga} .
    \label{11.1b}
     \end{eqnarray}

\vskip .5cm
Let $\mu^{\pm,\q}_{\ga,\La}$ denote  the ``dilute''
Gibbs measure in $\La$ for a $\pm$ b.c.\ $\q$ and parameters
$\beta$, $\ga$, $\la(\beta,\ga)$, namely the  Gibbs measure
in $\La $ (for a $\pm$ b.c.\ $\q$) conditioned on
$\{\Theta(\cdot;r)=1$ for all $r \in \delta_{\text{\rm
out}}^{\ell_{+,\ga}}[\La]\}$, cf. for partition function \eqref{a4.7}.
Given a contour $\Ga$ let us call
        \begin{eqnarray}
&&      \phi(\Ga) :=( \text{sp}(\Ga) \cup \text{int}(\Ga))
        \tobias{\setminus K^\pm}{}.
    \label{11.2}
     \end{eqnarray}
 Then \eqref{11.1b} implies that the probability w.r.t $\mu^{\pm,\q}_{\ga,\La}$ of
occurrence of a contour $\Ga$ is bounded by $ e^{-2c_f \ga^{-1}
N_\Ga}$. Summing over all $\Ga$ with $\phi(\Ga) \ni
0$, we get for small enough $\ga$ for space dimension $\geq 2$:
        \begin{eqnarray}
&&  \mu^{\pm,\q}_{\ga,\La}\Big(\;\{ \text{there is $\Ga$ so
that $ \phi(\Ga) \ni 0$}\;\}\Big) \le  e^{-c_f \ga^{-1}},
    \label{11.3}
     \end{eqnarray}
because in $d>1$ the number of contours which contain zero grows only
exponentially in $N_\Ga$ with a rate just depending on the dimension.
In other words for $d>1$ we have that
       \begin{eqnarray}
&& \hskip-1cm \mu^{+,\q}_{\ga,\La}\Big(\;\{
\eta(\cdot;0)=1\;\}\Big) \ge 1-e^{-c_f \ga^{-1} };\qquad
\mu^{-,\q}_{\ga,\La}\Big(\;\{ \eta(\cdot;0)=1\;\}\Big) \le
e^{-c_f \ga^{-1} }
    \label{11.4}
     \end{eqnarray}
and hence we have shown that the effect of the boundary conditions persists in
the thermodynamic limit.

\vskip .5cm

In the remainder of the section, we will prove what is still missing to
establish Theorem~\ref{thm2.1} and \ref{thm2.1q}, namely we show
that  the  dilute, finite volume,
Gibbs measures $\mu^{\pm,\q}_{\ga,\La}$ have limit points
$\mu^{\pm}_{\ga}$ in the thermodynamic limit
and that these limit points  are DLR measures.
The information gathered
in the previous sections yields such conclusions and much
more, it yields representations of dilute  Gibbs measures which make
quantitative the idea that the
configurations consist of  ``an ocean of plus'' (in the plus case)
perturbed by a collection of ``islands'' which are
typically rare and small. Finally, we establish the convergence of the
quantum states.

\vskip1cm

  \subsection{A representation of dilute Gibbs measures}

For simplicity, let us just consider the measure
$\mu^{+,\bar \q}_{\ga,\La}$, with plus boundary conditions.
Let $p_{\ga,\La}^{+,\bar \q}(d\q)$ be the Gibbs measure on
$\Q_+$ (supported on $\bar\q_{\La^c}$ outside $\La$) with
Hamiltonian $H_\ga(\q_\La|\bar
\q_{\La^c})
\tobias{+K^+_{\La}(\q_{\La})}
{+K^+_{\La\setminus A}(\q_{\La\setminus A})}$, the
latter defined in \eqref{10a.5}; recall that as a consequence of
\eqref{11.1b} the cutoff
weights ${\hat W}_{\ga,\la(\beta,\ga)}^{\pm}$ coincide with the
original weights $W_{\ga,\la(\beta,\ga)}^{\pm}$ of the LMP
  model, cf.~Theorem~\ref{thma4.2} for the values of the parameters
under consideration.  Hence
due to the representation \eqref{a4.8}, the partition
function associated to $p_{\ga,\La}^{+,\bar \q}(d\q)$ and the
partition function associated to
$\mu^{+,\bar \q}_{\ga,\La}$ coincide, where the latter is the partition function
just with respect to the energy  $H_\ga(\q_\La|\bar \q_{\La^c})$, but without the
restriction that the configurations are in the restricted
ensemble.

The measure $p_{\ga,\La}^{+,\bar \q}(d\q)$ does not contain all the
information contained in $\mu^{+,\bar
\q}_{\ga,\La}$, but the latter can be reconstructed
from the former, as we are going to show.  To this end we
denote by
$\pi^{+}_{\ga,\La}(\cdot|\q)$, $\q \in\Q_+$,
the law on $\mathcal C^+_\La$, defined as
       \begin{eqnarray}
&& \pi^{+}_{\ga,\La}\Big(\{\Ga_i\}\Big|\q_\La\Big)
 \;=\;
\frac{\dis{\prod_i
W^+_\tobias{{\ga,\la(\beta,\ga)}}{\ga}(\Ga_i,\q_\La)}}{\dis{\sum_{\{\Ga'_j\}\in \mathcal
C^+_\La}\prod_j W^+_\tobias{{\ga,\la(\beta,\ga)}}{\ga}(\Ga'_j,\q_\La)}},\quad \{\Ga_i\}
\in \mathcal C^+_\La .
    \label{11.5}
     \end{eqnarray}
The skew product of $p_{\ga,{\La}}^{+,\bar \q}(d\q_\La)$
and
$\pi^{+}_{\ga,\La}(\cdot|\q_\La)$
defines a measure on $\Q_+ \times
\mathcal C^+_\La$, denoted by
       \begin{eqnarray}
&& m_{\ga,\La}^{+,\bar\q} (d\q, \{\Ga_i\}):=
\tobias{\pi^{+}_{\ga,\La}\Big(\{\Ga_i\}\Big|\q_\La\Big)p_{\ga,\La}^{+,\bar
\q}(d\q_\La)}
{\pi^{+}_{\ga,\La}\Big(\{\Ga_i\}|\q_\La)\Big)p_{\ga,\La}^{+,\bar
\q}(d\q)} .
    \label{11.6}
     \end{eqnarray}
In words, the measure $m_{\ga,\La}^{+,\bar\q} (d\q, \{\Ga_i\})$ is
obtained by sampling $\q$ according to $p_{\ga,\La}^{+,\bar \q}(d\q)$,
this is the ``ocean of plus'' mentioned in
the previous heuristic description.  Then, given $\q$, we ``throw
independently except for exclusion'', contours $\Ga$ taken from
$\mathcal C^+_\La$, with rate $W^+_{\ga,\la(\beta,\ga)}(\Ga,\q_\La)$.
The contours describe the aforementioned ``islands''. The
  internal structure of the ``islands'' is subsumed in the weights.

The last step is to introduce a transformation
$\tobias{T^{+}_{\q,\{\Ga_i\}}}
{T^{+,\La}_{\q,\{\Ga_i\}}}$
which maps
\tobias{bounded measurable functions on $\Q$ into bounded
measurable functions on $\Q_+ \times \mathcal C^+$ such
that the following formula holds} {$L^{\infty}(\Q)$ into
$L^\infty (\Q_+ \times \mathcal C^+_\La)$ and such that}
       \begin{eqnarray}
&& \mu^{+,\bar \q}_{\ga,\La}(f)= \sum_{\{\Ga_i\}\in
\mathcal C^+_\La}
\tobias{\int_{\Q_+^\La} T^{+}_{\q,\{\Ga_i\}}(f)\,}
{\int_{\Q_+} T^{+,\La}_{\q,\{\Ga_i\}}(f)}
m_{\ga,\La}^{+,\bar\q} (d\q, \{\Ga_i\})
    \label{11.7}
     \end{eqnarray}
\tobias{where $\mathcal C^+= \Big\{ \{\Ga_i\}: |\mathrm{sp}(\Ga_i)|< \infty
\quad \mathrm{dist}(\mathrm{sp}(\Ga_i),\mathrm{sp}(\Ga_j))
\ge \ell_{+,\ga}\Big\}$ denotes the projective limit of the
sets of contours $\mathcal C^+_\La$.
}{}
Shorthanding    $\dis{\phi(\{\Ga_i\})= \bigsqcup_i
\phi(\Ga_i)}$ and
   \[
   \{\Ga_i\}_{\text{ext}}=\{\Ga_i \in
\{\Ga_\tobias{j}{i}\}: \phi(\Ga_i) \text{ \, is not contained in any }
  \phi(\Ga_j), j\ne i\}
  \]
we set \notes{in our definition $\phi(\{\Ga_i\})$ is an unbounded
  set. The sum in the definition of $\pi^+$ contains also the contours
inside of contours but $T$ is insensitive to them, this looks like an
overcounting. The first problem one can solve by looking at cylinder
sets $f$ and only the contours touching the support. To avoid
overcounting we can either sum only over external contours or
condition also on the internal contours.}
      \begin{equation}
          \label{11.8}
\tobias{T^{+}_{\q,\{\Ga_i\}}(f)}{T^{+,\La}_{\q,\{\Ga_i\}}(f)}
= \int
f(\q_{\phi(\{\Ga_i\})^c},\q'_{\phi(\{\Ga_i\})})
\, \mu_{\ga,\phi(\{\Ga_i\})}\Big(d\q' \Big|
\q_{\phi(\{\Ga_i\})^c}; \{\und
\Ga(\q_{\phi(\{\Ga_i\})^c},
\tobias{\q'_{\phi(\{\Ga_i\})}}
{\q'_{\phi(\und \Ga)}})\tobias{=}{\supset}
\{\Ga_i\}_{\text{ext}}\} \Big)
    \end{equation}
\notes{the condition in the definition of $T$ is not a set of contours
and collides thus with our initial condition.}
where the measure on the r.h.s.\ is the Gibbs measure in
$\phi(\{\Ga_i\})$ with b.c.\ $\q$ and conditioned on the set of
loops configurations which have among their contours all the
contours $\{\Ga_i\}_{\text{ext}}$.

The equality \eqref{11.7}
holds for $T^{+}_{\q,\{\Ga_i\}}(f)$ as defined in \eqref{11.8},
because conditioned on the
external contours $\{\Ga_i\}_{\text{ext}}$ the measures
$m^{+,{\bar \q}}_{\ga,\La}$ and $p^{+,{\bar\q}}_{\ga,\La}$
are identical outside of
$\Phi\left(\{\Ga_i\}_{\text{ext}}\right)$. The definition \eqref{11.8} just states
that inside of
$\Phi\left(\{\Ga_i\}_{\text{ext}}\right)$ we use the
conditional expectation of the Gibbs measure instead of the
conditional expectation of $m^{+,{\bar \q}}_{\ga,\La}$. The partition function associated to
 $m^{+,{\bar\q}}_{\ga,\La}$ coincides also with the partition functions of
$\mu^{+,{\bar \q}}_{\ga,\La}$ due to \eqref{a4.8}.
If $f$ is cylindrical w.r.t. $B$, i.e. depends only
on $\q_B$, then $T^+_{\q, \{\Ga_i\}}(f)$ depends only on
$\q_{B \tobias{\cup \delta_{\text{\rm
out}}^{\ell_{+,\ga}}[B]}{} \cup \phi(\{\Ga_i\}_B)}$, where
\[
\{\Ga_i\}_B = \{ \Ga_i |\, \mathrm{sp}(\Ga_i) \cap B \neq \emptyset \}.
\]

\vskip1cm

  \subsection{Couplings and thermodynamic limit}

Let $\La'$ and $\La$ be bounded, $\mathcal
D^{(\ell_{+,\ga})}$-measurable  sets with $\La'\supset
\La$,\notes{$\La'$ not used}
 $\q$ and $\q'$ be plus b.c.\ relative to $\La$ and
$\La'$. Define $m^{+,{\bar \q}}_{\ga,\La}$ and  $m^{+,{\bar
      \q'}}_{\ga,\La}$ as in \eqref{11.6}.  In Appendix C of
\cite{BKMP2} we prove that for any bounded set $B$ of
$\La$, there exists a coupling $P(d\xi,d\xi')$ of $\tobias{m^{+,{\bar
      \q}}_{\ga,\La}}{m}$ and
$m^{+,{\bar \q'}}_{\ga,\La}$, where $\xi= (\q,\{\Ga_i\})$, $\xi'=(\q',\{\Ga'_j\})$, with the
following properties:

\vskip.5cm

$\bullet$\;  A  set $\Delta$ is called  a set of agreement for
$(\xi,\xi')$ if $\q_\Delta=\q'_\Delta$, all $\phi(\Ga_i)$
and $\phi(\Ga'_j)$ are either in $\Delta$ or in $\Delta^c$
and,
      \begin{equation}
          \label{11.9}
\Big\{ \Ga_i \in\{\Ga_i\}: \text{sp}(\Ga_i)\subset \Delta\Big\}=
\Big\{ \Ga'_j \in\{\Ga'_j\}: \text{sp}(\Ga'_j)\subset
\Delta\Big\}
   \end{equation}

\vskip.5cm

$\bullet$\; There are positive constants $c_0$, $c_2$ such that for
given bounded subset $B \subset \mathbb{R}^d$ one has
      \begin{equation}
          \label{11.10}
P\Big( \{ \text{$B\subset \Delta$, $\Delta$ is a set of
agreement}\} \Big) \ge 1 -c_2 |B| e^{-c_0\ga
\text{dist}(B,\La^c)}
   \end{equation}

\vskip.5cm
\noindent
We call a function  $f$ {\em quasi-cylindrical} in $B$, if for any
pair of configurations $\q, \q'$ which have an agreement set
containing $B$, one has: $f(\q)=f(\q')$. In particular, if $f$ is
quasi-cylindrical in $B$ then $T^+_{\q, \{\Ga_i\}}(f)$ is
quasi-cylindrical in $B \cup \delta_{\text{\rm
out}}^{\ell_{+,\ga}}[B]$, because the extra
dependence via the Gibbs
measure in \eqref{11.7} is only at the boundary of
$\phi(\{\Ga_i\}_B){\cup \delta_{\text{\rm
out}}^{\ell_{+,\ga}}[B]}$
and the agreement set implies agreement on this boundary.

\vskip.5cm

$\bullet$\;
As a consequence of \eqref{11.10}, if $f$ is a bounded function,
cylindrical in $B$ (i.e.\ which depends only on $\q_B$) [or quasi-cylindrical in $B$], then
      \begin{equation}
          \label{11.11}
\big| \tobias{m^{+,{\bar \q}}_{\ga,\La}}{m}(f)-\tobias{m^{+,{\bar \q'}}_{\ga,\La}}{m'}(f)\big| \le \tobias{2}{} \|f\|_\infty  P\Big(
\{ \text{$B\subset \Delta$, $\Delta$ is a set of
agreement}\}^c \Big) \le \tobias{2}{} c_2 |B| \|f\|_\infty
 e^{-c_0\ga
\text{dist}(B,\La^c)}.
   \end{equation}
The same bound also holds for the dilute Gibbs measures {$\mu^{+,\bar
  \q}_{\ga,\La}$ and $\mu^{+,\bar \q'}_{\ga,\La}$}, because
$T^+_{\q, \{\Ga_i\}}(f)$ is
quasi-cylindrical and bounded if $f$ if quasi-cylindrical and bounded.

\vskip .5cm

The estimate \eqref{11.11}
implies that  in the limit $\La \nearrow \mathbb R^d$
(for $\mathcal{D}^{(\ell_{+,\ga})}$ measurable $\La$), the measures
$m_{\ga,\La}^{\pm,\bar\q} (d\q, \{\Ga_i\})$ converge
weakly, independently of the particular  boundary condition from
$\Q^\pm$ to  limit measures
$m^{\pm}_\ga$ on the space $\Q^{\pm} \times \mathcal C^{\pm}$. By \eqref{11.7},
also the Gibbs measures $ \mu^{\pm,\q}_{\ga,\La}$ converge
weakly to  limit measures $ \mu^{\pm}_{\ga}$, more precisely for all cylinder functions [quasi-cylinder functions] $f$:
\begin{equation}\label{11.11c}
    \lim_{\La\to \Rr^d} \mu^{\pm,\q}_{\ga,\La}(f)=\mu^{\pm}_{\ga}(f).
\end{equation}
Note that due to \eqref{11.4}, the two measures $\mu^{\pm}_{\ga}$ are
distinct from each other.

\vskip .5cm

Finally, let us remark that a representation like \eqref{11.7} holds also in
infinite volume. In fact, the measures
$p_{\ga,\La}^{+,\bar \q}(d\q)$ converge in the
thermodynamic limit, as the coupling property is a fortiori
{satisfied.}
Call the limit measure $p_{\ga}^{+}(d\q)$. For
 each $\q\in
 \Q^+$ a process on $\mathcal C^+$ can be defined as follows: each contour
$\Ga\in \{\Ga_i\}^+$ is placed with rate $W_\tobias{{\ga,\la(\beta,\ga)}}{\ga}(\Ga;\q)$
independently except for exclusion (well-definedness follows from
Kolmogorov's theorem for projective limits of probability
measures).

\tobias{The skew product of the process
$p_{\ga}^{+}(d\q)$ with the latter is just}{}
$m^+_\ga$
and, analogously to \eqref{11.7}, holds
       \begin{eqnarray}
&& \mu^{+ }_{\ga }(f)= \sum_{\{\Ga_i\}\in \mathcal C^+} \int
T^{+}_{\q,\{\Ga_i\}}(f)
~ m_{\ga }^{+ } (d\q, \{\Gamma_i\}).
    \label{11.23}
     \end{eqnarray}

\vskip1cm

  \subsection{The DLR property}\label{sec:11dlr}
To conclude the proof of Theorem
\ref{thm2.1} we thus need to prove that $\mu^{\pm}_{\ga}$
fulfills the DLR property. For the sake of simplicity we consider just
$\mu^{+}_{\ga}$. The proof for $\mu^{-}_{\ga}$ is exactly analogous. We have to prove that for every bounded
Borel sets $A \subset \mathbb R^d$, $A' \subset \mathbb
R^d$ and every bounded function $f$ cylindrical in $A'$ holds
    \begin{equation}
       \label{11.12}
\int \int f(\q')\mu_{\ga,A}(d\q'|\q_{A^c}) \mu_\ga^+(d\q)=
\int f(\q)\mu^+_\ga(d\q)
     \end{equation}
where $\mu_{\ga,A}(d\q'|\q_{A^c}) $ is the Gibbs measure in
$A$ with   b.c. $\q_{A^c}$ (for the parameters $\ga$,
$\beta$,  and $\la(\ga,\beta)$).

The standard proof is based on the fact that $\mu_\ga^+$ is usually
the limit of finite volume Gibbs measures,that is, measures which
satisfies \eqref{11.12}. Then one could use that the equality is preserved
under weak convergence and hence
holds in the infinite volume limit.  The difficulty here is that
neither $\mu_{\ga,A}(d\q'|\q_{A^c}) $ is cylindrical in $\q_{A^c}$,
nor do dilute Gibbs measures  satisfy the analogue of \eqref{11.12} in
finite volume. Instead, they satisfy:
    \begin{equation}
       \label{11.13}
\int \int f(\q')\mu^+_{\ga,\La;A}(d\q'|\q_{A^c})
\mu_{\ga,\La}^{+,\bar\q}(d\q)= \int
f(\q)\mu_{\ga,\La}^{+,\bar\q}(d\q)
     \end{equation}
where $\mu^+_{\ga,\La;A}(d\q'|\q_{A^c})$ is the conditional
probability of $\mu_{\ga,\La}^{+,\bar\q}(d\q)$ given
$\q_{A^c}$ outside $A$, namely
    \begin{equation}
       \label{11.14}
\mu^+_{\ga,\La;A}(d\q'|\q_{A^c})=
\mu_{\ga,A}\big(d\q'|\q_{A^c}; \{\Theta(\q';r) = 1 , r\in
\delta_{\text{\rm out}}^{\ell_{+,\ga}}[\La] \} \big) .
     \end{equation}
For $\La$ large enough, $A \subset (\La \setminus
\delta_{\text{\rm in}}\tobias{^{\ell_{+,\ga}}}{}[\La])$ and
the condition $\Theta(\q';r) = 1$, for $r\in \delta_{\text{\rm
out}}^{\ell_{+,\ga}}[\La]$, is a condition both on
$\q'_{A^c}=\q_{A^c}$ (which  is irrelevant in the following because it
is satisfied almost surely w.r.t.\ $
\mu_{\ga,\La}^{+,\bar\q}$) and a condition on $\q'_A$. The
latter is the requirement that no long loop in $\q'_A$
reaches $\delta_{\text{\rm in}}^{\ell_{+,\ga}}[\La]$.

{Condition~\ref{2.a13} follows from the fact that
$\mu_\ga^+$-a.s. all contours are of finite size. Hence any bounded region intersect $\mu_\ga^+$-a.s. only a finite number of loops.}
%
%
%

\vskip.5cm
We postpone the proof that there is $c,c'>0$ so that for all $A$, $\La$ and all $\q_{A^c}$ as above
    \begin{equation}
       \label{11.15}
\mu_{\ga,A}\Big(  \{  \q'(t) \cap \delta_{\text{\rm
in}}^{\ell_{+,\ga}}[\La] \ne \emptyset,\; \text{for some\;}
t\in [0,\beta]\} \Big|\q_{A^c} \Big) \le c e^{-c'
\;\text{dist}(A,\La^c)}.
     \end{equation}
The bound \eqref{11.15} implies that
    \begin{equation}
       \label{11.16}
\lim_{\La\nearrow \mathbb R^d}\int \Big( \int
f(\q')\mu_{\ga,A}(d\q'|\q_{A^c}) -f(\q)\Big)
 \mu_{\ga,\La}^{+,\bar\q}(d\q)=0.
     \end{equation}
As the expression in the bracket is a quasi-cylindrical function, the
result \eqref{11.11c} yields \eqref{11.12}.

\vskip.5cm

\noindent
{\it Proof of \eqref{11.15}.} {Using that at least one loop has to be long and the stability property \eqref{2.a11} we obtain}
    \begin{eqnarray*}
       \label{11.22}
&&\mu_{\ga,A}\Big(  \{  \q'(t) \cap \delta_{\text{\rm
in}}^{\ell_{+,\ga}}[\La]  \ne \emptyset,
t\in [0,\beta]\} \Big|\q_{A^c} \Big)\le \tobias{e^{\beta b}}{c''}
 W_{0|0}^\beta \big(\sup_{0\le t\le
\beta}|\om(t)-\om(0)|> \text{dist}(A,\delta_{\text{\rm
in}}^{\ell_{+,\ga}}[\La])\big) \\
\tobias{}
{&& c''=
  \sum_{n\ge 1} \frac{|A|^n}{n!}  e^{-\beta
(b+\la(\ga,\beta))n} n}.
     \end{eqnarray*}
By \eqref{a7.12} for $h=1$, one obtains
    \begin{eqnarray*}
 W_{0|0}^\beta \big(\sup_{0\le t\le
\beta}|\om(t)-\om(0)|> R\big) \leq  \tobias{c e^{- R^2/(2\beta)}}
{\Big(c e^{- R^2/(2\beta)}\Big)^{p(h)}}
     \end{eqnarray*}
{which together with the previous inequality implies} \eqref{11.15}.

\vskip1cm

  \subsection{Remarks about the structure of DLR measures\label{subsecQuant}}

We have proved so far that there are two distinct DLR measures,
namely the measures $\mu_{\ga}^{\pm}$ which are limits of
 finite volume, dilute, $\pm$ Gibbs measures.  We have also derived
uniqueness, in the sense that any sequence of dilute
plus Gibbs measures converges to $\mu_\ga^+$ and any
sequence of minus measures converges to $\mu_\ga^-$.
{This directly implies that }
measures $\mu_{\ga}^{\pm}$ are invariant under translations
by integer multiples of $\ell_{+,\ga}$ and, by the same
argument as used in \cite{BMP}, one can show that any translational invariant
DLR measure is a convex combination of $\mu_{\ga}^{\pm}$. \notes{really?}

\vskip .2cm
\subsection{Convergence of quantum states}
\label{sec:quantum}

\vskip 1.5cm
Finally, it remains to establish the convergence of the quantum states
themselves. The precise definition of the quantum state and hence of
the associated finite volume measure on configurations of loops was
left open in Section~\ref{susecAlg}. The exact structure of the system
at the boundary was left open. Two
kind of boundary conditions can be treated for the quantum
system without adding any additional considerations. First, we imagine
that the system interact with the outside given not by particles but
by a continuous constant density $\rho_\beta^\pm$. More precisely outside of $\La$ we
extend the configuration $\q$ by a constant density function with
value $\rho_\beta^\pm$, that is we replace in \eqref{2.a5} the configuration
$\bar q$ by the constant function $\rho_\beta^\pm$. The corresponding
finite volume Gibbs measure is denoted by
$\mu^{\pm,\rho_\beta^\pm}_{\ga,\La}(d\q)$

Second, consider boundary conditions $\bar q$ such
that
\begin{equation}
 \Big| \ell_-^{-d} |\bar q\cap C^{(\ell_-)}_r|
- \rho^{\pm}_{\beta}\Big|\le \ga^{a''}
\end{equation}
for $a'' \gg \alpha$.
For both type of boundary condition we will show in the following that all the
results holds as well.

The reason for the introduction of the ``dilute'' Gibbs measures,
partition function respectively, is
the inductive procedure used to compare ${\hat W}_{\ga,\la(\beta,\ga)}^{\pm}$
and $W_{\ga,\la(\beta,\ga)}^{\pm}$, cf. Theorem~\ref{thma4.1}. In all
steps in the induction with the exception of the final one, ``dilute''
partition functions appear naturally.

Due to finite interaction range
there is no direct interaction over the boundary for distances greater
than $2\gamma^{-1}$ except by long loops. Hence for contours $\Ga$
completely contained in $\La
\setminus \delta_{\text{\rm in}}^{\ell_{+,\ga}}[\La]$ the main bound \eqref{11.1b}
follows directly from the previous consideration just by conditioning on $\La^c
\cup \delta_{\text{\rm in}}^{\ell_{+,\ga}}[\La]$, because the
induction in the size of the contour in Theorem~\ref{thma4.1} goes
from small to larger contours. Hence only contours touching
$\delta_{\text{\rm in}}^{\ell_{+,\ga}}[\La]$ have to be considered
separately. We have to consider two cases. First, a block in a contour
with $\Theta=1$ is in $\delta_{\text{\rm in}}^{\ell_{+,\ga}}[\La]$.
Then well inside this block, that is more that $2\gamma^{-1}$ away
from the boundary, there is no direct influence of neither the density
outside nor the Dirichlet boundary condition, because all loops are
are short. That a contour contains an outer layer of blocks with
$\Theta=1$ is only used in Subsection~\ref{subsec:6.5}. The corridors used
there are well inside the block and hence all the arguments still
holds. Second,  a block in a contour with $\Theta=0$ is in
$\delta_{\text{\rm in}}^{\ell_{+,\ga}}[\La]$. Blocks of such type are
treated by the comparison with the non-local functional $F_{\beta,\lambda_\beta}$,
cf. Subsection~\ref{subsec:6.6t}. The Dirichlet boundary condition does not
alter the underlying classical proof by Lebowitz and Penrose,
cf. Appendix~\ref{app:aC}, because the effect of the Dirichlet boundary
condition is of range 1 and hence the change of the density in a
block of sidelength $\ell_{-,\ga}$ is arbitrary small compared to the
precision $\zeta = \ga^a$. In more details, the Dirichlet boundary condition is
described by the additional condition that loops cannot leave the
considered volume $\La$. For a Brownian bridge the probability that a
given loop will leave $\La$ decays exponentially with the distance to
the boundary of $\La$ and is independent of $\ga$. Hence the number of
loops leaving through the boundary $\La$ from a given block is
Poisson distribution where the parameter is proportional to the
surface of the block. Hence the density of the volume will at most
change by $1/\ell_{-,\ga} \ll \zeta$. The restriction on the boundary
conditions were chosen such that they give rise to a surface
correction much small than the gain term obtained from the variational
problem. For simplicity, choose $a'' \gg a'$. These considerations
yield that all contours fulfill the bound \eqref{11.1b} and hence also
all the other consequences hold.

We showed in Subsection~\ref{susecAlg} that is sufficient to show
the convergence of quantum states for fixed $\q'$ to show convergence
of the following expression
\begin{equation}\label{Q1}
   \lim_{\La \to \Rr^d} 
    \inttwo{\Q^\La}{\q_\Delta=\emptyset}
    e^{-H_{\beta,\ga,\la_{\beta,\ga}}(\q'_\Delta|\q ,\bar
      q)}\,\,\mu^{+,\bar q}_{\ga,\La}(d\q)=
    \inttwo{\Q}{\q_\Delta=\emptyset}
    e^{-H_{\beta,\ga,\la_{\beta,\ga}}(\q'_\Delta|\q ,\bar
      q)}\,\,\mu^{+}_{\ga}(d\q)
\end{equation}
The function $g$ defined for $\q \in \Q$ with $\q_\Delta$ as
\begin{equation}\label{Q2}
    g(\q):=e^{-H_{\beta,\ga,\la_{\beta,\ga}}(\q'_\Delta|\q ,\bar q)}
\end{equation}
and zero otherwise
is quasi-cylindrical in $C \cup \delta_{\text{\rm
out}}^{\ell_{+,\ga}}[C]$, where $C$ is a cube from $\mathcal
D^{(\ell_{+,\ga})}$ which covers all (finite many) pathes from
$\q'_\Delta$. The quasi-cylindricity holds, because only long loops could
lead to a long range interaction, but they cannot extend out of an
agreement set.  Hence the convergence of \eqref{Q1}
follows directly from \eqref{11.11c}.


\subsection*{Acknowledgement:} This research has been partially supported by
MURST, NATO Grant PST.CLG.976552, NSF Grant DMR 01-279-26. T.K want to thank for the support via a DAAD PostDoc and
A.v. Humboldt Feodor-Lynen fellowship.

The authors declare that no  artificial intelligence tools were used in the research at any stage, that includes, but is not exclusive, to writing, editing, projecting, preparing, discussing revising and correcting this manuscript. With the exception of this acknowledgement.
Hectic times require hectic measures.
\newpage

\appendix

\vskip3cm

 \setcounter{equation}{0}

\section {\bf Properties of the Hamiltonian}
\label{app:d0}

 \vskip1cm

\subsection {Assumptions on the interaction.}
\label{subsec:appd0.1}

We suppose that $j(r,r')$ is a bounded, symmetric
probability kernel on $\mathbb{R}^d$, translation
invariant, i.e.\ $j(r,r')=j(0,r'-r)$ for all $r,r'$, and
supported by the unit ball, i.e.\  $j(r,r')=0$ if
$|r-r'|>1$.  Letting
    \begin{equation}
         \label{d0.3}
A^{(\ell)}(r;r'):= \;\;\sup_{r_1,r_2\in \,
C^{(2\ell)}_{r'}\cup \,\delta_{\text{\rm out}}^{2\ell}
[C^{(2\ell)}_{r'}]} \;\;\big|j(r,r_1)-j (r,r_2)\big|,\qquad
\ell= 2^n, n\in \mathbb Z
     \end{equation}
we also suppose that there is a constant $c$ so that for
any $r$, $r'$ and $\ell$ as above,
   \begin{equation}
     \label{d0.1}
\int A^{(\ell)}(r; r') dr' \le c \ell,\qquad
 \int A^{(\ell)}(r; r') dr \le c \ell
     \end{equation}

\eqref{d0.1} is obviously satisfied if $j(r,r')$ is
differentiable with bounded derivative or if $j(r,r')=
c\text{\bf 1}_{|r-r'|\leq 1}$ ($c$ the normalization
constant), the case considered in \cite{LMP}.

 \vskip1cm

   \subsection {Stability of energy.}
   \label{subsec:d0.2}
We will next prove \eqref{2.a11}. Let $q \in Q_n$ and $\bar
q \in Q$. Recalling that $b=\inf_{x\ge 0} e'_\la(x)$,
     \begin{equation*}
e_\la(x+y) - e_\la(x) \geq b \ y,\qquad x,y\geq 0
     \end{equation*}
we obtain
      \begin{equation*}
e_\la\left( j_\ga * [q + \bar{q}]\right)- e_\la\left( j_\ga
* \bar{q}(r)\right) \geq b\   j_\ga * q(r) = b \sum_{i=1}^n
j_\ga(r,q_i)
      \end{equation*}
so that
       \begin{equation*}
h_{\ga,\la}(q|\bar{q}) \geq \int_{\mathbb{R}^d} b
\sum_{i=1}^n j_\ga(r,q_i)dr = bn
     \end{equation*}
because $j_\ga(r,r')=\ga^dj(\ga r,\ga r')$  is a symmetric
probability kernel. \qed

 \vskip1cm

\subsection {Energy coarse-graining.}  For any $\ell = 2^n, n\in
\mathbb Z$, we set
      \begin{equation}
        \label{d0.2}
\text{\rm Av}^{(\ell)}(f;r):= \ell^{-d}
\int_{C^{(\ell)}_{r}} f(r') dr',\quad j^{(\ell)}(r,r'):=
\text{\rm Av}^{(\ell)}( j(r ,\cdot);r')
     \end{equation}

\vskip1cm

     \begin{lemma}
    \label{lemmad0.1}
Let $\rho\in L^\infty (\mathbb R^d,\mathbb R_+)$,
$\ell=2^n, n\in \mathbb Z$, $c$ as in \eqref{d0.1}, then,
calling $\rho^{ (\ell)}(r)=\text{\rm Av}^{(\ell)}(\rho;r)$
 \begin{eqnarray}
         \label{d0.2b}
&&|j^{(\ell)}(r,r')- j(r ,r')| \leq A^{(\ell)}(r; r')
 \\
   \label{d0.5}
&&j * \rho^{ (\ell)} =  j^{ (\ell)}  * \rho
\\&& \hskip-.2cm
\Big|  j   * \rho(r)- j^{ (\ell)}  * \rho(r)\Big|
 \leq \; c  \ell \;\sup_{|r'-r|\le 1+10\ell}|\text{\rm
Av}^{(\ell)}(\rho;r')|
    \label{d0.4}
     \end{eqnarray}

    \end{lemma}

 \vskip.5cm

{\bf Proof.}   \eqref{d0.2b} is an immediate consequence of
\eqref{d0.2} and \eqref{d0.3}.
 \eqref{d0.5} follows from
   \begin{equation*}
j * \rho^{ (\ell)} =  j^{ (\ell)}  * \rho^{ (\ell)}= j^{
(\ell)}
* \rho
     \end{equation*}
 and
 \eqref{d0.4}
   from \eqref{d0.2b} and  \eqref{d0.1}.

   \qed

 \vskip1cm

The Lebowitz-Penrose limit involves a coarse graining of
the Hamiltonian, the coarse-grained version of the
Hamiltonian with mesh $\ell$ being the Hamiltonian with
kernel
      \begin{equation}
        \label{d0.7}
j_\ga^{(\ell)}(r,r'):=  \ga^d j^{(\ga \ell)}(\ga r,\ga r')
     \end{equation}
namely
   \begin{equation}
     \label{d0.6}
H^{(\ell)}_{\ga,\la}(\q) = \int_0^\beta
h^{(\ell)}_{\ga,\la}(\q(t)) dt ,\qquad
h^{(\ell)}_{\ga,\la}(q) := \int e_{\la_\beta}(j^{
(\ell)}_\ga * q) dr
     \end{equation}

In  Appendix \ref{app:aC}, we will actually use the  mesh
$\ell=\ga^{-1/2}$, and replace  $H_{\ga,\la}(\q|\bar\q)$ by
$\beta h_{\ga,\la}^{(\ga^{-1/2})}(\q(0)|\bar\q(0))$ with
$\q=\q_\La$, $\bar\q = \bar\q_{\La^c}$;
 $\La$ is a bounded
$\mathcal D^{(\ell_{-,\ga})}$-measurable region,
$\si(\q+\bar\q;\cdot)\equiv 1$ (i.e.\ all loops are short)
and there is a constant, namely
   \begin{equation}
     \label{d0.6aa}
X_0 = \max\{ \sqrt 2, e^{-\beta b}\}
     \end{equation}
such that
    \begin{equation}
            \label{d0.8}
|\q(0)\cap C ^{(\ell_{-,\ga})}| \le X_0
\ell_{-,\ga}^d,\quad | \bar\q(0)\cap C ^{(\ell_{-,\ga})}
| \le X_0 \ell_{-,\ga}^d,\qquad \text{for all $C
^{(\ell_{-,\ga})} \in \mathcal D ^{(\ell_{-,\ga})}$}
     \end{equation}
Given $\La$ in $\mathbb R^d$ and $R>0$, we write
     \begin{equation}
       \label{d0.9}
\partial_R\La
=\{r \in \La: \text{\rm dist}(r,\La^c) \leq R\} \cup \{r
\in \La^c: \text{\rm dist}(r,\La) \leq R\}
          \end{equation}

 \vskip1cm

 \begin{prop}
 \label{propaA.1}
There are $c$ and $\ga^*$ positive so that, for any $\ga\le
\ga^*$, $\La$, $\q$  and $\bar \q$ as above,
    \begin{equation}
         \label{d0.10}
 \Big| H_{\ga,\la}(\q|\bar \q ) - \beta h^{(\ga^{-1/2})}_{\ga,\la}(\q(0)|\bar \q(0) ) \Big|
 \le c  \ga^{1/2}\big(|\La| +|\partial_{4\ga^{-1}}\La|\big)
     \end{equation}

 \end{prop}

\vskip.5cm

{\bf Proof.} By assumption, the loops of $\q$ and $\bar \q$
are all short, then, if $\ga$ is small enough, for all $r$
such that dist$(r,\La)> 2\ga^{-1}$, and for all $t\in
[0,\beta]$,
    \begin{equation}
        \label{d0.11}
   j_\ga *[\q(t)+\bar \q(t)](r) =    [j_\ga * \bar \q(t)](r),\qquad
   j^{(\ga^{-1/2})}_\ga *[\q(t)+\bar \q(t)](r) =   [j^{(\ga^{-1/2})}_\ga * \bar
   \q(t)](r)
     \end{equation}
where $j_\ga *[\q+\q']= j_\ga *\q+j_\ga *\q'$, which is the
same as defining $\q+\q'$ as the configuration which
collects all loops of $\q$ and $\q'$.

Recalling that $j$ is bounded and supported by the unit
ball, by \eqref{d0.8} there is a constant $c$ so that, for
any $t\in [0,\beta]$,
    \begin{equation}
        \label{d0.12}
   j_\ga *[\q(t)+\bar \q(t)]  \le c,\qquad
   j^{(\ga^{-1/2})}_\ga *[\q(t)+\bar \q(t)]   \le c
     \end{equation}
Since $|e'_\la(x)|$ is bounded on the compacts, there is a
new constant $c$ so that
    \begin{eqnarray}
&&\text{l.h.s.\ of \eqref{d0.10}} \le c \int_0^\beta
\int_{\La\cup
\partial_{2\ga^{-1}}\La}  \Big(\Big| j_\ga*[\q(t)+\bar \q(t)] -
j^{(\ga^{-1/2})}_\ga*[\q(0)+\bar \q(0)]\Big| \nonumber
\\&&\hskip4cm + \Big|j_\ga* \bar q(t)  -
j^{(\ga^{-1/2})}_\ga* \bar \q(0) \Big| \Big) dr dt
        \label{d0.11aa}
     \end{eqnarray}
Since all the loops of $\q$ and $\bar \q$ are short,
recalling the definition \eqref{d0.3} of $A^{(\ell)}$, we
get
    \begin{eqnarray*}
&& \int_{\La\cup \partial_{2\ga^{-1}}\La}  \Big|
j_\ga*[\q(t)+\bar \q(t)] - j^{(\ga^{-1/2})}_\ga*[\q(0)+\bar
\q(0)] \Big|dr
\\&& \hskip2cm \le
\sum_{q_i\in \q(0)+\bar \q(0)} \ga^d\int_{\La\cup
\partial_{2\ga^{-1}}\La}
 A^{(\ga^{1/2})}(\ga r,\ga q_i)\, dr
\\&& \hskip2cm
\le \; c \ga^{1/2} |(\q(0)+\bar \q(0))\cap (\La\cup
\partial_{4\ga^{-1}}\La)|
     \end{eqnarray*}
because $ A^{(\ga^{1/2})}(\ga r,\ga q_i)=0$, $r\in
\La\cup
\partial_{2\ga^{-1}}\La$,
 if $q_i\notin \La\cup
\partial_{4\ga^{-1}}\La$.


The same procedure is used for the last term  with  $\bar
\q$ alone in \eqref{d0.11aa} and the proposition is proved.
\qed

  \vskip1cm

The same proof applies for the following variants:

\vskip.5cm

\begin{corol}
\label{Ku129}
 There are $c$ and $\ga^*$ positive so that for any
 $\ga \leq \ga^*$, $\La$, $\q$ and $\bar \q$ as above

   \begin{equation}
 \Big| H_{\ga,\la}(\q|\bar \q )
 - H_{\ga,\la}\big(\mathrm{Av}^{(\ga^{-1/2})}(\q(0))\big|\ \bar \q\big)\Big|
 \le c  \ga^{1/2}\big(|\La| +|\partial_{4\ga^{-1}}\La|\big)
     \end{equation}
 \end{corol}

 \begin{corol}
 \label{corolaA.1}
Let $\La$ be a cube of side $2^n\ell_{+,\ga}$, $\q=\q_\La$
and $\Theta\big(\q + \rho_\beta^{+}\text{\bf 1}_{\La^c}
\big)=1 $ then,
    \begin{equation}
    \label{aA.15}
 \Big| H_{\ga,\la}(\q|
 \rho^+_{\beta,\la}{\bf 1}_{\La^c} ) -
 \beta h^{(\ga^{-1/2})}_{\ga,\la}(\q(0)|\rho^+_{\beta,\la}{\bf 1}_{\La^c} ) \Big|
 \le c  \ga^{1/2}|\La|
     \end{equation}
[the analogous result holds for the minus case].
 \end{corol}

\vskip1cm

    \begin{lemma}
    \label{lemmaa5.2}
There is a constant $c$ so that if $\eta(\q(0);\cdot)\ne 0$
and $\si(\q;\cdot)\equiv 1$, then
      \begin{equation}
      \label{a5.12}
j_\ga*\q(t) \le c \rho_{\beta}^+,\qquad 0\le t\le \beta
      \end{equation}

     \end{lemma}

       \vskip.5cm

{\bf Proof.}  Since $[j_\ga*\q(t) ](r)  \leq \ga^d
\|j\|_\infty \big|\q(t)\cap B(r,\ga^{-1})\big|$,
$B(r,\ga^{-1})$ the ball with center $r$ and radius
$\ga^{-1}$, for $\ga$ small enough, $[j_\ga*\q(t) ](r) $
is bounded by the sum of $\big|\q(0)\cap
C^{(\ell_{-,\ga})}_{r'}\big|$ over all cubes with
$|r'-r|\le 3\ga^{-1}$ (because, by assumption, all loops
are short). Such a sum is bounded by $(3\ga^{-1})^d
(\rho_{\beta}^++\zeta)$, because
$\eta^{(\zeta,\ell_{-,\ga})}(\q(0);\cdot)\neq 0$.  The
lemma is proved.  \qed

  \vskip1cm

The next lemma is used in Section \ref{sec:a5}  in the
following context: $\La_n$ is the cube of side
$2^n\ell_{+,\ga}$, it is union of $2^d$ cubes
$\La_{n-1}(i)$, $i=1,..,2^d$, of side
$2^{n-1}\ell_{+,\ga}$.   $\q \in \Q^{\La_n}$ is such that
$\si(\q;\cdot)\equiv 1$ and $\eta(\q(0);r)=1$ for all $r\in
\La_n$.  We then set $\q^{(i)}= \q_{\La_{n-1}(i)}$ and
      \begin{equation}
      \label{a5.13}
U^{\pm}_\ga(\q) := H_{\ga,\la}(\q|\rho_\beta
^{\pm}\text{\bf 1}_{\La_n^c}) - \sum_{i=1}^{2^d}
H_{\ga,\la}(\q^{(i)}|\rho_\beta ^{\pm}\text{\bf
1}_{\La_{n-1}(i)^c})
      \end{equation}

\vskip1cm

    \begin{lemma}
    \label{lemmaa5.3}
There is a constant $c>0$ so that with the above notation
      \begin{equation}
      \label{a5.14}
\big |U^{\pm}_\ga(\q) \big| \le c \big(2^{n}
\ell_{+,\ga}\big)^{d-1} \ga^{-1}
      \end{equation}

     \end{lemma}

       \vskip.5cm

{\bf Proof.}  Using repeatedly the relation
      \begin{equation*}
 H_{\ga,\la}(\q + \q'|\q'') = H_{\ga,\la}(\q |\q'+ \q'')+
H_{\ga,\la}(\q'|\q'')
      \end{equation*}
we have
      \begin{equation}
      \label{a5.15}
H_{\ga,\la}(\q|\rho_\beta ^{\pm}\text{\bf 1}_{\La_n^c})=
\sum_{i=1}^{2^d} H_{\ga,\la}(\q^{(i)}|\q^{(i+1)}+\cdots +
\q^{(2^d)} + \rho_\beta ^{\pm}\text{\bf 1}_{\La_n^c})
      \end{equation}
Writing  $\q$ for $\q^{(i)}$ and $\q'$ for
$\q^{(i+1)}+\cdots + \q^{(2^d)} + \rho_\beta
^{\pm}\text{\bf 1}_{\La_n^c}$, we have, calling $\Si$ the
boundary of $\La_{n-1}(i)$ and using the fact that all
loops are short,
      \begin{eqnarray}
H_{\ga,\la}(\q|\q')-H_{\ga,\la}(\q) &=& \int_0^\beta
dt\int_{\text{dist} (r,\Si) \le \ga^{-1}+\ga^{-1/2}} \Big\{
e_\la \big( j_\ga*[\q(t)+\q'(t)](r)\big)\nonumber \\&&
\hskip1.5cm -e_\la \big(j_\ga*\q'(t)(r)\big)-e_\la
\big(j_\ga*\q(t)(r)\big)\Big\}dr
      \label{a5.16}
      \end{eqnarray}
Thus, by Lemma \ref{lemmaa5.2},
      \begin{equation*}
\big| H_{\ga,\la}(\q|\q')-H_{\ga,\la}(\q) \big | \le c'
|\Si| \ga^{-1}
      \end{equation*}
Analogously
      \begin{equation*}
\big|  H_{\ga,\la}(\q^{(i)}|\rho_\beta ^{\pm}\text{\bf
1}_{\La_{n-1}(i)^c}) - H_{\ga,\la}(\q^{(i)})\big| \le c'
|\Si| \ga^{-1}
      \end{equation*}
and the lemma is proved.  \qed

\vskip3cm

\setcounter{equation}{0}

\section{\bf Bounds on long loops}
       \label{sec:a7}

In this Appendix we will prove \eqref{b6.4}.   We shorthand
    \begin{equation}
    \label{a7.1}
 \nu_{\La}^>(d\q )= \nu_{\La}(d\q ) \,\text{\bf 1}_{\q=\q^>}
     \end{equation}
where, for any $\q$, $\q^>$ is the collection of long loops
in $\q$,  so that $\nu_{\La}^>$ is supported by
configurations with only long loops. Calling $\La= \{r\in
\text{sp}(\Ga):\si_\Ga(r) = 0\}$,  \eqref{b6.4} becomes a
consequence of the following proposition:

\vskip1cm

    \begin{prop}
    \label{propa7.1}
There is $c_1>0$ so that for any    bounded, $\mathcal
D^{(\ell_{+,\ga})}$-measurable region $\La$ and any
$\bar{\q}\in Q_{\text{\rm fin}}$  (i.e.\ $\bar{\q}$ is a
configuration in $\mathbb R^d$ with finitely many loops
not necessarily starting from $\La^c$)
     \begin{equation}
     \label{a7.2}
\int_{\si(\q;\cdot) = 0 \text{ on $\La$}} \nu^>_{\La}(d\q)
e^ {- H_{\ga}(\q| \bar{\q})} \; \leq \; e^{-c_1
\ga^{-1}N_\La }
      \end{equation}
where $N_\La= |\La|/\ell_{+,\ga}^d$

    \end{prop}

\vskip.5cm

{\bf Proof.}   By \eqref{2.a11},
    \begin{eqnarray}
    \label{a7.3}
  \text{l.h.s.\ of \eqref{a7.2}}  \leq
  \int_{\si(\q;\cdot) = 0 \text{ on $\La$}} \nu^>_{\La}(d\q)
  e^ {- \beta b |\q |}
     \end{eqnarray}
Denoting by $Q_1$ the space of single loops, namely the
space of periodic, continuous, $\mathbb R^d$-valued
functions $\om(t)$, $t\in   [0,\beta]$, we define
iteratively an increasing sequence of stopping times $T_k$,
$k\ge 0$, with values in $[0,\beta]\cup\{+\infty\}$, as
follows. We set  $T_0=0$ and, for $k\ge 1$, $T_k=\infty$ if
$T_{k-1} \geq \beta$, while, if $T_{k-1}<\beta$,
    \begin{equation}
    \label{a7.4}
T_k  =
  \begin{cases}
   t \in (T_{k-1},\beta] &
   \text{if $|\om(s)-\om(T_{k-1})| < \gamma^{-1/2}$ for all $s\in [T_{k-1},t)$,}
   \\ \; &  \hskip1cm \text{and $|\om(t)-\om(T_{k-1})| =
   \gamma^{-1/2}$}
   \\ +\infty &\text{if $|\om(s)-\om(T_{k-1})| < \gamma^{-1/2}$  for all $s\in [T_{k-1},\beta]$ }
  \end{cases}
   \end{equation}
Written differently: $T_k = \inf\{t  > T_{k-1}| \ \sup_{s
\in (T_{k-1},t]}|\omega(s)-\omega(T_{k-1})| \geq
\gamma^{-1/2}\}$
%
%
%

By \eqref{a7.1}, $\nu^>_{\La}(d\q)$ is supported by
configurations $\q=(\q_1,..\q_n)$ with $T_1(\q_i)<\infty$,
$i=1,,.,n$.  We decompose
     \begin{equation}
     \label{a7.5}
 \{T_1<\infty\} = \bigsqcup_{k\geq 2} \{T_k=\infty,T_{k-1}<\infty\}
      \end{equation}
hence
    \begin{equation}
    \label{a7.6}
 \nu_{\La}^>(d\q )=  \sum_{n,k_1,..,k_n} \, \frac{1}{n!} \prod_{i=1}^n W_{x_i,x_i}(d
 \om_i)\,dx_i
  \text{\bf
 1}_{x_i\in \La}\;
 \text{\bf 1}_{\{T_{k_i}(\om_i)=\infty,T_{k_i-1}(\om_i)<\infty\}}
     \end{equation}
where $W_{x,x}(d \om)$ is the law on $Q_1$ of a Brownian
bridge starting at $x$.

We are going to show that the condition $\{\si(\q;\cdot) =
0 \text{ on $\La$}\}$ in \eqref{a7.2} implies that
    \begin{equation}
    \label{a7.7}
 k_1+..+k_n \geq 2^{-d} N_{\La}
     \end{equation}
In fact, if $|\om(s)-\om(a)| \leq \ga^{-1/2}$ for all
$a\leq s\leq
b$, %
%
 then the set $\{\om(s), a\leq s
\leq b\}$ is contained in a ball of radius $\ga^{-1/2}$,
which is covered by at most $2^d$ cubes of $\mathcal
D^{(\ga^{-1/2})}$, hence by at most $2^d$ cubes of
$\mathcal D^{(\ell_{+,\ga})}$. Thus if $\om \in
\{T_k=\infty,T_{k-1}<\infty\}$, then $\{\om(t), 0\leq t
\leq \beta\}$ touches at most $2^d k$ cubes of $\mathcal
D^{(\ell_{+,\ga})}$ and \eqref{a7.7} follows.

By  \eqref{a7.3}
    \begin{equation}
    \label{a7.8}
  \text{l.h.s.\ of \eqref{a7.2}}  \leq
  \sum_{n=1}^\infty
 \sumtwo{k_i\geq 2, i=1,..,n}{k_1+,..+k_n \geq 2^{-d} N_{\La}}  \, \frac{(|\La| e^{-\beta b})^n}{n!}
   \prod_{i=1}^n W_{0,0}\big( \{T_{k_i} =\infty,T_{k_i-1} <\infty\}\big)
     \end{equation}
We bound $W_{0,0}\big( \{T_{k_i} = \infty,T_{k_i-1}
<\infty\}\big)\le W_{0,0}\big( \{ T_{k_i-1}
<\infty\}\big)$.  We claim that for $h\ge 2$
    \begin{eqnarray}
&& W_{0,0}\big(   T_{h} <\infty\big) \leq W_{0,0}\big(
T_{[\frac{h}{2}]} <\frac{\beta}{2}\big)+
 W_{0,0}\big(  \{ T_{[\frac{h}{2}]}\ge \frac{\beta}{2}\}\cap \{ T_{h} <\infty\}\big) \le
 2 W_{0,0}\big(  S_{[\frac{h}{2}]} <\frac{\beta}{2}\big)
 \nonumber\\
      \label{a7.9}
     \end{eqnarray}
where $S_k$ are the stopping times defined as in
\eqref{a7.4} but having replaced $\ga^{-1/2}$ by
$\ga^{-1/2}/2$.

Let  $\tilde \om (s)=\om(\beta-s)$, then $\beta-S_1(\tilde
\om)\geq T_{h-1}(\om)$ and, by iteration, $\beta-S_k(\tilde
\om)\geq T_{h-k}(\om)$.  Thus, if $ T_{ [{h}/{2}]}(\om)\ge
{\beta}/{2}$, then  $\beta-S_{h- [{h}/{2}]}(\tilde \om)\geq
{\beta}/{2}$,  and $ S_{h- [{h}/{2}]}(\tilde \om)\leq
{\beta}/{2}$. Then the last inequality in \eqref{a7.9}
follows the symmetry under time reflection of the law of
the Brownian bridge. We also have
    \begin{eqnarray}
W_{0,0}\big(   T_{1} <\infty\big)  \le
 2 W_{0,0}\big(   T_{1} <\frac{\beta}{2}\big) \le
 2 W_{0,0}\big(   S_{1} <\frac{\beta}{2}\big)
      \label{a7.10}
     \end{eqnarray}
We can now reduce  to a probability of a Brownian motion
(starting from 0), whose law will be denoted by $W_0$. In
fact, if $f(\om)$ is a bounded function measurable on
$\{\om(s), s\le \beta/2\}$, then
    \begin{eqnarray*}
\int W_{0,0} (d\om) f(\om) = (2\pi \beta)^{d/2} \int P_{0}
(d\om) f(\om) e^{-\om(\beta/2)^2/(2\beta/2)}(2\pi
\beta/2)^{-d/2}
     \end{eqnarray*}
so that, for $h\ge 1$,
    \begin{eqnarray}
W_{0,0}\big(   T_{h} <\infty\big) \leq 2^{d/2+1} W_{0}\big(
S _{p(h)} <\frac{\beta}{2}\big),\quad p(h)=\max\{[h/2], 1\}
      \label{a7.11}
     \end{eqnarray}
By the strong Markov property of Brownian motion, the
difference of the stopping times $S_k-S_{k-1}$ is
independent of $S_l -S_{l-1}$, then, by classical
properties of Brownian motion, see \cite{simon},
    \begin{eqnarray}
W_{0,0}\left(  T_h < \infty \right) \leq 2^{d/2+1} W_0\big(
S_1 \leq \frac{\beta}{2} \big)^{p(h)} \leq \Big(c e^{-
\ga^{-1}/(2\beta)} \Big)^{p(h)}
      \label{a7.12}
     \end{eqnarray}
for a suitable constant $c$.

Going back to \eqref{a7.8}, we get
    \begin{eqnarray*}
  \text{l.h.s.\ of \eqref{a7.2}}  &\leq &
  \sum_{n=1}^\infty
 \sumtwo{k_i\geq 2, i=1,..,n}{k_1+..+k_n \geq 2^{-d} N_{\La}} \, \frac{(|\La| e^{-\beta b})^n}{n!}
    \Big(c e^{-  \ga^{-1}/(2\beta)}  \Big)^{p(k_1-1)+\cdots p(k_n-1)}
   \nonumber  \\
    &\leq &
  \sum_{n=1}^\infty
 \sumtwo{k_i\geq 1, i=1,..,n}{k_1+..+k_n +n\geq 2^{-d} N_{\La}} \, \frac{(|\La| e^{-\beta b})^n}{n!}
    \Big(c e^{-  \ga^{-1}/(2\beta)}  \Big)^{p(k_1)+\cdots +p(k_n)}
     \end{eqnarray*}
and since $k_1+..+k_n \geq n$,
    \begin{eqnarray}
  \text{l.h.s.\ of \eqref{a7.2}}  &\leq &
  \sum_{n=1}^\infty
 \sumtwo{k_i\geq 1, i=1,..,n}{k_1+..+k_n \geq 2^{-d-1} N_{\La}} \, \frac{(|\La| e^{-\beta b})^n}{n!}
    \Big(c e^{-  \ga^{-1}/(2\beta)}  \Big)^{p(k_1)+\cdots +p(k_n)}
       \label{a7.13}
     \end{eqnarray}
Moreover
    \begin{eqnarray}
 p(k_1)+\cdots +p(k_n) \geq \frac 14 ( k_1 +\cdots + k_n)
       \label{a7.14}
     \end{eqnarray}
hence, calling
    \begin{eqnarray}
 N = 2^{-d-1} N_{\La} /8
       \label{a7.15}
     \end{eqnarray}
and supposing $\ga$ small enough,  (in particular $\ga$
such that $ce^{- \ga^{-1}/(16 \beta ) }< 1/2$)
    \begin{eqnarray}
  \text{l.h.s.\ of \eqref{a7.2}}  &\leq & [ce^{-\ga^{-1}/(2\beta)}]^N
  \sum_{n=1}^\infty  \frac{(|\La| e^{-\beta b})^n}{n!} \Big(
 \sum_{ k \geq 1 } \,
   \big(c e^{-  \ga^{-1}/(2\beta)}  \big)^{  k /8}\Big)^n \nonumber \\
    &\leq &  \exp\Big\{ - N [\ga^{-1}/(2\beta) - \log c] +
|\La| e^{-\beta b} 2ce^{- \ga^{-1}/(16 \beta )}\Big\}
       \label{a7.16}
     \end{eqnarray}
which proves \eqref{a7.2}.   \qed

\vskip3cm

\setcounter{equation}{0}

\section {\bf Basic properties of mean field}
\label{app:dA}

\vskip1cm

\subsection {Mean field free energy density.}

The mean field free energy density $f_{\beta,\la}(x)$,
$x\ge 0$, defined in \eqref{2.a14}, has a phase diagram
which exhibits a phase transition. There is in fact a
critical inverse temperature $\beta_c=(3/2)^{3/2}$ such
that, for $\beta<\beta_c$, $f''_{\beta,\la}(x)>0$, for all
$\la$ and $x$, while, for any $\beta>\beta_c$ there is a
unique value of $\la$, $\la=\la_\beta$, where
$f_{\beta,\lambda_\beta}$ has two distinct minimizers
$\rho_{\beta}^{\pm}$, elsewhere the minimizer is unique.
However there is an interval of values of $\la$ containing
$\la_\beta$ where there are still two local minima at
$\rho_{\beta,\la}^{\pm}$ with
$\rho_{\beta,\la}^{+}>\rho_{\beta,\la}^{-}$.
For $\la>\la_\beta$ the global
minimizer is $\rho_{\beta,\la}^{+}$ which is interpreted as the
 density of the liquid
phase. For $\la>\la_\beta$ it is the only  equilibrium
phase. For $\la < \la_\beta$ instead,
$\rho_{\beta,\la}^{-}$ is the global minimizer which is
interpreted as the density of the gas phase. The two
values $\rho_{\beta}^{+}>\rho_{\beta}^{-}$ are then the
densities of liquid and vapor at phase coexistence,
occurring at $\la=\la_\beta$.

Minimizers are solutions of  the mean field equation
$f'_{\beta,\la}(x)=0$, which, at $\la=\la_\beta$,  reads as
    \begin{equation}
     \label{dA.4}
x= K_\beta(x):=e^{-\beta e'_{\la_\beta}(x)} .
     \end{equation}
When $\beta>\beta_c$, it has three solutions:
$0<\rho^-_{\beta}<\rho^0_\beta<\rho^+_{\beta}$, and
    \begin{equation}
     \label{dA.5}
f''_{\beta,\la_\beta}(\rho_{\beta}^{\pm}) >0.
     \end{equation}
Notice that
    \begin{equation}
     \label{dA.5a}
\rho^+_{\beta}= e^{-\beta e'_{\la_\beta}(\rho^+_{\beta})}
\leq e^{-\beta b},
     \end{equation}
where
    \begin{equation}
     \label{1e.7}
   \inf_{x\ge 0} e'_\la(x) = e'_\la(\sqrt 2)=:  b.
     \end{equation}

There is $\beta_0>\beta_c$ such that
    \begin{equation}
     \label{dA.6}
K'_{\beta}(\rho_{\beta}^{+}) \in (-1,1)\quad \text{for all
$\beta\in (\beta_c,\beta_0)$}
     \end{equation}
while $K'_{\beta}(\rho_{\beta}^{-}) \in (0,1)$ for all
$\beta>\beta_c$. Therefore the linearization of the map
$x\to K_\beta(x)$, at $x=\rho_\beta^{\pm}$ is a contraction
when $\beta\in (\beta_c,\beta_0)$, this is the reason for
restricting in the paper of $\beta$ to $(\beta_c,\beta_0)$.

Because of \eqref{1e.7} the maximum of $K_{\beta}(x)$ is reached at
$x=\sqrt 2$.For any
    \begin{equation}
     \label{1e.7z}
  X \geq X_0:=\max\{\sqrt 2,K_\beta(\sqrt 2)\}, \qquad K_\beta(\sqrt
  2)= e^{-\beta b}
     \end{equation}
holds that
    \begin{equation}
     \label{A.9}
 K_{\beta}(x) : [0,X] \to [a,X]
 \qquad \text{ $a =\min \{K_\beta(0), K_\beta(X)\}$}.
     \end{equation}

In Section \ref{sec:10} we have considered the plus and
minus  free energy densities, defined for $\beta>1$ by
    \begin{equation}
     \label{aAa.1}
 f_{\beta}^{0,\pm}(x) = -\la_{\text{eff}}^{\pm} x - \frac {s(x)}{\beta}, \quad
 \la_{\text{eff}}^{\pm} = - e'_{\la_\beta}(\rho_\beta^{\pm}).
     \end{equation}
The function $f_{\beta}^{0,+}(x)$, [resp.\ $ f_{\beta}^{0,-}(x)$] is  strictly convex
 with minimum at $x= \rho_\beta^+$, [resp.\ $x= \rho_\beta^-$].
The interpolating mean field free energy is defined as
    \begin{equation}
     \label{aAa.1a}
f^{\pm}_{\beta,u}(x):= uf_{\beta,\la_\beta}
(x)+(1-u)f_{\beta}^{0,\pm}(x),\quad u\in [0,1],
     \end{equation}
which, for $u<1$ has $\rho_\beta^+$ [resp.\ $ \rho_\beta^-$] as its unique
minimizer.

\vskip1cm

\subsection {Mean field free energy functional.}

In terms of  statistical mechanics, $f_{\beta,\la}$ is the
limit canonical free energy density with a mean field
interaction.  For Kac potentials,  in the ``the Lebowitz-Penrose
limit'' (i.e.\ one takes first the thermodynamic
 limit and then the limit $\ga\to 0$, cf. Appendix~\ref{app:aC}, the
 relevant quantity is a free
energy functional, which, in the model considered in this
paper, is given by
    \begin{equation}
     \label{aB.1}
F_{\beta,\la}(\rho) = \int e_{\la}(j*\rho) -
\frac{s(\rho)}{\beta}\; dr.
     \end{equation}
We regard $F_{\beta,\la}(\rho)$ as a [non local free energy] functional on
$L^\infty_0(\mathbb R^d;\mathbb R_+)$, the space of non
negative valued, bounded measurable functions with compact
support. The integral on the r.h.s.\ is then well defined
as $e_\la(0)=s(0)=0$.

Notice also that on a torus $\La$,
$F_{\beta,\la}(x \text{\bf 1}_\La)= f_{\beta,\la}(x)|\La|$.

\vskip.5cm

It is often convenient to rewrite $F_{\beta,\la}(\rho)$  as
    \begin{equation}
     \label{aB.2}
F_{\beta,\la}(\rho) = \int f_{\beta,\la}(j*\rho) +
\frac{1}{\beta} \Big( s(j*\rho)- j*s(\rho)\Big)\; dr.
     \end{equation}
In fact, the second term is non negative for any
$\rho$, by the concavity of $s(\cdot)$, so that the
minimizer of the functional on a torus is a function
constantly equal to the minimizer of
$f_{\beta,\la}(\cdot)$.

The conditional free energy defined as
    \begin{equation}
     \label{aB.3}
F_{\beta,\la}(\rho|\bar\rho) =
F_{\beta,\la}(\rho+\bar\rho)-F_{\beta,\la}(\bar\rho)
     \end{equation}
is  the free energy of $\rho\in L^\infty_0(\mathbb
R^d;\mathbb R_+)$  ``conditioned on'' $\bar\rho\in
L^\infty_0(\mathbb R^d;\mathbb R_+)$.

\vskip.5cm

Associated to the  free energy densities  $f_{\beta}^{0,\pm}(\cdot)$
and $f^{\pm}_{\beta,u}(x)$
defined in \eqref{aAa.1}  and \eqref{aAa.1a}, we  consider
their corresponding free energy functionals, defined for
$\beta>1$ on $L^\infty_0(\mathbb R^d;\mathbb R_+)$ by
    \begin{equation}
     \label{aAa.2}
F_\beta^{0,\pm}(\rho):= \int f_{\beta}^{0,\pm}(\rho(r))\, d
r
     \end{equation}
and
    \begin{equation}
     \label{aAa.2a}
F_{\beta,u}^{ \pm}(\rho) =uF_{\beta,\la}(\rho)+(1-u)
F_\beta^{0,\pm}(\rho)= \int f_{\beta,u}^{ \pm}(\rho(r))\, d
r .
     \end{equation}

\vskip1cm

\subsection {Equilibrium.}  We will prove here  that the
conditional free energy ``has range 2'',  in Lemma~\ref{lemmaaB.1}
below, and that equilibrium outside a
region propagates inside, in Lemma~\ref{lemmaaB.2} below.  We
will use throughout the sequel the following notation:
given a function $\rho$ and a set $\La$,
     \begin{equation}
       \label{aB.4}
\rho_\La(r) = \rho(r) \text{\bf 1}_{r\in \La}.
          \end{equation}

\vskip.5cm

       \begin{lemma}
       \label{lemmaaB.1}
 Let $\La$ be a bounded measurable region and  $\bar\rho_{\La^c} \in L^\infty_0(\La^c;\mathbb
 R_+)$. Then
$F_{\beta,\la}(\rho_\La|\bar\rho_{\La^c})$  is   continuous
on $L^\infty(\La;\mathbb R_+)$, it does not depend on
$\bar\rho(r)$,
 $r\in  \La^c\setminus \partial_2 \La $ and there is a constant $c$
independent of $\La$ and $\bar\rho_{\La^c}$ so that
     \begin{equation}
       \label{aB.6}
F_{\beta,\la}(\rho_\La|\bar \rho_{\La^c}) \ge \int_\La
f_{\beta,\la}(j*[\rho_\La+\bar\rho_{\La^c}]) dr  - c
(1+\|\rho_\La+\bar\rho_{\La^c}\|^4) |\partial_1\La|
          \end{equation}

     \end{lemma}

\vskip.5cm

{\bf Proof.}  Continuity follows from the continuity of
$t\to f_{\beta,\la}(t)$ and $t\to s(t)$.  To prove
\eqref{aB.6},  we use equation \eqref{aB.2} to rewrite
$F_{\beta,\la}$ in the following way, ($\partial_R\La$ has
been defined in \eqref{d0.9}),
    \begin{eqnarray}
&& \hskip-.5cm F_{\beta,\la}(\rho_\La|\bar\rho_{\La^c}) =
\int_{\La\cup
\partial_1\La} f_{\beta,\la}(j*[\rho_\La+\bar\rho_{\La^c}]) dr  \nonumber
 \\&& \hskip2cm  + \frac{1}{\beta}
\int_{\La \cup \partial_1\La}  s(j*[\rho_\La +
\bar\rho_{\La^c}])- j*s([\rho_\La + \bar\rho_{\La^c}]) \;
dr \nonumber
 \\&& \hskip2cm  -\int_{\partial_1\La}
 f_{\beta,\la}(j* \bar\rho_{\La^c}) dr
- \frac{1}{\beta} \int_{\partial_1\La}
s(j*\bar\rho_{\La^c})- j*s(\bar\rho_{\La^c}) \; dr ,
     \label{aB.5}
     \end{eqnarray}
where we have used that $f_{\beta,\la}(0)=s(0)=0$.

The r.h.s.\ of \eqref{aB.5} does  not change if we vary
$\bar\rho_{\La^c}$ outside $\La\cup \partial_2\La$. By
concavity of $s(\cdot)$, the second term on the r.h.s.\ of
\eqref{aB.5} is non negative.  Since there is $c$ so that
for any $t$, $|f_{\beta,\la}(t)| \leq c(1+t^4)$ we obtain the estimate
\eqref{aB.6}.   The lemma is proved. \qed

\vskip.5cm


\vskip1cm

       \begin{lemma}
       \label{lemmaaB.2}
Let $\La$ be a bounded measurable region and $\La'=\La
\cup
\partial_2\La$. Then, for $\la\geq \la_\beta$, [resp.\ $\la\leq
\la_\beta$], $\rho_{\beta,\la}^+{\bf 1}_\La$
[resp.\ $\rho_{\beta,\la}^-{\bf 1}_\La$] is the unique minimizer
of $F_{\beta,\la}(\rho_\La|\rho_{\beta,\la}^+{\bf
1}_{\La'\setminus \La })$
[resp.\ $F_{\beta,\la}(\rho_\La|\rho_{\beta,\la}^-{\bf
1}_{\La'\setminus \La })$].

     \end{lemma}

\vskip.5cm

{\bf Proof.} Suppose $\la\geq\la_\beta$; the other case is
analogous and omitted.
Calling $\bar\rho_{\La^c} = \rho_{\beta,\la}^+{\bf 1}_{\La^c }$, 
the first   term on the r.h.s.\ of   \eqref{aB.5} is
bounded from below by $|\La \cup
\partial_1\La| f_{\beta,\la}(\rho_{\beta,\la}^+)$, the   inequality being strict
unless $\rho_\La= \rho_{\beta,\la}^+{\bf 1}_{\La}$. The
second term is non negative, the other ones do not depend
on $\rho_\La$, thus the minimizer is unique and given by
$\rho_\La= \rho_{\beta,\la}^+{\bf 1}_{\La}$.    \qed

\vskip1cm

\subsection {Surface corrections.} Here we suppose $\la=\la_\beta$.
We have proved in  Lemma \ref{lemmaaB.2}, that, if outside
a region $\La$ the density $\bar\rho(r)$ has its equilibrium value
$\rho_\beta^+$, then the conditional minimal
free energy is obtained when $\rho(r)=\rho_\beta^+$ in the
whole of $\La$ (the analogous property holds in the minus
case). This does not mean that the corresponding free
energy is equal to the equilibrium free energy density
$f_{\beta,\la_\beta}(\rho_\beta^+)$ times the volume,
because
   \begin{equation}
     \label{dA.9}
I_{\La,\La^c}^{\pm}
: =F_{\beta,\la_\beta}\Big(\rho_\beta^{\pm}\text{\bf 1}_\La
\Big|\rho_{\beta}^{\pm}{\bf 1}_{\La^c}\Big) -
f_{\beta,\la_\beta}(\rho_\beta^{\pm})|\La|
     \end{equation}
is not 0.   $I_{\La,\La^c}^{\pm}$ is a ``surface
term''.  In fact using expreesion \eqref{aB.1} instead of \eqref{aB.2}
in the derivation of \eqref{aB.5}
we obtain

    \begin{eqnarray*}
&& \hskip-.5cm F_{\beta,\la}(\rho_\La|\bar\rho_{\La^c}) =
\int_{\La\cup
\partial_1\La} e_{\beta,\la}(j*[\rho_\La+\bar\rho_{\La^c}]) dr - \frac{1}{\beta}
\int_{\La }  s(\rho_\La) \; dr \nonumber
 \\&& \hskip2cm  -\int_{\partial_1\La}
 e_{\beta,\la}(j* \bar\rho_{\La^c}) dr .
     \end{eqnarray*}
Using this expression in \eqref{dA.9}, we get
     \begin{eqnarray}
&& I_{\La,\La^c}^{\pm} 
     =  \int_{\partial_1 \Lambda}
    \text{\bf 1}_{\La^c} \ e_{\la_\beta}(\rho_\beta^{\pm})
    - e_{\la_\beta}(j*\rho_\beta^{\pm}\text{\bf
    1}_{\La^c}) \; dr \nonumber\\
      \label{B.18a}
\\&&\hskip2cm = h(\rho_\beta^{\pm}\text{\bf
1}_{\La}|\rho_\beta^{\pm}\text{\bf
    1}_{\La^c})-|\La| e_{\la_\beta}(\rho_\beta^{\pm}) .
\nonumber
      \end{eqnarray}
 An analogous formula, with $\Delta$ replacing
$\La^c$, defines $I_{\La,\Delta}^{\pm}$. Note that
$I_{\La,\Delta}^{\pm}\ne I_{\Delta,\La}^{\pm}$.


\vskip1cm

       \begin{prop}
       \label{propdB.1}
Let $\La$ be a bounded measurable region,  then
   \begin{equation}
     \label{dA.10}
F_{\beta,\la_\beta}\Big(\rho_\beta^{-}\text{\bf 1}_\La\Big)
+ I_{\La^c,\La}^{-}=
F_{\beta,\la_\beta}\Big(\rho_\beta^{+}\text{\bf 1}_\La\Big)
+ I_{\La^c,\La}^{+}=
f_{\beta,\la_\beta}(\rho_\beta^{\pm})|\La| .
     \end{equation}
     \end{prop}

\vskip.5cm

{\bf Proof.} Using equation \eqref{aB.1} we can write
$F_{\beta,\la_\beta}\Big(\rho_\beta^{\pm}\text{\bf
1}_\La\Big)$ as
\begin{eqnarray*}
&=&\int e_{\la_\beta}(j*(\rho_\beta^{\pm}\text{\bf
1}_\La))dr
- \frac{s(\rho_\beta^{\pm})}{\beta}\ |\La|\\
&=&\int_\La e_{\la_\beta}(\rho_\beta^{\pm}) dr -
\frac{s(\rho_\beta^{\pm})}{\beta}|\La|+\int_{\partial_1
\La}
\Big(e_{\la_\beta}(j*(\rho_\beta^{\pm}\text{\bf 1}_\La))
-\text{\bf 1}_\La \
e_{\la_\beta}(\rho_\beta^{\pm})\Big)dr \\
&=& f_{\beta,\la_\beta}(\rho_\beta^{\pm} ) -
I^{\pm}_{\La^c,\La} .
\end{eqnarray*}
\notes{plus has beend changed to minus}
\qed

\vskip.5cm

{\bf Remark.}  If in Proposition \ref{propdB.1} we take
$\La$  equal to $C_i^{-}$, to $C_i^+$, or to $C_i^{-} \cup T_i \cup C_i^+$ we obtain equations
\eqref{b6.15}.

\vskip 2cm
\subsection {Dynamics.}

 For any bounded measurable region $\La$, we consider the evolution equation
    \begin{equation}
     \label{B.12}
      \begin{cases}
\dis{\frac{d \rho }{d t}}=-\rho+ \Phi_\beta(\rho;\cdot)
&\text{on $\La$}
\\ \rho(r,t)=\rho_0(r)  &\text{on $\La^c\times \mathbb R_+$}
\\ \rho(r,0)=\rho_0(r)  &\text{on $ \mathbb R^d$} \end{cases}
     \end{equation}
where
   \begin{equation}
     \label{dA.9an}
 \Phi_\beta(\rho;r) = e^{-\beta j* e'_{\la_\beta}(j*\rho(r))}
     \end{equation}
and   $\rho_0 \in L^\infty(\mathbb R^d, \mathbb R_+)$.

The evolution equation \eqref{B.12} has a unique global solution $\rho(r,t)$,
more precisely, for each $r$, the solution $\rho(r,t)$ is differentiable for
$t>0$ and converges to $\rho_0(r)$ as $t\to 0$. Moreover,
for each $t$, the solution $\rho(\cdot,t) \in L^\infty(\mathbb R^d,
\mathbb R_+)$. By setting
$T_t^\Lambda(\rho_0)=\rho(\cdot,t)$ we thus define a
semigroup $T_t^\Lambda$ on $L^\infty (\mathbb R^d,\mathbb
R_+)$.

The solution $\rho(r,t)$ of \eqref{B.12} solves also the
integral version of \eqref{B.12}, namely
    \begin{equation}
     \label{Ku8}
 \rho(r,t) = e^{-t}\rho_0(r) + \int_0^t e^{-(t-s)}
\Phi_\beta(\rho(\cdot,s);r) ds,
    \end{equation}
from where we deduce that
  \begin{equation}
     \label{Ku128}
e^{-t} \rho_0(r) \leq \rho(r,t) \leq e^{-t} \rho_0(r) +
(1-e^{-t}) e^{-\beta b}
    \end{equation}
with $b$ defined as in Subsection \ref{subsec:d0.2}.

The upper bound in \eqref{Ku128} allows to improve the
lower bound.  In fact,
 since $|e'_{\lambda_\beta}(x)| \leq
|\lambda_\beta|+|x| + \frac{|x|^3}{3!}$, we can  bound
$\Phi_\beta$ from below by
    \begin{equation*}
\Phi_\beta(\rho(t,\cdot);r)\geq e^{-\beta \left(
|\lambda_\beta|+\|\rho_0\|_{L^\infty}+ e^{-\beta b}+D_\beta
+\frac{1}{3!}(\|\rho_0\|_{L^\infty}+e^{-\beta b})^3
\right)}
 =: { {D}(\beta,\|\rho_0\|_{L^\infty} )}>0,
   \end{equation*} \notes{$e^{-\beta b}$ added}
which implies
    \begin{equation}
        \label{Ku128a}
        \rho(r,t)\geq e^{-t} \rho_0(r) +(1-e^{-t}){
{D}(\beta,\|\rho_0\|_{L^\infty} )}.
  \end{equation}

Our main  interest for dynamics is that it makes the free
energy decrease. A direct calculation shows that \vskip1cm

        \begin{prop}
        \label{propB.2}
Let $\rho \in L^\infty(\mathbb R^d)$ and $\La$ as above.
Then for any $t\ge 0$
  \begin{equation}
     \label{B.15}
F_{\beta,\la_\beta}\Big(T^\La_t(\rho)\,\text{\bf
1}_\La\Big|\rho\,\text{\bf 1}_{\La^c}\Big)-
F_{\beta,\la_\beta}\Big(\rho\,\text{\bf
1}_\La\Big|\rho\,\text{\bf 1}_{\La^c}\Big)= - \int_0^t\,
\mathcal I_\beta(T^\La_s(\rho);\La)ds,
    \end{equation}
where
 \begin{equation}
     \label{B.16}
\mathcal I_\beta(\rho;\La)= \int_\La\,\Big(-\rho+
\Phi_\beta(\rho;\cdot) \Big)\frac 1\beta \Big(  -\log\rho
+ \log \Phi_\beta(\rho;\cdot)\Big) dr.
     \end{equation}
The functional $\mathcal I_\beta(\rho;\La)\ge 0$ and equality holds iff
    \begin{equation}
     \label{B.12a}
\rho(r)= \Phi_\beta(\rho;r),\qquad \text{for all\ $r\in
\La$}.
     \end{equation}

         \end{prop}

\vskip1cm

       \begin{prop}
       \label{propB.6}
Let $\rho \in L^\infty(\mathbb R^d)$ be non-negative and $\La$
as above, then, as $t\to \infty$, $T_t^\La(\rho)$ converges
along subsequences in $L^\infty(\mathbb R^d)$. Moreover,  any
limit point $u$ of $T^\La_t(\rho)$ satisfies equality \eqref{B.12a}
and $F_\beta(u|\rho_{\La^c})\le F (\rho_\La|\rho_{\La^c})$.

     \end{prop}

    {\bf Proof.}
For each non-negative $\rho \in L^\infty(\mathbb R^d)$ the
set $\{T_t^\La(\rho) - e^{-t} \rho | \ t \in [0,\infty)\}$
is relative compact in the uniform norm. This is true,
because the following estimate shows that the set is equi-continuous.
     \begin{eqnarray}
        &&\left|\left(T_t^\La(\rho)(r)-e^{-t} \rho(r)\right)
        -\left(T_t^\La(\rho)(r')-e^{-t} \rho(r')\right)\right|\\
        &\leq &\int_0^t e^{-(t-s)} \left| \Phi_\beta(T_s^\La(\rho);r)
        - \Phi_\beta(T_s^\La(\rho);r') \right|\\
        &\leq& \beta b e^{-\beta b} \int_0^t e^{-(t-s)} \|T_s^\La(\rho) \|_{L^\infty}\  ds
        \int |j(r,x)-j(r',x)| dx.
     \end{eqnarray}
Let  $\rho\in L^\infty(\mathbb R^d,\mathbb R_+)$ and let
$\left(T_{t_n}^\La(\rho)\right)_{n \in \mathbb N}$ be a
converging subsequence. Recalling the bounds \eqref{Ku128} and
\eqref{Ku128a}, the integrand in
$F_{\beta,\la_\beta}\Big(T^\La_t(\rho)\,\text{\bf
1}_\La\Big|\rho\, \text{\bf 1}_{\La^c}\Big)$ is uniformly
integrable. Hence the equation \eqref{B.15} holds also for the limit point and
$\mathcal I_\beta(T_t^\La(\rho);\La)$ is continuous in $t$. Thus
$\mathcal I_\beta(\cdot;\La)$ is $0$ for the limit point.
     \qed

\vskip1cm

For the interpolating free energy functional, dynamics is
defined by replacing $ \Phi_\beta(\rho;r)$ with
   \begin{equation}
     \label{dA.9a}
 \Phi^{\pm}_{\beta,u}(\rho;r) =
 e^{-\beta \big(u j* e'_{\la_\beta}(j*\rho(r))+(1-u)e'_{\la_\beta}( \rho_\beta^{\pm})\big)}.
     \end{equation}
Recall that, in fact, by definition $\la^{\pm}_{\text{eff}}=
-e'_{\la_\beta}( \rho_\beta^{\pm})$. The solutions to the
evolution equation define a semigroup $T_t^{\La,\pm
,u}(\rho)$ for which the analogues of Proposition
\ref{propB.2} and  Proposition \ref{propB.6} hold.

\vskip3cm

\setcounter{equation}{0}

\section {\bf Non local functionals, small deviations}
\label{sec:dB}

\vskip1cm

The main result in this appendix is that if a density
profile is in a neighborhood of the equilibrium value in a
region $\La$, then by decreasing its free energy it can be
made closer to equilibrium at an exponential rate from the
boundary of $\La$, Theorem \ref{thmdB.1} below.

By default, throughout the sequel $\ell$ denotes an element
of $ \{2^n, n\in \mathbb Z\}$.  The ``coarse grained
image'' $\text{Av}^{(\ell)}(\rho;r)$, $\rho\in
L^\infty(\mathbb R^d)$, see \eqref{d0.2},   is a bounded
function constant on the cubes of $\mathcal D^{(\ell)}$.
For any $\zeta>0$ we then set
    \begin{equation}
      \label{dB.2}
\eta^{(\zeta,\ell)}(\rho;r) =\begin{cases} \pm 1
 &\text{if $|\text{Av}^{(\ell)}(\rho;r)- \rho_{\beta,\pm}|\le \zeta$}
\\ 0 &\text{otherwise}\end{cases}.
     \end{equation}

\vskip1cm

       \begin{lemma}
       \label{lemmadB.1}

There are $\zeta'_0$, 
$d_0$ and $\eps_0$ all positive such that for any $\zeta\le
\zeta'_0$,  any $\ell \le d_0 \zeta $, any bounded $\mathcal D^{(\ell)}$-measurable
region $\La$ and any $\rho \in L^\infty(\mathbb R^d,\mathbb
R_+)$ such that $\eta^{(\zeta,\ell)}(\rho;r)=\pm 1$,
dist$(r,\La)\le 10$, it holds that
     \begin{equation}
           \label{dB.3}
 \big| \Phi_\beta(\rho;r)- \rho^{\pm}_{\beta} \big|\le   (1
 -\eps_0)\zeta,\qquad r\in \La
          \end{equation}
and
     \begin{equation}
     \label{dB.4}
 \big| j*\rho (r) -  \rho^{\pm}_{\beta} \big|
 \le 2\zeta,\qquad \text{{\rm dist}}(r, \La) \le 2 ,
          \end{equation}
where $ \Phi_\beta$ is as defined below in
\eqref{dA.9an}.
     \end{lemma}

\vskip.5cm

{\bf Proof.} As the proofs for the $+$ and  the $-$ cases
are the same, we will only consider the former. By
\eqref{dA.6} one has  for $\beta \in (\beta_0,\beta_c)$
    \begin{equation*}
\left|\frac{d}{ds}e^{-\beta e'_{\la_\beta}( s)
}\Big|_{s=\rho^+_\beta} \right| <1 .
     \end{equation*}
Then, for any $\epsilon>0$ small enough, there
 is $\delta>0$ such that, for all $s>0$
with $|\rho^+_\beta - s| \leq \delta$, it holds
    \begin{equation}
    \label{ada.4}
\beta |e''_{\la_\beta}(s)| \leq (1-\epsilon) e^{\beta
e'_{\la_\beta}( \rho^+_{\beta})} = \frac{1-\epsilon}{
\rho^+_\beta} .
    \end{equation}
Take $\zeta_0 \leq \delta/2$. The bound \eqref{d0.4} yields for
$\ell<1$,
    \begin{equation*}
\big|j*\rho(r) - j^{(\ell)}*\rho(r) \big| \le
(\rho_\beta^++\zeta)c\ell, \qquad \text{dist}(r,\La) \leq 2 .
    \end{equation*}
The equality \eqref{d0.5} impplies that $j^{(\ell)}*\rho \in
[\rho_\beta^+-\zeta,\rho_\beta^++\zeta]$ such that one has
    \begin{equation}
    \label{Ku103}
\big|j*\rho(r) - \rho_\beta^+\big| \le c\ell
(\rho_\beta^++\zeta)+ \zeta \le (1+\epsilon) \zeta
    \end{equation}
for the choice $0<d_0<\frac{\epsilon}{c(\rho^+_\beta +
\zeta_0)}$.  We have
 thus
  proved \eqref{dB.4}. The bound \eqref{ada.4} implies that
\begin{equation}
\label{Ku106} \beta
|j*e_{\la_\beta}'(j*\rho)-e_{\la_\beta}'(\rho_\beta^+)|
\leq (1-\epsilon) e^{\beta e'_{\la_\beta}(\rho^+_\beta)}
j*|j*\rho-\rho_\beta^+|.
\end{equation}
Using \eqref{Ku103}, we can bound this by
\begin{equation}
\label{Ku104}(1-\epsilon) e^{\beta
e_{\la_\beta}'(\rho^+_\beta)} (1+\epsilon) \zeta
\end{equation}
which, in particular, is  bounded by
\begin{equation}
\label{Ku105} e^{\beta e_{\la_\beta}'(\rho^+_\beta)} \zeta .
\end{equation}
By the inequality $|e^x-1|\leq e^{|x|} |x|$ applied for
$x$ equalt to the l.h.s. of \eqref{Ku106},
using the bounds \eqref{Ku104} and \eqref{Ku105} we get
\begin{eqnarray}
| \Phi_\beta(\rho;r) - \rho^+_\beta |
&\leq& e^{\zeta e^{\beta e_{\la_\beta}'(\rho^+_\beta)}}
(1-\epsilon^2) \zeta .\label{Ku127}
\end{eqnarray}
Choose $\zeta_0 < e^{-{\beta e_{\la_\beta}'(\rho^+_\beta)}}
\ln(1+\epsilon^2) $ then \eqref{Ku127} is bounded by $
(1-\epsilon^4) \zeta$. Choosing $\epsilon$ small enough
the lemma is proved with $\zeta_0'=\min\{\delta/2,
e^{-{\beta e_{\la_\beta}'(\rho^+_\beta)}} \ln(1+\epsilon^2)
\}$.  \qed

\vskip1cm

Given $\zeta$, $\ell$ and a $\mathcal
D^{(\ell)}$-measurable region $\La$, we set
   \begin{equation}
N^{\pm}_{\zeta,\ell,\La}:= \Big\{\rho\in L^\infty(\mathbb
R^d,\mathbb R^+): \eta^{(\zeta,\ell)}(\rho;r)=\pm 1,
\text{whenever dist$(r,\La) \le 10$}\Big\} .
      \label{dB.8}
     \end{equation}

\vskip.5cm

      \begin{lemma}
      \label{lemmadB.2}
Let $\zeta$ and $\ell$ as in Lemma \ref{lemmadB.1}, $\La$ a
bounded $\mathcal D^{(\ell)}$-measurable region and $\rho
\in N^{\pm}_{\zeta,\ell,\La}$, then $T^\La_t(\rho) \in
N^{\pm}_{\zeta,\ell,\La}$ for all $t\ge 0$.

     \end{lemma}

     \vskip.5cm

{\bf Proof.}  We just consider the $+$ case.
 Let $\tau>0$, let $v_0\in N^+_{\zeta,\ell,\La}$ and let
 \begin{eqnarray*}
\mathcal S_{\tau,v_0}= \Big\{v\in
C\big([0,\tau];L^\infty(\mathbb R^d,\mathbb R^+)\big):
v(\cdot,0)=v_0(\cdot) \; \text{ and $v(r,t)=v_0(r)$ on
$\La^c\times [0,\tau]$} \;\Big\} .
  \end{eqnarray*}
Let then $\Psi$ be the map on $\mathcal S_{\tau,v_0}$
defined by setting for $t\in [0,\tau]$ and  $r\in \La$:
      \[
\Psi(v)(r,t)= e^{-t}v_0(r) + \int_0^t ds\, e^{-(t-s)}
\Phi_\beta(v(\cdot,s) ;r) .
      \]
As in \eqref{Ku128} we have that
$\Psi(v):[0,\tau]\rightarrow L^\infty(\mathbb R^d, \mathbb
R^+)$ is continuous and
   \[
e^{-t} v_0(r) \leq \Psi(v)(r,t) \leq e^{-t} v_0(r) +
(1-e^{-t})e^{-\beta b} .
   \]
Hence $\Psi$ maps $\mathcal S_{\tau,v_0}$ onto itself. If
$\tau>0$ is small enough, $\Psi$ is a contraction on
$\mathcal S_{\tau,v_0}$ w.r.t.\ the sup norm, because $0
\leq \Phi_\beta(v(\cdot,s)) \leq e^{-\beta b}$. The fixed
point of $\Psi$ is the orbit $T^\La_t(v_0)$, $0\le t \le
\tau$.

If $v\in N^+_{\zeta,\ell,\La}$, then the bound \eqref{dB.3} yields
   \begin{equation}
 \rho_\beta^+ - (1-\eps_0)\zeta \le \Phi_\beta(v;r) \le  \rho_\beta^+
 + (1-\eps_0)\zeta .
      \label{dB.8aa}
     \end{equation}
Hence  the set
      \[
Y =\left\{v\in \mathcal S_{\tau,v_0}: v(\cdot,t)\in
N^+_{\zeta,\ell,\La}\;
 \text{ for all  $t\in [0,\tau]$} \right\}
      \]
is invariant under the action of $\Psi$.  Since $Y$ is closed, the
fixed point of $\Psi$ is in $Y$, hence $T^\La_t(v_0)\in
N^+_{\zeta,\ell,\La}$, $0\le t\le \tau$. By induction the
statement remains valid for all $t$.  \qed

  \vskip1cm

{\bf Remark.}  The same conclusion holds for the semigroup
$T^{\La,\pm,u}_t$ defined below \eqref{dA.9a}, namely
 $T^{\La,\pm,u}_t(\rho) \in N^{\pm}_{\zeta,\ell,\La}$ for
all $t\ge 0$.  The   proof is completely analogous with the exception
of the derivation of \eqref{dB.8aa} we observe that
$\Phi^{\pm}_{\beta,u}(\rho;r)$, which is defined in
\eqref{dA.9a}, satisfies
   \begin{equation*}
 \Phi^{+}_{\beta,u}(v;r) =  \big(\Phi_{\beta}(v;r)\big)^u
 \big(\rho_\beta^{+}\big)^{1-u} .
     \end{equation*}
Hence, if  $v\in N^+_{\zeta,\ell,\La}$, then one can estimate
   \begin{equation*}
  \Phi^{\pm}_{\beta,u}(v;r)\le  \big(\rho_\beta^{+}\big)^{1-u}
 \big(\rho_\beta^{+}+(1-\eps_0)\zeta\big)^{u} \le \rho_\beta^{+}+(1-\eps_0)\zeta
     \end{equation*}
with equality only if $u=1$.  The lower bound is proved
analogously, hence also $T^{\La,\pm,u}_t(\rho) \in
N^{\pm}_{\zeta,\ell,\La}$ for all $t\ge 0$.

\vskip1cm

With the above notation, we set for any $\rho \in
N^{\pm}_{\zeta,\ell,\La}$,
     \begin{equation}
     \label{dB.9}
\mathcal X^{\pm}_{\La,\rho}=\Big\{ v\in
N^{\pm}_{\zeta,\ell,\La}: v_{\La^c}= \rho_{\La^c}\Big\}
     \end{equation}
where $\psi_\Delta$ stands for the restriction of a
function $\psi$ to a set $\Delta$.

\vskip.5cm

     \begin{thm}
      \label{thmdB.1}
There are  $\zeta_0$ ($\zeta_0<\zeta'_0$, $\zeta'_0$ as in
Lemma \ref{lemmadB.1}) $\om$ and $c_\om$ all positive.
For any $\zeta<\zeta_0$, $\ell<d_0
\zeta$ ($d_0$ as in Lemma \ref{lemmadB.1}) , $\La$ a
bounded, $\mathcal D^{(\ell)}$-measurable region, $\rho$
and $\hat\rho$ in $N^{\pm}_{\zeta,\ell,\La}$ the following holds:
 \begin{itemize}

\item   In  $\mathcal
X^{\pm}_{\La,\rho}$, there is a unique minimizer
$\psi^{\pm}$ of $ F_{\beta,\la_\beta}(\cdot
|\rho_{\La^c})$; more precisely, the minimizer $\psi^{\pm}(r)$ is the
unique solution of the equation  $\psi^{\pm}(r)=\Phi_\beta(\psi^{\pm};r)$,
$r\in \La$, in $\mathcal X^{\pm}_{\La,\rho}$. Moreover
     \begin{eqnarray}
&& \psi^{\pm}(r) \in [\rho^{\pm}_\beta-\zeta
+\eps_0\zeta,\rho^{\pm}_\beta+\zeta-\eps_0\zeta],\quad r
\in \La
       \label{p4.3.1.10a}
\\&&
 |\psi_\La(r)-\rho^{\pm}_\beta| \le c_\om e^{-\om \,
\text{dist}(r, \La_{\ne}^c)} ,
   \label{p4.3.1.10}
     \end{eqnarray}
where $\La_{\ne}^c=\{r\in \La^c:$ dist$(r,\La)\le 2;
\rho_{\La^c}(r)\ne \rho^{\pm}_{\beta})\}$ and $\eps_0$ is as in
Lemma \ref{lemmadB.1}).

\item   Finally, if $\hat \psi$ is the minimizer of $
F_{\beta,\la_\beta}(\cdot|\hat\rho_{\La^c})$ in $\mathcal
X_{\La,\hat\rho}$, and calling, by a slight abuse of notation,
$\La_{\ne}^c=\{r\in \La^c:$ dist$(r,\La)\le 2;
\hat\rho_{\La^c}(r)\ne \rho_{\La^c}(r)\}$, then it holds that
     \begin{equation}
      \label{p4.3.1.11}
|\hat \psi_\La(r)-\psi_\La(r)| \le c_\om e^{-\om \,
\text{dist}(r, \La_{\ne}^c)} .
     \end{equation}
 \end{itemize}

     \end{thm}
\vskip.5cm

{\bf Proof.} To simplify notation we consider only the $+$
case. By Lemma \ref{lemmadB.2}, the following is known:
the semi-group $T^\La_t$ leaves $\mathcal X^+_{\La,\rho}$ invariant and
since $\mathcal X^+_{\La,\rho}$ is
closed under uniform convergence on compacts, 
for any $u\in \mathcal X^+_{\La,\rho}$, the solution $T_t^\La u$
converges by subsequences to an element $\psi$ of
    \[
X^0_{\La,\rho}=\Big\{ \psi\in \mathcal X^+_{\La,\rho}:
\;\text{$\psi$ solves \eqref{B.12a} in $\La$} \Big\}
   \]
and $F_{\beta,\la_\beta}(v_\La|\rho_{\La^c})\ge
F_{\beta,\la_\beta}(\psi_\La|\rho_{\La^c})$, the inequality
being strict unless $v\in X^0_{\La,\rho}$:
    \begin{equation}
      \label{p4.3.1.12}
  F_{\beta,\la_\beta}(v_\La|\rho_{\La^c}) >\inf_{\psi\in X^0_{\La,\rho}}
F_{\beta,\la_\beta}(\psi_\La|\rho_{\La^c}),\quad \text{for
any $v\in \mathcal X_{\La,\rho}\setminus X^0_{\La,\rho}$} .
     \end{equation}
By Lemma \ref{lemmadB.1}, any $\psi \in X^0_{\La,\rho}$
satisfies the first condition \eqref{p4.3.1.10a}.

We will show next that  for $\zeta$ small enough,
$X^0_{\La,\rho}$ consists of only one element $\psi^+$,
which is therefore the strict minimizer of
$F_{\beta,\la_\beta}(v_\La|\rho_{\La^c})$. Suppose that
$\psi$ and $\psi'$ are  in $\mathcal X^0_{\La,\rho}$, then,
for any $r \in \La$, and writing $\psi_s=
s\psi+(1-s)\psi'$, $0\le s \le 1$,
 \begin{eqnarray}
  \psi(r) - \psi'(r) &= &   \Phi_\beta(\psi;r)- \Phi_\beta(\psi';r) \nonumber
  \\ &=& -\beta \int_0^1  e^{-\beta
j* e'_{\la_\beta}(j*\psi_s)}
j*\Big(e''_{\la_\beta}(j*\psi_s)j*(\psi-\psi')\Big) ds .
  \label{aeD.17}
  \end{eqnarray}
Since $|e^{-\beta e'_{\la_\beta}(\rho^+_\beta)}
e''_{\la_\beta}(\rho_{\beta}^+)|< 1$ and  by bound \eqref{dB.4} and
since the expressions
$j*\psi(r)$ and $j*\psi'(r)$ are both in
$[\rho_\beta^+-2\zeta,\rho_\beta^++2\zeta]$ for all
dist$(r,\La)\le 2$, it then follows that if   $\zeta_0$ is
small enough, one has
  \begin{equation}
      \label{p4.3.1.13}
e^{-\om}:=  \sup_{ |a|,|b|\le 2\zeta_0} \beta  e^{-\beta
e'_{\la_\beta}(\rho^+_\beta+a)}
e''_{\la_\beta}(\rho^+_\beta+b) < 1 .
   \end{equation}
Therefore, one can estimate \eqref{aeD.17} by
 \begin{eqnarray*}
 \big| \psi(r) - \psi'(r) \big| &\le &     e^{-\om}
j*   j*\big|\psi-\psi'\big|
  \end{eqnarray*}
and since $\psi(r)=\psi'(r)=\rho(r)$ for $r\notin \La$, it
then follows that $\psi=\psi'$.

Finally, let $\psi \in X^0_{\La,\rho}$ and $\hat\psi \in
X^0_{\La,\hat\rho}$, then, for $r\in \La$,
     \begin{eqnarray}
&& |\psi (r)-\hat\psi (r)| \le e^{-\om}  j*j*(|\psi -
\hat\psi| \text{\bf 1}_\La) + c j*j* \text{\bf
1}_{\La_{\ne}^c} .
     \label{p4.3.1.14}
     \end{eqnarray}
By iterating \eqref{p4.3.1.14} and calling $n_0$
the largest integer such that $2n_0 \le$ dist$(r,
\La_{\ne}^c)$, we get
       \[
|\psi (r)-\hat\psi (r)| \le c\sum_{n\ge n_0} e^{-\om n} ,
       \]
which yields  \eqref{p4.3.1.11} with
   \begin{equation}
      \label{p4.3.1.15}
c_\om:=\frac{c\,e^\om}{1-e^{-\om}} .
   \end{equation}
By choosing $\hat\rho(r)=\rho^+_\beta$, we have $\hat\psi
=\rho^+_\beta$  and \eqref{p4.3.1.10} follows from
\eqref{p4.3.1.11}. \qed

\vskip1cm

{\bf Remark.} The same conclusions hold for the
interpolating free energy functionals
$F_{\beta,u}^{\pm}(\rho_\La|\rho_{\La^c})$. In this case the r.h.s.\ of
\eqref{aeD.17} becomes
 \begin{eqnarray*}
  -\beta u\int_0^1  e^{-\beta \big(
u j* e'_{\la_\beta}(j*\psi_s)
+(1-u)e'_{\la_\beta}(\rho_\beta^+)\big)}
j*\Big(e''_{\la_\beta}(j*\psi_s)j*(\psi-\psi')\Big) ds
  \end{eqnarray*}
for which the same bounds as in the previous proof apply.

\vskip3cm

\setcounter{equation}{0}

\section {\bf Non local functionals, large deviations}
\label{sec:D}

       \begin{lemma}
       \label{lemmaD.1}
For any $\rho \in L^\infty(\mathbb R^d;[0, X_0])$, $X_0>0$,
e.g. as in \eqref{1e.7z},
    \begin{equation}
     \label{D.1}
  s(j*\rho(r))- j*s(\rho)(r)\; \geq\; \frac{1}{2X_0} \int
j(r,r')\Big(\rho(r')-j*\rho(r)\Big)^2dr'
     \end{equation}

     \end{lemma}

\vskip.5cm

{\bf Proof.}  According to the integral form of the Taylor
reminde,r one can express for any $x,y \in \mathbb{R}$
\[
s(y)-s(x)=s'(y)(y-x)+\int_x^y s''(z) (x-z) dz
\]
For $x,y \in [0,X_0]$ one can bound this integral below by
\[
\geq s'(y)(y-x)- \frac{1}{X_0} \int_x^y (x-z) dz =
s'(y)(y-x) + \frac{1}{2 X_0} (x-y)^2
\]
Hence putting $y=j*\rho(r),\ x=\rho(r')$ one obtains the
required result having in mind that
\[
\int j(r,r') s'(j*\rho(r)) \left( j*\rho(r) - \rho(r')
\right) dr' =0.
\]
\qed

\vskip1cm

In the next theorem  we will use the following notation.

$\bullet$\;  Given $\zeta$ and $\ell$. Let $\La$ denotes a
bounded, $\mathcal D^{(\ell)}$-measurable set, such that
the maximal connected components of $\La^c$ are at mutual
distance $>2$. By $\chi_{\La^c}(r)$, we denote  a
function on $\La^c$, which is constant on each of the maximal connected components
of $\La^c$ and this constant is either equal to
$\rho_{\beta}^{+}$ or to $\rho_{\beta}^{-}$. We will then
call $\La_{\pm}^c$ the regions where
$\chi_{\La^c}(r)=\rho_{\beta}^{\pm}$.

$\bullet$\; Given $\rho_\La\in L^\infty(\La;\mathbb R_+)$,
we associate families $\{C^0_i, i\le N_0\}$ and $
\{C^{\pm}_j, j\le N_{\pm}\}$, with $N_0, N_{\pm} \geq 0$, as follows
    \begin{eqnarray}
     \label{D.8b}
&& \{C^0_i\} := \Big\{ C^0_i\in \mathcal D^{(\ell)}:
\text{$C^0_i\subset \La$;\; $\int _\La j(r,r')\,dr = 1$\;
for all $r'\in C^0_i$;} \nonumber \\&& \hskip4cm \text{
$\eta^{(\zeta,\ell)}(\rho_\La;r)=0$, $r\in C^0_i$}\Big\}
     \end{eqnarray}
and
    \begin{eqnarray}
 \{C^{\pm}_j\} &:=& \Big\{ (C^+_j,C^-_j)\in \mathcal D^{(\ell)} \times
  \mathcal D^{(\ell)}: \text{ $[C_j^{+}\cup C_j^-] \subset \La$;  $C_j^-
\subset \delta_{\text{{\rm out}}}^\ell[C_j^+]$} . \nonumber
\\&& \hskip.5cm
 \text{
 $\eta^{(\zeta,\ell)}(\rho_\La;r)=\pm
1$ and $\int _\La j(r',r)\,dr' = 1$, %
for all $r\in C_j^{\pm}$;}
 \nonumber
\\&& \hskip.5cm
 \text{ $[C_j^{+}\cup C_j^-]\cap [C_k^+\cup C_k^-]=\emptyset$ if $j\ne k
 \Big \}$}
     \label{D.8c}
     \end{eqnarray}

 \vskip1cm

       \begin{thm}
       \label{thmD.1}
There are $\zeta_1 $, $d_1$  and $c$ all positive such that
for any $\zeta<\zeta_1$ and $\ell\le d_1\zeta^2$ the
following holds: Let $\La$ be a bounded, $\mathcal
D^{(\ell)}$-measurable region,
$\rho_\La \in L^\infty(\La;[0, X_0])$, then
    \begin{equation}
     \label{D.5}
F_{\beta,\la_\beta}(\rho_\La|\chi_{\La^c}) \ge
f_{\beta,\la_\beta}(\rho_\beta^+) |\La|+ I_{\La,\La^c_-}^-
+I_{\La,\La^c_+}^+
  + c \zeta^2
\ell^d\big( N_0 + N_{\pm} \big) ,
     \end{equation}
where $I^{\pm}_{\La,\Delta}$ is defined in \eqref{dA.9}.
     \end{thm}

Recall that $f_{\beta,\la_\beta}(\rho_\beta^+)=f_{\beta,\la_\beta}(\rho_\beta^-)$).

\vskip.5cm

{\bf Proof.}  We will use the following shorthands
     \begin{equation}
     \label{D.7}
 \mathcal R:=j*(\rho_\La+\chi_{\La^c}^+),\qquad
  D:= \Big\{ r\in \La: |\mathcal R(r) - \rho_\beta^{\pm}| >\frac \zeta
  2\Big\} .
     \end{equation}
Recalling that $F_{\beta,\la_\beta}(\rho_\La|\chi_{\La^c})=
F_{\beta,\la_\beta}(\rho_\La+\chi_{\La^c})-F_{\beta,\la_\beta}(\chi_{\La^c})$,
we write
     \begin{eqnarray*}
&&F_{\beta,\la_\beta} (\rho_\La|\chi_{\La^c}) = F_1+F_2 +
F_3,
\end{eqnarray*}
where
\begin{eqnarray*}
&& F_1:= \int_{\La} f_{\beta,\la_\beta}( \mathcal R)\; dr,\qquad F_2
:= \frac 1 \beta\int_{\La} s\big( \mathcal R\big) - j*
s(\rho_\La+\chi_{\La^c})\;dr
\\ && F_3:= \int_{\La^c} f_{\beta,\la_\beta}( \mathcal R)-
f_{\beta,\la_\beta}({j*\chi_{\La^c}})\; dr+  \frac 1
\beta\int_{\La^c} s\big( \mathcal R\big) - j*
s(\rho_\La+\chi_{\La^c})\;dr
\\&&\hskip1cm -\int_{\La} f_{\beta,\la_\beta}({j*\chi_{\La^c}})\; dr-
 \frac 1 \beta\int s(j*\chi_{\La^c})- j* s(\chi_{\La^c})\;dr .
     \end{eqnarray*}
We claim that
    \begin{equation}
     \label{D.5c}
F_{3}  \geq I_{\La,\La^c_-}^- +I_{\La,\La^c_+}^+,
      \end{equation}
and that there are constants 
$a$ and $a'$ positive so that
    \begin{equation}
     \label{D.5d}
F_{2}  \geq  \frac{a}{16X_0} \zeta^2 N_0\ell^d - c_1\zeta^2
|D| + \frac{a'}{16X_0} \zeta^2 N_{\pm}\ell^d.
      \end{equation}
This will be proven below. 
Since there is a positive constant $c_1$ such that
    \begin{equation}
     \label{D.8}
F_1 \ge  f_{\beta,\la_\beta}(\rho_\beta^+)|\La|+ c_1\zeta^2
|D|
     \end{equation}
the bound \eqref{D.5} follows.

\noindent {\it Proof of \eqref{D.5c}.}  Using that: 

\noindent $\bullet$\; the second
integral in the definition of $F_3$ is non negative;

\noindent $\bullet$\; $f_{\beta,\la_\beta}( \mathcal R)
\geq f_{\beta,\la_\beta}(\rho_\beta^{\pm})$;  

\noindent $\bullet$\;
$s(j*\chi_{\La^c})=s(j*\rho_\beta^+\text{\bf 1}_{\La_+^c})
+ s(j*\rho_\beta^-\text{\bf 1}_{\La_-^c})$,

\noindent $\bullet$\; the sets $\La_+^c$ and $\La_-^c$ are disjoint;

\noindent we can bound 
    \begin{eqnarray*}
F_3 &\ge & \int_{\La^c_+}
f_{\beta,\la_\beta}(\rho_\beta^+)-
f_{\beta,\la_\beta}({j*\rho_\beta^+\text{\bf
1}_{\La_{+}^c}}) \;dr + \int_{\La^c_-}
f_{\beta,\la_\beta}(\rho_\beta^-)-
f_{\beta,\la_\beta}({j*\rho_\beta^-\text{\bf
1}_{\La_{-}^c}})\;dr \\&& - \int_{\La}
f_{\beta,\la_\beta}({j*\rho_\beta^+\text{\bf
1}_{\La_{+}^c}})\;dr - \int_{\La}
f_{\beta,\la_\beta}({j*\rho_\beta^-\text{\bf
1}_{\La_{-}^c}})\;dr\\&& - \frac{1}{\beta} \int
s(j*\rho_\beta^+\text{\bf 1}_{\La_+^c}) -j*s(
\rho_\beta^+\text{\bf 1}_{\La_+^c})\;dr  - \frac{1}{\beta}
\int s(j*\rho_\beta^-\text{\bf 1}_{\La_-^c}) -j*s(
\rho_\beta^-\text{\bf 1}_{\La_-^c})\;dr
     \end{eqnarray*}
which yields \eqref{D.5c} by the definition
\eqref{dA.9}.

\vskip.5cm

\noindent{\it Proof of \eqref{D.5d}.} Lemma \ref{lemmaD.1} implies that 
    \begin{eqnarray}
 F_2 &\ge &  \sum_{i\le N_0} S(C_i^0) + \sum_{j\le N_{\pm}}
 S(C_j^{\pm}),
 \nonumber
\end{eqnarray}
where
\begin{eqnarray}
S(C_i^0) &=&
 \frac{1}{2X_0} \int_\La \;dr \int_{C_i^0}
j(r,r')\Big(\rho(r')-\mathcal R(r)\Big)^2 \,dr'   \nonumber
\\  S(C_j^{\pm}) &=&
\frac{1}{2X_0} \int_\La \;dr \int_{C_j^+ \cup C_j^-}
j(r,r')\Big(\rho(r')-\mathcal R(r)\Big)^2 .
     \label{D.8a}
     \end{eqnarray}

Let us first prove the bound for $\sum_{i\le N_0} S(C_i^0)$.
Recalling the notations \eqref{d0.3} and \eqref{d0.2}, Lemma~\ref{lemmad0.1} yields
$|j(r,r')-j^{(\ell)}(r,r')|\le A^{(\ell)}(r,r')$. Hence 
Assumption~\eqref{d0.1} implies that 
    \begin{equation*}
 \int dr A^{(\ell)}(r,r') \le c_3 \ell .
     \end{equation*}
Since $(\rho(r')-\mathcal R(r))^2\le X_0^2$, we can estimate 
    \begin{equation*}
S(C_i^0) \ge  \frac{1}{2X_0} \int_\La dr \int_{C_i^0}
j^{(\ell)}(r,r')\Big(\rho(r')-\mathcal R(r)\Big)^2 dr' -
c_3 \frac {X_0}2\ell \ell^d .
     \end{equation*}
Recalling that $j^{(\ell)}(r,r')$ is constant as function of $r'$ in
$C_i^0$, one obtains  by Cauchy-Schwartz 
    \begin{equation*}
|C_i^0| \int_{C_i^0} \Big(\rho(r')-\mathcal R(r)\Big)^2 dr' \ge
\Big( \int_{C_i^0} [\rho(r')-\mathcal R(r)]dr' \Big)^2 .
     \end{equation*}
For $r\in D^c$, it holds
    \begin{equation*}
\Big( \int_{C_i^0} \rho(r')-\mathcal R(r)\Big)^2 dr' \ge \big
(|C_i^0| \frac \zeta 2\big)^2 .
     \end{equation*}
Hence, calling $r_i$ the center of $C_i^0$, we get 
    \begin{equation}
     \label{D.9}
 S(C_i^0)\ge  \frac{\zeta^2}{8X_0}  \ell^d \int_{ D^c\cap \La}
 j^{(\ell)}(r,r_i)dr- c_3\frac {X_0}2\ell \ell^d .
     \end{equation}

Then one can estimate below
    \begin{eqnarray}
     \label{D.9kk}
  \frac{ \zeta^2}{8X_0}  \ell^d \int_{ D^c\cap \La}
 j^{(\ell)}(r,r_i)dr 
&\geq & \frac{\zeta^2}{8X_0}\ell^d  -  \frac{\zeta^2}{8X_0}
 \int_D dr \int_{C_i^0}
 j^{(\ell)}(r,r') dr',
     \end{eqnarray}
where we used that $\int_{\La}
 j^{(\ell)}(r,r')\,dr =1$, $r'\in C^0_i$, which holds by the
 definition of the family
$\{C^0_i\}$.  
Let $a\in (0,1)$ be such that
   \begin{equation}
     \label{D.10}
 \frac{a }{8X_0}  \le \frac{c_1}{2},\qquad
 \text{$c_1$ as in \eqref{D.8}}.
     \end{equation}
Then \eqref{D.9kk} can be bounded by 
\begin{equation}
\frac{a}{8X_0} \zeta^2 \ell^d - \frac{c_1}{2} \zeta^2|D| \int_{C_i^0}
 j^{(\ell)}(r,r') dr'.
\end{equation}

Summarizing, we obtain
    \begin{equation}
     \label{D.11}
 \sum_{i=1}^{N_0} S(C_i^0)\ge \Big( \frac{a}{8X_0} \zeta^2 -
 c_3\frac {X_0}2\ell\Big) N_0 \ell^d   - \frac{c_1}{2} \zeta^2|D|.
     \end{equation}
Choose $d_1$ such that
    \begin{equation*}
 c_3\frac {X_0}2d_1\zeta^2\le \frac{a}{16X_0} \zeta^2
     \end{equation*}
the first part of \eqref{D.5d} is proven.

To bound $S(C_j^{\pm})$ we proceed as before.  By an
 abuse of notation, we denote by
$j^{(\ell)}(r,r')$ the quantity defined by averaging
$j(r,r'')$ over $r''\in C_j^{+}\cup C_j^{-}$ instead of a
single cube.  By assumption \eqref{d0.1}, there is a
constant $c_4$ so that
    \begin{equation*}
 \int dr A^{(\ell)}(r,r') \le c_4 \ell,\qquad r' \in C_j^{+}\cup
 C_j^{-} .
     \end{equation*}
Hence, we get
    \begin{equation*}
S(C_j^{\pm}) \ge  \frac{1}{2X_0} \int dr \int_{C_j^+\cup
C_j^- } j^{(\ell)}(r,r')\Big(\rho(r')-\mathcal R(r)\Big)^2
dr' - c_4 \frac {X_0}2\ell 2\ell^d .
     \end{equation*}
Using that  $j^{(\ell)}(r,r')$ is constant when $r'$ varies in
$C_j^+\cup C_j^-$, one can move it out of the integral with respect to
$r'$. Denoting by $r_j^{\pm}$ the
centers of $C_j^{\pm}$ and calling, one can bound
 $r^*= r_j^--r_j^+$,
    \begin{equation*}
 \int_{C_j^+\cup C_j^- }
\Big(\rho(r')-\mathcal R(r)\Big)^2 dr'\ge \frac 12
\int_{C_j^+} \Big(\rho(r')-\rho(r'+r^*)\Big)^2 dr' .
     \end{equation*}
The Cauchy-Schwartz inequality implies 
    \begin{equation*}
|C_j^+| \int_{C_j^+} \Big(\rho(r')-\rho(r'+r^*)\Big)^2 dr'
\ge \Big( \int_{C_j^+} \rho(r')dr' -\int_{C_j^-} \rho(r')
dr' \Big)^2 .
     \end{equation*}
The analysis of $S(C_j^{\pm})$ proceeds hereafter as in the case of 
$S(C_i^0)$ yielding
    \begin{eqnarray}
 \sum_{j\le N_{\pm}}S(C_j^{\pm}) &\geq &   \frac{a'}{16X_0} \zeta^2
N_{\pm}\ell^d {-\frac{c_1}{2}\zeta^2|D|}
     \label{D.12}
     \end{eqnarray}
for a suitable $a'$.

Thus
\eqref{D.5d} and hence the theorem is proven.
%
%

 \qed

 \vskip3cm

\setcounter{equation}{0}
\section {\bf Reduction to variational problems}
\label{app:aC}

Following  the classical proof by Lebowitz and Penrose,
\cite{LP}, we will derive here upper and lower bounds on
the partition function in terms of a variational problem
which involves a non local free energy functional.

We fix a $\beta\in (\beta_c,\beta_0)$ and  a $\la \in
(\la_\beta- 1,\la_\beta+ 1)$. We will frequently drop $\beta$
from the notation. We consider constrained partition
functions generically denoted by
   \begin{equation}
     \label{aC.1}
Z_{\ga,\la}(\La; \mathcal A |\bar\q)
    \end{equation}
where $\La$ is a bounded $\mathcal
D^{(\ell_{+,\ga})}$-measurable region and $\bar\q$ a
boundary condition.  We will suppose that all loops in
$\bar\q$ are short, that $\bar\q(0) \subset \La^c$ and
   \begin{equation}
     \label{aC.2}
\ell_{-,\ga}^{-d} \Big| \bar\q(0)\cap
C_r^{(\ell_{-,\ga})}\Big| = \text{{\rm
Av}}^{(\ell_{-,\ga})}(\bar \q(0); r) \le X_{0}.
     \end{equation}
The constraint $\mathcal A$ is determined by a $\mathcal
D^{(\ell_{-,\ga})}$-measurable, $\{0,\pm 1\}$-valued
function $\eta_{\text{\rm cs}}(r)$, $r\in \La$, as follows
   \begin{equation}
     \label{aC.3}
\mathcal A= \Big\{\q(0): \eta^{(\zeta,\ell_{-,\ga})}(
\q(0); r) = \eta_{\text{\rm cs}}(r); \;\;\text{{\rm
Av}}^{(\ell_{-,\ga})}(\q(0); r) \le X_{0},\;\; r\in \La
\Big\} .
    \end{equation}
The transition to continuum involves two spatial scales,
whose lengths are $\ga^{-1/2}$ and $\ga^{-1}$.  We set
   \begin{equation}
     \label{aC.4}
\bar\rho(r):= \text{{\rm Av}}^{(\ga^{-1/2})}(\bar q(0);
\ga^{-1}r), \qquad r \in (\ga \La)^c
    \end{equation}
and
   \begin{eqnarray}
&&\hskip-1cm\mathcal A^*:= \Big\{ \rho \in L^\infty(\ga
\La,\mathbb R_+): \eta^{(\zeta,\ga\ell_{-,\ga})}(\rho; r)=
\eta_{\text{\rm cs}}(\ga^{-1}r), \nonumber\\&& \hskip5cm
 \text{{\rm Av}}^{(\ga\ell_{-,\ga})}(\rho;r) \le
X_{0}, \; r\in \ga\La  \Big\} .
     \label{aC.5}
    \end{eqnarray}

\vskip1cm

  \begin{prop}
 \label{propaB.1}
There is a constant $c$ such that one has
    \begin{eqnarray}
&& \hskip-1cm  \log Z_{\ga,\la}(\La; \mathcal A |\bar\q)
\le - \inf_{\rho \in \mathcal A^*} \beta \ga^{-d}
F_{\beta,\la, \ga\La }(\rho|\bar\rho)  + c\ga^{1/2}
\big(|\La|+ |\partial _{4\ga^{-1}}\La|\big),
     \label{aC.6}
    \end{eqnarray}
where $\partial_R\La$ is defined as in \eqref{d0.9}; while,
for any $\rho \in \mathcal
A^*$ it holds that 
    \begin{eqnarray}
&& \hskip-1cm  \log Z_{\ga,\la}(\La; \mathcal A |\bar\q)
\ge - \beta \ga^{-d} F_{\beta,\la, \ga\La }(\rho|\bar\rho)
- c\ga^{1/2} (|\La|+|\partial _{4\ga^{-1}}\La|) .
     \label{aC.7}
    \end{eqnarray}

  \end{prop}

\vskip.5cm

{\bf Proof.}  Upper bound.  By Proposition \ref{propaA.1}
one can show that
     \begin{eqnarray*}
&&  \log Z_{\ga,\la}(\La; \mathcal A |\bar\q) \le  \log
Z^{(\ga^{-1/2})}_{\ga,\la}(\La; \mathcal A |\bar\q(0)) +
c\ga^{1/2} |\Delta|
    \end{eqnarray*}
where $\Delta = \La \cup  \partial
_{4\ga^{-1}}\La$ and
$Z^{(\ga^{-1/2})}_{\ga,\la}$ denotes the partition
function for the classical model, i.e.\ for particles and
not for loop configurations with Hamiltonian
$h^{(\ga^{-1/2})}_{\ga,\la}(\q(0)|\bar\q(0))$, that is the same as
\eqref{2.a2} with $j_\ga$ replaced by $j_\ga^{(\ga^{-1/2})}(r,r')$; the latter obtained from
$j_\ga(r,r')$ by averaging the second variable over the
cubes of $\mathcal D^{(\ga^{-1/2})}$. The Hamiltonian
$h^{(\ga^{-1/2})}_{\ga,\la}(q|\bar\q(0))$ depends on $q$
and $\bar\q(0)$ only via the number of particles in each
of the cubes of $\mathcal D^{(\ga^{-1/2})}$. We then
integrate over the positions of the particles $\q(0)$
keeping  fixed the particle numbers in the cubes $\mathcal D^{(\ga^{-1/2})}$.  By the
Stirling formula,
     \begin{eqnarray*}
&& \big| \log n! - n(\log n -1) \big| \leq c \log n  
    \end{eqnarray*}
with $c$ a suitable constant.  The number of particles in a
cube of $\mathcal D^{(\ga^{-1/2})}$ in  $\Delta$   is
bounded by $X_0 \ell_{-,\ga}^d$, due to the constraint
$\mathcal A$ and the assumptions on $\bar \q(0)$.  Thus,
calling $ \mathcal A^*_d$ the collection of all $\rho\in
\mathcal A^*$ which are $\mathcal
D^{(\ga^{1/2})}$-measurable and such that for all $r$,
$\rho(r) \ga^{-d/2}$ is an integer, there is a constant $c$
so that,
     \begin{eqnarray*}
&&  Z_{\ga,\la}(\La; \mathcal A |\bar\q) \le  \sum_{\rho
\in
 \mathcal A^*_d} \exp\Big\{-\beta\ga^{-d} F^{(\ga^{1/2})}_{\beta,\la,\ga\La}(\rho|\bar \rho) +
c\Big(\ga^{1/2} + \ga^{d/2} \log
\ga^{-1}\Big)|\Delta|\Big\} ,
    \end{eqnarray*}
where $F^{(\ga^{1/2})}$ is the functional $F$ where the kernel $j_\ga(r,r')$
is replaced by $j^{(\ga^{1/2})}(r,r')$.  Since
$F^{(\ga^{1/2})}_{\beta,\la,\ga\La}= F_{\beta,\la,\ga\La} $
on $\mathcal D^{(\ga^{1/2})}$-measurable functions, the
last inequality holds with $F_{\beta,\la,\ga\La} $ as well.

Since
     \begin{eqnarray*}
&& \text{Card
 $(\mathcal A^*_d)$}  \leq  [X_0  \ell_{-,\ga}^d]^{|\La|/
 \ga^{-d/2}}
    \end{eqnarray*}
we have, for  a suitable  constant $c$,
    \begin{eqnarray*}
&& \log Z_{\ga,\la}(\La; \mathcal A |\bar\q) \le
-\beta\ga^{-d} \inf_{\rho \in \mathcal A^*}
F_{\beta,\la,\ga\La}(\rho|\bar \rho) + c\Big(\ga^{1/2} +
\ga^{d/2} \log \ga^{-1}\Big)|\Delta|
    \end{eqnarray*}
which proves \eqref{aC.6}.

\vskip.5cm

Lower bound.  Let $\rho \in \mathcal A^*$, $\rho\leq X_0$
and associate $\hat \rho\in \mathcal A^*_d$, where $\hat \rho(r)=n
\ga^{d/2}$ whenever $\rho(r)\in [(n-1/2) \ga^{d/2},(n+1/2)
\ga^{d/2})$.  Then, as before
    \begin{eqnarray*}
&& \log Z_{\ga,\la}(\La; \mathcal A |\bar\q) \ge  -\beta
\ga^{-d} F_{\beta,\la,\ga\La}(\hat\rho|\bar \rho) -
c\Big(\ga^{1/2} + \ga^{d/2} \log \ga^{-1}\Big)|\Delta| .
    \end{eqnarray*}
The bound  \eqref{aC.7} then
follows from
    \begin{eqnarray*}
&& \ga^{-d}\Big| F_{\beta,\la,\ga\La}(\hat\rho|\bar \rho) -
F_{\beta,\la,\ga\La}( \rho|\bar \rho)\Big| \le c \ga^{d/2}
\log \ga^{-1}|\Delta| .
    \end{eqnarray*}
\qed

\vskip1cm

{\bf Remark.} Proposition \ref{propaB.1} extends
 to the case of the interpolating Hamiltonian
of Section \ref{sec:10} because the reference Hamiltonian
is one-body and for it, the transition to continuum is
trivial.

\vskip1cm

 Using instead of Proposition~\ref{propaA.1} either Corollary~\ref{Ku129} or
Corollary~\ref{corolaA.1} we obtain the following variants
of Proposition~\ref{propaB.1}:

\begin{corol}
\label{Ku130} There is a constant $c$ such that,
    \begin{eqnarray}
&& \hskip-1cm  \log Z_{\ga,\la}(\La; \mathcal A |\bar\q)
\le - \inf_{\rho \in \mathcal A^*} \beta \ga^{-d}
F_{\beta,\la, \ga\La }(\rho|\bar\q)  + c\ga^{1/2}
\big(|\La|+ |\partial _{4\ga^{-1}}\La|\big),
    \end{eqnarray}
where $\partial_R\La$ is defined in \eqref{d0.9}; while,
for any $\rho \in \mathcal
A^*$, it holds that
    \begin{eqnarray}
    \label{Ku132}
&& \hskip-1cm  \log Z_{\ga,\la}(\La; \mathcal A |\bar\q)
\ge - \beta \ga^{-d} F_{\beta,\la, \ga\La }(\rho|\bar\q)  -
c\ga^{1/2} (|\La|+|\partial _{4\ga^{-1}}\La|) .
    \end{eqnarray}
\end{corol}

\begin{corol}
\label{Ku131} There is a constant $c$ such that,
    \begin{eqnarray}
    \label{Ku133}
&& \hskip-1cm  \log Z_{\ga,\la}(\La; \mathcal A
|\rho_{\beta,\la}^\pm) \le - \inf_{\rho \in \mathcal A^*}
\beta \ga^{-d} F_{\beta,\la, \ga\La
}(\rho|\rho_{\beta,\la}^\pm)  + c\ga^{1/2} \big(|\La|+
|\partial _{4\ga^{-1}}\La|\big) ,
    \end{eqnarray}
where $\partial_R\La$ is defined in \eqref{d0.9}; while,
for any $\rho \in \mathcal
A^*$, it holds that
    \begin{eqnarray}
&& \hskip-1cm  \log Z_{\ga,\la}(\La; \mathcal A
|\rho_{\beta,\la}^\pm) \ge - \beta \ga^{-d} F_{\beta,\la,
\ga\La }(\rho|\rho_{\beta,\la}^\pm)  - c\ga^{1/2}
(|\La|+|\partial _{4\ga^{-1}}\La|).
    \end{eqnarray}
\end{corol}

 \vskip3cm

     \setcounter{equation}{0}

    \section {\bf Cluster expansion}
    \label{app:F}

In this appendix we recall some basic facts of cluster
expansion in the context of the contour models we have used
in this paper.  For simplicity we just refer to the plus
case, and consider the family $\{\Ga\}^+$ of all possible
bounded plus contours and the associated non negative weights  $w^+(\Ga)$,
for $\Ga\in \{\Ga\}^+$, which are bounded by $e^{-2c_f \ga^{-1} N_\Ga}$.
For instance,the special choice $w^+(\Ga)= {\hat
W}_{\ga,\la}^{\pm}(\Ga_i;\q)$, for $\q\in \Q_+$.

 \vskip1cm

\subsection{The abstract setup}

We regard  $\{\Ga\}^+$ as a graph, calling connected two
elements, $\Ga$ and $\Ga'$ of $\{\Ga\}^+$ if sp$(\Ga)\cap
$ sp$(\Ga')\ne \emptyset$.  We then define for any finite
set $\mathcal C \subset \{\Ga\}^+$:
      \begin{equation}
      \label{gg.1}
Z(\mathcal C,w):=\;\;\sum_{ \{\Ga_i\} \subset \mathcal C:
\Ga_i \text{is disconnected from} \;\Ga_j, i\ne j} \;\;
\prod_{i} w(\Ga_i),
      \end{equation}
which we regard as a function of the variables $w(\Ga)$.

Denote by $\mathcal I$ the collection of all multi-indices
$I$ on $\{\Ga\}^+$, i.e.\ $I$ is a $\mathbb N_+$-valued
function on  $\{\Ga\}^+$ such that $|I|:=\dis{\sum_{\Ga\in
\{\Ga\}^+} I(\Ga)<\infty}$.   Let $I\in\mathcal I$ and call
${mathcal C}_I :=\{\Ga: I(\Ga)>0\}$, and define
      \begin{equation}
      \label{gg.2}
\om_I = \frac{1}{I!} D^I \log Z(\mathcal C,w)\Big |_{w=0}
\; \prod_{\Ga\in \mathcal C}w(\Ga)^{I(\Ga)} ,
      \end{equation}
where $D^I$ denoted the product of partial derivatives with respect to
$w(\Ga)$, where one
differentiates $I(\Ga)$-times with respect to $w(\Ga)$, while $I! := \prod I(\Ga)!$.

 \vskip1cm

\subsection{The main result}

Using the above notation, and define $\om_I := 0$ unless ${\mathcal C}_I
=\{\Ga: I(\Ga)>0\}$ is connected.  Moreover, calling
$\dis{R(I):= \bigcup_{\Ga: I(\Ga)>0} \text{sp}(\Ga)}$,
then if $\ga$ is small enough,
      \begin{equation}
      \label{gg.3}
  \;\;\sum_{I: R(I) \ni r}
|\,\om_I| e^{ c_f \ga^{-1} |I|} \le 1
      \end{equation}
and, for any finite set $\mathcal C \subset \{\Ga\}^+$, it holds
      \begin{equation}
      \label{gg.4}
\log Z(\mathcal C,w) =\;\;\sum_{ I: R(I) \subset \mathcal
C }  \;\; \om_I .
      \end{equation}

 \vskip1cm

 Note that the number $\#(\Delta)$
 of contours $\Ga$ with sp$(\Ga)=\Delta$ is
bounded by $\big(3^{\ga^{-2\alpha d}}2\big)^{N_\Delta}$ and
since $2\alpha \ll 1$, we have that $\#(\Delta) e^{-2c_f
\ga^{-1} N_\Delta } \ll 1$ for $\ga$ small enough. Hence the inequality
\eqref{gg.3} follows from the general theory of cluster expansion, see for
instance
\cite{BZ2}.

%
%

 \vskip1cm

\subsection{Effective Hamiltonian}

Let $\La$ be a bounded $\mathcal
D^{(\ell_{+,\ga})}$-measurable region and $\q\in \Q_+$.
Then
     \begin{equation}
      \label{gg.7}
\sum_{ \{\Ga_i\} \subset \mathcal C^+_\La} \;\; \prod_{i}
{\hat W}_{\ga,\la}^{+}(\Ga_i;\q) =: e^{-K^{+m}_\La(\q_\La)}
      \end{equation}
defines an Hamiltonian $K^{+}_\La $ whose potentials
$U^{\pm}_\Delta$ are given by
     \begin{equation}
      \label{gg.8}
U^{\pm}_\Delta(\q_\Delta) =- \;\;\sum_{ I: R(I)=  \Delta }
\;\; \om_I ,
      \end{equation}
if $\Delta$ is connected, otherwise $U^{\pm}_\Delta=0$.
The bound \eqref{gg.3} implies that
     \begin{equation}
      \label{gg.9}
\|U^{\pm}_\Delta(\q_\Delta)\|_\infty = e^{-c_f \ga^{-1}
N_\Delta} 2^{N_\Delta} .
      \end{equation}

 \vskip1cm

In the remaining part of the appendix we state and prove
results used in the text.

 \vskip.5cm

    \begin{lemma}
    \label{lemmaa5.1}

There is $\ga^*>0$ so that  for any $\ga \leq \ga^*$ and
any bounded, $\mathcal{D}^{(\ell_{+,\ga})}$-measurable set
$\Delta$
      \begin{equation}
      \label{a5.5}
\log\left(\sum_{\{\Ga_i\}, \text{\rm sp}(\Ga_i) \cap
\Delta \neq \emptyset} \prod_{i} \left\| {\hat
W}_{\ga,\la}^{\pm}(\Ga_i;\cdot) \right\|_\infty \right)
\leq N_\Delta e^{- c_f\ga^{-1}}
      \end{equation}
where $N_\Delta$ is the number of
$\mathcal{D}^{(\ell_{+,\ga})}$
 cubes in $\Delta$, i.e.\ $N_\Delta \ell_{+,\ga}^d :=|\Delta|$.
     \end{lemma}

      \vskip.5cm

{\bf Proof.}  By definition \eqref{a4.19} it holds
      \begin{equation}
      \label{a5.6}
\left\| {\hat W}_{\ga,\la}^{\pm}(\Ga_i;\cdot)
\right\|_\infty \leq
 e^{- 2c_f\ga^{-1}N_{\Ga_i}} .
      \end{equation}
Moreover, for any  bounded,
$\mathcal{D}^{\ell_{+,\ga}}$-measurable set $D$ using that
$\ell_{+,\ga}/\ell_{-,\ga}= \ga^{-2\alpha}$, we obtain that
the cardinality of  $\{\Ga: \text{sp}(\Ga)=D\}$ is bounded by
      \begin{equation}
      \label{a5.7}
  \leq
\Big(3^{\ga^{-2\alpha d}}2\Big)^{N_D} .
      \end{equation}
Thus one can estimatte the exponent of the left hand side of \eqref
{a5.5} by
      \begin{equation}
      \label{a5.8}
\sum_{\{\Ga_i\}, \text{\rm sp}(\Ga_i) \cap
\Delta \neq \emptyset} \prod_{i} \left\| {\hat
W}_{\ga,\la}^{\pm}(\Ga_i;\cdot) \right\|_\infty \le \Big( 1+
\sum_{D\ni 0} \big(3^{\ga^{-2\alpha d}}2\big)^{N_D}  e^{-2
c_f\ga^{-1}N_{D}}\Big) ^{N_\Delta} .
      \end{equation}
Since $2\alpha d<1$ and if $\ga$ is small enough, we have
      \begin{equation}
      \label{a5.10}
 \sum_{D\ni 0} 3^{N_D
\ga^{-2\alpha d}}2^{N_D}  e^{- 2c_f\ga^{-1}N_{D}}\le  e^{-
c_f\ga^{-1}}
      \end{equation}
and one yields
      \begin{equation}
      \label{a5.11}
\sum_{\{\Ga_i\}, \text{\rm sp}(\Ga_i) \cap
\Delta \neq \emptyset} \prod_{i} \left\| {\hat
W}_{\ga,\la}^{\pm}(\Ga_i;\cdot) \right\|_\infty \le \Big( 1+  e^{-
c_f\ga^{-1} }\Big) ^{N_\Delta} \le \exp\{N_\Delta e^{-
c_f\ga^{-1} }\} .
      \end{equation}

The lemma is proved.            \qed

 \vskip1cm

Several times in the text the following
corollary of the above lemma is used.  We give
here a general formulation which covers all the different cases.
Let$\La$ is a bounded, $\mathcal
D^{(\ell_{-,\ga})}$-measurable region, and  considering,
for the sake of definiteness, the plus restricted ensemble,
let $\bar \q$ be a plus boundary condition outside $\La$
(the configuration $\bar \q$ may as well be replaced by
$\rho_\beta^+\text{\bf 1}_{\La^c}$). Let $\mathcal A$ be a set of a ``decreasing constraint'' on the contours, meaning that
if $\{\Ga_i\}\in \mathcal A$ and if $\{\Ga_{i_k}\}$ a subset
of $\{\Ga_i\}$ then $\{\Ga_{i_k}\} \in \mathcal A$; we also
suppose that the cardinality of $\mathcal A$ and the cardinality of its
elements are finite. Finally, the set $\Delta$ is a bounded,
$\mathcal D^{(\ell_{+,\ga})}$-measurable subset of $\La$.

 \vskip1cm

    \begin{lemma}
    \label{lemmaF.2}

Let  $\ga^*>0$ as in Lemma \ref{lemmaa5.1}. For all $\ga \leq
\ga^*$ and all $\La$, $\bar\q$, $\Delta$ and $\mathcal A$ as
above, one has
      \begin{equation}
      \label{F.8}
\frac{{\hat Z}^+_{\ga,\la}\big(\La;\{\mathcal A, \text{\rm
sp}(\und \Ga)\cap \Delta=\emptyset\}\big|\bar \q\big)}{
{\hat Z}^+_{\ga,\la}\big(\La; \mathcal A\big|\bar \q\big)}
\geq e^{-N_\Delta e^{- c_f\ga^{-1} }} .
      \end{equation}
     \end{lemma}

       \vskip.5cm

{\bf Proof.}  Denote by  $\Xi$ the denominator on the l.h.s.\ of
\eqref{F.8}.  Its expression involves a sum over
$\{\Ga'_j\}\in \mathcal A$. For each element $\{\Ga'_j\}\in
\mathcal A$,  call $\{\Ga_i\}$ the subset of all contours with
 sp$(\Ga_i) \cap \Delta \ne \emptyset$. One splits the sum over contours into
$\{\Ga_i\}$ and $\{\Ga_i'\} \setminus \{\Ga_i\}$. As $\mathcal{A}$ is
decreasing one has that $\{\Ga_i'\} \setminus \{\Ga_i\} \in \mathcal
A$ and hence we obtain
\eqref{F.8}.  Thus
     \begin{equation*}
\Xi \leq  \Big(\sum_{\{\Ga_i\}, \text{\rm sp}(\Ga_i) \cap
\Delta \neq \emptyset} \prod_{i} \left\| {\hat
W}_{\ga,\la}^{\pm}(\Ga_i;\cdot) \right\|_\infty \Big)\;
{\hat Z}^+_{\ga,\la}\big(\La;\{\mathcal A, \text{sp}(\und
\Ga)\cap \Delta=\emptyset\}\big|\bar \q\big)
      \end{equation*}
which, by \eqref{a5.5}, yields \eqref{F.8}.  The lemma is
proved. \qed


\newpage

\bibliographystyle{amsalpha}

\end{document}